\documentclass[hidelinks,11pt, letterpaper]{article}
\author{Mark Braverman \thanks{Department of Computer Science, Princeton University.
Research supported in part by the NSF Alan T. Waterman Award, Grant No. 1933331, a Packard Fellowship in Science and Engineering, and the Simons Collaboration on Algorithms and Geometry.}
\qquad Subhash Khot \thanks{Department of Computer Science, Courant Institute of Mathematical Sciences, New York University. Supported by
the NSF Award CCF-1422159, the Simons Collaboration on Algorithms and Geometry, and the Simons Investigator Award.}
\qquad Noam Lifshitz \thanks{Einstein Institute of Mathematics, Hebrew University, Jerusalem, Israel. Research supported in part by ERC advanced grant 834735.}
\qquad Dor Minzer \thanks{Department of Mathematics, Massachusetts Institute of Technology, Cambridge, USA. Supported by a Sloan Research
Fellowship, NSF CCF award 2227876 and NSF CAREER award 2239160.}}
\date{\vspace{-5ex}}
\usepackage{fullpage}
\usepackage{amsthm}
\usepackage{amsmath,amssymb,amsfonts,nicefrac}
\usepackage{xspace}
\usepackage{color}
\usepackage{url}
\usepackage{hyperref}
\usepackage{bm}
\usepackage{bbm}
\usepackage{times}
\usepackage[T1]{fontenc}

\usepackage{enumitem}

\newtheorem{thm}{Theorem}[section]

\newtheorem{rconj}[thm]{Conjecture}
\newtheorem{lemma}[thm]{Lemma}
\newtheorem{corollary}[thm]{Corollary}
\newtheorem{claim}[thm]{Claim}
\newtheorem{proposition}[thm]{Proposition}

\newtheorem{definition}[thm]{Definition}
\newtheorem{remark}[thm]{Remark}

\newtheorem{fact}[thm]{Fact}

\newcommand\card[1]{\left| {#1} \right|}
\newcommand\sett[2]{\left\{ \left. #1 \;\right\vert #2 \right\}}

\newcommand\set[1]{{\left\{ #1 \right\}}}
\newcommand\Prob[2]{{\Pr_{#1}\left[ {#2} \right]}}

\newcommand\cProb[3]{{\Pr_{#1}\left[ \left. #3 \;\right\vert #2 \right]}}

\newcommand\Expect[2]{{\mathop{\mathbb{E}}_{#1}\left[ {#2} \right]}}

\newcommand\cExpect[3]{{\mathop{\mathbb{E}}_{#1}\left[ \left. #3 \;\right\vert #2 \right]}}
\newcommand\norm[1]{\left\| #1 \right\|}
\newcommand\power[1]{\set{0,1}^{#1}}

\newcommand\half{{1\over2}}

\newcommand\skipi{{\vskip 10pt}}

\newcommand\inner[2]{\langle{#1},{#2}\rangle}
\newcommand\eps{\varepsilon}

\renewcommand\geq{\geqslant}
\renewcommand\leq{\leqslant}

\makeatletter
\newtheorem*{rep@theorem}{\rep@title}
\newcommand{\newreptheorem}[2]{%
\newenvironment{rep#1}[1]{%
\def\rep@title{\bf #2 \ref*{##1} \text{(Restated)} }%
\begin{rep@theorem} }%
{\end{rep@theorem} } }

\newtheorem*{rep@claim}{\rep@title}
\newcommand{\newrepclaim}[2]{%
\newenvironment{rep#1}[1]{%
\def\rep@title{\bf #2 \ref*{##1} \text{(Restated)} }%
\begin{rep@claim} }%
{\end{rep@claim} } }

\newtheorem*{rep@lemma}{\rep@title}
\newcommand{\newreplemma}[2]{%
\newenvironment{rep#1}[1]{%
\def\rep@title{\bf #2 \ref*{##1} \text{(Restated)} }%
\begin{rep@lemma} }%
{\end{rep@lemma} } }

\makeatother
\newreptheorem{theorem}{Theorem}
\newrepclaim{claim}{Claim}
\newreplemma{lemma}{Lemma}

\title{An Invariance Principle for the Multi-slice, with Applications}
\begin{document}
\maketitle
\begin{abstract}
Given an alphabet size $m\in\mathbb{N}$ thought of as a constant, and $\vec{k} = (k_1,\ldots,k_m)$ whose entries sum of up $n$,
the $\vec{k}$-multi-slice is the set of vectors $x\in [m]^n$ in which each symbol $i\in [m]$ appears precisely $k_i$ times.
We show an invariance principle for low-degree functions over the multi-slice, to functions over the product space $([m]^n,\mu^n)$
in which $\mu(i) = k_i/n$. This answers a question raised by~\cite{FKMW}.

As applications of the invariance principle, we show:
\begin{enumerate}
  \item An analogue of the ``dictatorship test implies computational hardness'' paradigm for problems with perfect completeness,
  for a certain class of dictatorship tests. Our computational hardness is proved assuming a recent strengthening of the Unique-Games
  Conjecture, called the Rich $2$-to-$1$ Games Conjecture.

  Using this analogue, we show that assuming the Rich $2$-to-$1$ Games Conjecture, (a) there is an $r$-ary CSP $\mathcal{P}_r$
  for which it is NP-hard to distinguish satisfiable instances of the CSP and instances that are at most $\frac{2r+1}{2^r} + o(1)$
  satisfiable, and (b) hardness of distinguishing $3$-colorable graphs, and graphs that do not contain an independent set of
  size $o(1)$.
  \item A reduction of the problem of studying expectations of products of functions on the multi-slice
  to studying expectations of products of functions on correlated, product spaces. In particular, we are able to
  deduce analogues of the Gaussian bounds from~\cite{MosselGaussian} for the multi-slice.
  \item In a companion paper, we show further applications of our invariance principle in extremal combinatorics,
  and more specifically to proving removal lemmas of a wide family of hypergraphs $H$ called $\zeta$-forests, which is
  a natural extension of the well-studied case of matchings.
\end{enumerate}
\end{abstract}
\section{Introduction}
The invariance principle of Mossel, O'Donnell and Oleszkiewicz~\cite{MOO} is a powerful analytical tool with wide range of applications
throughout theoretical computer science, discrete mathematics and combinatorics.\footnote{Earlier versions of the invariance
principle have been proved earlier by Rotar~\cite{Rotar} as well as by Chatterjee~\cite{Chatterjee}, however they are less relevant
with respect to the applications to PCP's, which are a key point of this paper.} Initially motivated by proving optimal hardness of approximation result
for Max-Cut~\cite{KKMO} (assuming the Unique-Games Conjecture~\cite{Khot}), this technique has become ubiquitous in analysis of PCP constructions,
and also has many other significant applications in different areas of discrete mathematics.

The invariance principle has been since extended and generalized in several ways: for general product spaces and operators~\cite{MosselGaussian},
under weaker assumptions~\cite{MosselResilient}, and to a limited extent also beyond product spaces~\cite{FKMW,FM2019}. These extensions
are crucial in the analysis of many PCP constructions (especially based on the Unique-Games Conjecture),
a prime example being Raghavendra's result regarding the optimality of semi-definite algorithms for CSPs~\cite{Rag}. They
are also useful in other areas such as coding theory~\cite{coding}, learning theory~\cite{learning}, property
testing~\cite{testing}, as well as in combinatorics~\cite{DFR,Lifshitz,KLLMcodes}.

The main goal of this paper is to establish an invariance principle for the multi-slice, a generalization of the Boolean slice
for which an invariance principle was recently established~\cite{FKMW,FM2019}. We also give applications of this invariance principle
to hardness of approximation, and extremal combinatorics.

\subsection{The multi-slice}
Much of this paper is devoted to the study of multi-slices, defined below.
\begin{definition}
  Let $n\in\mathbb{N}$ and $\Sigma = [m]$ be a finite alphabet.
  For a tuple of non-negative integers $\vec{k} = (k_1,\ldots,k_m)$ that add up to $n$,
  we define the multi-slice $\mathcal{U}_{\vec{k}}\subseteq [m]^n$ as the set
  of vectors $x\in[m]^n$ that for all $i\in[m]$, have exactly $k_i$ of their
  coordinates equal to $i$.
\end{definition}
We will mostly be concerned with $\alpha$-balanced multi-slices, which are multi-slices
$\mathcal{U}_{\vec{k}}\subseteq[m]^n$ in which $k_i\geq \alpha n$ for all $i\in [m]$. Here and throughout,
the alphabet $m$ and the parameter $\alpha$ are thought of as constants.\footnote{There are natural extensions of our results
to the case that $m$ is not necessarily a constant but is allowed to grow with $n$, but we omit them from this paper.}
Given a multi-slice
$\mathcal{U}_{\vec{k}}$, it is natural to consider its product analogue, i.e.~$[m]^n$ along with the product
measure $\nu_{\vec{k}}^{\otimes n}$ defined by $\nu_{\vec{k}}(i) = \frac{k_i}{n}$, and ask how different do
the domains $(\mathcal{U}_{\vec{k}},{\sf Uniform})$ and $([m]^n,\nu_{\vec{k}}^{\otimes n})$ behave.
More precisely, we will be interested in the following type of invariance between the two domains:
given a function $f\colon \mathcal{U}_{\vec{k}}\to\mathbb{R}$, can one associate with $f$ a function $\tilde{f}\colon [m]^n\to\mathbb{R}$,
such that $f$ and $\tilde{f}$ are very close to each other? For the question to make sense, we must of course define what ``close to each other'' even
means -- these are functions over two different domains, so distance measures such as the $L^2$ distance do not make sense.
To define things more precisely, we note that a typical point ${\bf y}\sim\nu_{\vec{k}}^{\otimes n}$
is quite close to a point in the multi-slice -- its distance from it is typically $O_{m}(\sqrt{n})$. Thus, it makes sense to consider a coupling between
the two domains, say $({\bf x},{\bf y})$, such that ${\bf x}$ is (marginally) a uniform point from the multi-slice, ${\bf y}$ is (marginally) distributed
according to $\nu_{\vec{k}}^{\otimes n}$, and typically the Hamming distance between ${\bf x}$ and ${\bf y}$ is $O_{m}(\sqrt{n})$. In this language,
we will say that $f$ and $\tilde{f}$ are close if $f({\bf x}) \approx \tilde{f}({\bf y})$ where we sample ${\bf x},{\bf y}$ according to the above coupled
distribution.\footnote{For our applications, we will need additional natural properties from the associated function $\tilde{f}$.
These properties are mainly concerned with the notion of ``influence'' and ``noisy-influence'' of coordinates,
and we will want them to be roughly preserved.}

Unfortunately, one cannot hope for a statement of this great generality. The issue is that ``high-degree'' functions may be able to
distinguish the multi-slice from its product analogue. To be more concrete, take any $\mathcal{A}\subseteq {[n] \choose k_1}$ of size $\half {n\choose k_1}$,
and consider the function $f(x) = \sum\limits_{A\in\mathcal{A}} 1_{x_A = 1^{k_1}}$. Then there is no $\tilde{f}$ that is close to
$f$ in the above sense -- the issue is that there is no good choice for $\tilde{f}(y)$ when the number of $1$'s in
$\tilde{f}$ is different from $k_1$ (which happens with probability $1-o(1)$).

Our first result asserts that ``high degree'' functions are in fact the only obstruction to an invariance between the multi-slice and its
product analogue. Namely, we show that if $f$ is a low-degree function, then one may find $\tilde{f}$ as above. To make a formal statement, we require a few definitions.
First, we formally define the degree of a function on the multi-slice.
\begin{definition}
  For a multi-slice $\mathcal{U}_{\vec{k}}$ over alphabet $[n]$, we say a function $f\colon\mathcal{U}_{\vec{k}}\to\mathbb{R}$ is a $d$-junta if there exists a set of coordinates
  $A\subseteq[n]$ of size at most $d$, such that $f(x) = g(x_{A})$ for some function $g\colon[m]^A\to\mathbb{R}$.
\end{definition}

We can now define the space of degree $d$ functions. We will use expectation inner products throughout the paper, i.e.~for
$f,g\colon\mathcal{U}_{\vec{k}}\to\mathbb{R}$ we define $\inner{f}{g} = \Expect{{\bf x}\in_R \mathcal{U}_{\vec{k}}}{f({\bf x})g({\bf x})}$.
\begin{definition}
  For $d=0,1,\ldots,n$, we denote by $V_{d}(\mathcal{U}_{\vec{k}})\subseteq \set{f\colon\mathcal{U}_{\vec{k}}\to\mathbb{R}}$
  the span of $d$-juntas. We often refer to this space as the space of degree $d$ functions.
  We also define $V_{=d}(\mathcal{U}_{\vec{k}}) = V_{d}(\mathcal{U}_{\vec{k}})\cap V_{d-1}(\mathcal{U}_{\vec{k}})^{\perp}$.
\end{definition}

Thus, slightly informally, we will often refer to functions from $V_{d}(\mathcal{U}_{\vec{k}})$ for small $d$ as low-degree functions,
and refer to functions perpendicular to that space, i.e.~in $V_{d}(\mathcal{U}_{\vec{k}})^{\perp}$, as high-degree functions.

Next, we wish to define the notion of couplings that will be central in our paper, and we first need to define symmetric distributions.
\begin{definition}
  For $r\in\mathbb{N}$, we say a distribution $\mu$ over $([m_1]\times\ldots\times[m_r])^n$ is symmetric under the action of $S_n$
  (or simply symmetric) if for all $\pi\in S_n$, the distribution of $(\pi({\bf x}(1)), \ldots, \pi({\bf x}(r)))$, where $({\bf x}(1),\ldots,{\bf x}(r))\sim \mu$,
  is the same as $\mu$.
\end{definition}

We will often identify the domains $([m_1]\times\ldots\times[m_r])^n$ and
$[m_1]^n\times\ldots\times[m_r]^n$, choosing the precise view depending on the context.
Next, we define a specific type of couplings that will be useful for us in this paper.
\begin{definition}\label{def:coupling}
  Let $\alpha,\zeta>0$, $m\in\mathbb{N}$.  For symmetric distributions $\nu_1,\nu_2$ over $[m]^n$,
  a $(\alpha,\zeta)$-coupling between $([m]^n,\nu_1)$ and
  $([m]^n,\nu_2)$ is a jointly distributed $\mathcal{C} = ({\bf x}, {\bf y})$ satisfying the following properties.
  \begin{enumerate}
    \item The marginal distribution of ${\bf x}$ is $\nu_1$, and the marginal distribution of ${\bf y}$ is $\nu_2$.
    \item The distribution of $\mathcal{C}$ is symmetric under the action of $S_n$.
    \item For all $i\in [n]$, $\Prob{({\bf x},{\bf y})\sim\mathcal{C}}{{\bf x}_i\neq {\bf y}_i}\leq \zeta$.
    \item Tail bounds: for all $\eps>0$, it holds that $\Prob{({\bf x},{\bf y})\sim\mathcal{C}}{\card{\sett{i}{x_i\neq y_i}}\geq \eps n}\leq
    \frac{1}{\alpha} e^{-\alpha\eps^2 n}$.
%   For each $i\in [n]$, it holds that $\Prob{(x,y)\sim \mathcal{C}}{x_i \neq y_i}\leq \delta$.
%    \item The coupling commutes with the action of $S_n$: namely, for each $\pi\in S_n$, the distribution
%    of $(\pi(x),\pi(y))$ is identical to the distribution of $(x',y')$ where we sample $(a,b)\sim \mathcal{C}$,
%    set $x' = \pi(a)$, and then sample $(a',y')\sim\mathcal{C}$ conditioned on $a' = x'$.
  \end{enumerate}
\end{definition}
A basic example is given by Boolean slices. For example, setting $m=2$ and taking $\nu_1$ to be the uniform distribution over $\power{n}$,
$\nu_2$ to be the uniform distribution over the vectors from $\power{n}$ with Hamming weight $n/2$.
Indeed, a valid coupling in this case is simple to construct: $({\bf x}, {\bf y})$ are sampled by taking ${\bf x}\sim \nu_1$;
if the Hamming weight of ${\bf x}$ is at least $n/2$, we take ${\bf y}\leq {\bf x}$ of Hamming weight $n/2$ randomly, and
otherwise we take ${\bf y}\geq {\bf x}$ of Hamming weight $n/2$ randomly. One may show that this is an $(\alpha,\zeta)$-coupling for
$\alpha = \Omega(1)$ and $\zeta = O\left(\frac{1}{\sqrt{n}}\right)$. We omit the proof of this fact as in Proposition~\ref{prop:coupling_construction},
we give a more general construction that also applies to this case and gives the same parameters.

Earlier, we have seen that couplings allow us to define distance measures between functions on different domains.
More importantly however, they also allow us to naturally lift functions from one domain to another, as described below.
Suppose $\mathcal{C}$ is a coupling between $(\mathcal{U}_{\vec{k}}, {\sf Uniform})$ and $([m]^n,\nu_{\vec{k}}^{\otimes n})$.
We introduce the operators
$\mathrm{T}_{\mathcal{C}}\colon L^2(\mathcal{U}_{\vec{k}}, {\sf Uniform})\to L^2([m]^{n},\nu_{\vec{k}}^{\otimes n})$ and
its adjoint operator $\mathrm{T}_{\mathcal{C}}^{*}\colon L^2([m]^{n},\nu_{\vec{k}}^{\otimes n})\to L^2(\mathcal{U}_{\vec{k}}, {\sf Uniform})$,
defined as
\[
\mathrm{T}_{\mathcal{C}} f(y) = \cExpect{({\bf x}, {\bf y})\sim \mathcal{C}}{{\bf y} = y}{f({\bf x})},
\qquad
\mathrm{T}_{\mathcal{C}}^{*} g(x) = \cExpect{({\bf x},{\bf y})\sim \mathcal{C}}{{\bf x} = x}{g({\bf y})}.
\]
With this in mind, given a function $f\colon \mathcal{U}_{\vec{k}}\to\mathbb{R}$, our associated function on the product space will be
$\mathrm{T}_{\mathcal{C}} f$, where $\mathcal{C}$ is a coupling with sufficiently good parameters. Our invariance principle for low-degree functions reads:
\begin{lemma}\label{lem:inv_low_deg}
Let $d,m,n\in\mathbb{N}$, $\alpha, \zeta>0$, and let $\mathcal{U}_{\vec{k}}$ be multi-slice over alphabet $[m]$.
Suppose $f\colon \mathcal{U}_{\vec{k}}\to\mathbb{R}$ is a function of degree at most $d$, and $\mathcal{C},\mathcal{C}'$ are $(\alpha,\zeta)$-couplings
between $(\mathcal{U}_{\vec{k}})$ and $([m]^n, \nu_{\vec{k}}^{\otimes n})$. Then
\[
\Expect{({\bf x},{\bf y})\sim \mathcal{C}'}
{(f({\bf x}) - \mathrm{T}_{\mathcal{C}} f({\bf y}))^2} \leq  8 \sqrt{d\zeta}\norm{f}_2^2.
\]
\end{lemma}
We note that the bound on the right hand side of Lemma~\ref{lem:inv_low_deg} is independent of $\alpha$, and the proof only uses the first three items
in Definition~\ref{def:coupling}.

Next, to go beyond low-degree functions, we must consider settings in which the contributions of high-degree terms are vanishing
(in which case the question is, in essence, about low-degree functions). In~\cite{MOO,MosselGaussian}, it was shown that connected product distributions provide
an important such example, and we show that an appropriate version of this result continues to hold in the multi-slice. To formally define this
setting, we require a few definitions. First, we define the notion of admissible distributions.
\begin{definition}
  For a distribution $\mu$ over $([m_1]\times\cdots\times[m_r])^n$ and $\vec{a}\in[m_1]\times\cdots\times[m_r]$, we denote
  \[
  \mu_{\vec{a}} = \Prob{\substack{({\bf x}(1),\ldots,{\bf x}(r))\sim \mu\\ i\in_R [n]}}{{\bf x}(1)_i = a_1,\ldots, {\bf x}(r)_i = a_r}.
  \]
\end{definition}

\begin{definition}\label{def:admissible}
  Let $\alpha\in (0,1)$, $r\in\mathbb{N}$ and let $\mathcal{U}_{\vec{k}(1)},\ldots,\mathcal{U}_{\vec{k}(r)}$ be multi-slices over alphabets
  $[m_1],\ldots,[m_r]$ respectively.
  We say a distribution $\mu$ over $\mathcal{U}_{\vec{k}(1)}\times\ldots\times\mathcal{U}_{\vec{k}(r)}$ is
  $\alpha$-admissible if it satisfies the following properties.
  \begin{enumerate}
    %\item The distribution $\mu$ commutes with the action of $S_n$.
    \item The distribution $\mu$ is symmetric under $S_n$.
    \item For all $\vec{a}\in [m_1]\times\ldots\times [m_r]$ it either holds that $\mu_{\vec{a}} = 0$ or $\mu_{\vec{a}}\geq \alpha$.
  \end{enumerate}
\end{definition}

Next, we define the notion of connectedness for a distribution $\mu$.
\begin{definition}\label{def:connected}
Let $r,m_1,\ldots,m_r\in\mathbb{N}$, and let $\mu$ be a distribution over $[m_1]^n\times\ldots[m_r]^n$. We say $\mu$ is connected,
if for all $i$, the graph $H_i = (V_1\cup V_2, E)$ defined as follows is connected:
(1) $V_1 \subseteq [m_i]^n$, $V_2 = \prod\limits_{j\neq i} [m_j]^n$
are the support of the corresponding marginal distributions of $\mu$, and
(2) $(v_1,v_2)$ is an edge if $x\in [m_1]^n\times\ldots[m_r]^n$ whose $i$th component is in $v_1$, and
the rest of its components are according to $v_2$, is in the support of $\mu$.
\end{definition}

Finally, we define the product analogue $\tilde{\mu}$ of a given distribution $\mu$ over the multi-slice.
\begin{definition}\label{def:product_ver}
  Let $\mathcal{U}_{\vec{k}(1)},\ldots,\mathcal{U}_{\vec{k}(r)}$ be multislices over alphabets $[m_1],\ldots,[m_r]$,
  and let $\mu$ be a distribution over $\mathcal{U}_{\vec{k}(1)}\times\ldots\times\mathcal{U}_{\vec{k}(r)}$.
  For each $i\in[n]$, define the distribution $\tilde{\mu}_i$ over $[m_1]\times\ldots\times[m_r]$ as
  \[
  \tilde{\mu}_i(\vec{a}) = \Prob{({\bf x}(1),\ldots{\bf x}(r))\sim\mu}{{\bf x}(1)_i = a_1,\ldots,{\bf x}(r)_i = a_r}.
  \]
  Then, the product analogue of $\mu$ is defined as $\tilde{\mu} = \prod\limits_{i=1}^{n}\tilde{\mu}_i$.
\end{definition}
We remark that we will mostly be interested with admissible distributions $\mu$, hence the marginal distributions
$\mu_{i}$ (as well as of $\tilde{\mu}_i$) will be the same for all $i\in [n]$.
We can now state our main invariance principle.
\begin{thm}\label{thm:basic_inv_multi}
  For all $\alpha\in (0,1)$, $M, r\in\mathbb{N}$, $m_1,\ldots,m_{r}\in\mathbb{N}$, $\eps>0$
  there are $\zeta>0$, $N\in\mathbb{N}$ such that the following holds for $n\geq N$.
  Suppose that
  \begin{itemize}
    \item For $i=1,\ldots, r$, the multi-slice $\mathcal{U}_{\vec{k}(i)}\subseteq[m_i]^n$ is $\alpha$-balanced, and
    $\mathcal{C}_i$ is a $(\alpha,\zeta)$-coupling between $(\mathcal{U}_{\vec{k}(i)},{\sf Uniform})$ and $([m_i]^n, \nu_{\vec{k}(i)}^{\otimes n})$;
    \item $\mu$ is a connected, $\alpha$-admissible distribution over $\prod\limits_{i=1}^{r}\mathcal{U}_{\vec{k}(i)}$;
    \item $\tilde{\mu}$ is the product version of $\mu$ as in Definition~\ref{def:product_ver},
    and there is an $(\alpha,\zeta)$-coupling between $\mu$ and $\tilde{\mu}$.
  \end{itemize}
  Then for all $f_i\colon \mathcal{U}_{\vec{k}(i)}\to\mathbb{R}$ such that $\norm{f_i}_{2r}\leq M$ for all $i$, it holds that
  \[
  \card{
  \Expect{({\bf x}(1),\ldots,{\bf x}(r))\sim \mu}{\prod\limits_{i=1}^{r}f_i({\bf x}(i))} -
  \Expect{({\bf y}(1),\ldots, {\bf y}(r))\sim \tilde{\mu}}{\prod\limits_{i=1}^{r}\mathrm{T}_{\mathcal{C}_i}f_i({\bf y}(i))}
  }
  \leq \eps.
  \]
\end{thm}

A simple example application for Theorem~\ref{thm:basic_inv_multi} is given in the Boolean slice and the Kneser graph on it. Namely, we
consider the case that $r = 2$, $m_1 = m_2 = 2$, $\vec{k}(1) = \vec{k}(2) = \vec{k} = (\ell, n-\ell)$ for $\ell=0.49 n$ and $\mu$ is the uniform distribution
on $(x,y)$ from $\sett{(x,y)\in \mathcal{U}_{\vec{k}}\times \mathcal{U}_{\vec{k}}}{(x_i,y_i) \neq (2,2)~\forall i\in [n]}$. The conditions of the theorem
are easily seen to hold in this case, and hence Theorem~\ref{thm:basic_inv_multi} applies. In this case, if $f_1,f_2$ are indicator functions of subsets
of $\mathcal{U}_{\vec{k}}$, the first expectation in Theorem~\ref{thm:basic_inv_multi} counts the number of edges in the Kneser graph that go between
these two subsets, and the theorem allows us to transfer this problem into a related problem over a product domain. In the latter domain good bounds are
often already known (for example in~\cite{MosselGaussian} bounds for such expectations were proved by further moving to Gaussian space).

\begin{remark}
A few remarks are in order.
\begin{enumerate}
  \item We give a more general version of Theorem~\ref{thm:basic_inv_multi} in Section~\ref{sec:beyond?}.
  Morally speaking, this version works for distributions for which the contribution of high-degree
  functions to expectations such as
  \[
  \Expect{({\bf x}(1),\ldots,{\bf x}(r))\sim \mu}{\prod\limits_{i=1}^{r}f_i({\bf x}(i))}
  \]
  is negligible.
  See Section~\ref{sec:beyond?} for details. Thus, the majority of the effort in the proof of Theorem~\ref{thm:basic_inv_multi}
  lies in showing that high-degree terms do not contribute much to the expectation of the product provided $\mu$ is connected and admissible.
  \item We remark that for most distributions $\mu$ of interest, the existence of couplings as in Theorem~\ref{thm:basic_inv_multi} is easy to
  establish, and thus encourage the reader to think of these conditions as trivially holding.
  See Proposition~\ref{prop:coupling_construction} for a general construction of couplings (which captures all of the cases considered herein).
\end{enumerate}
\end{remark}

We note that our invariance principle is different from the one proved in~\cite{FKMW,FM2019}, even when specialized to the Boolean slice, i.e.~the alphabet
$m=2$. In particular, our results are incomparable to theirs.
The reason is that therein, the authors find a canonical basis of the space of functions $f\colon \mathcal{U}_{\vec{k}}\to\mathbb{R}$, which then
allows them to interpret them as functions over $\power{n}$ without any ambiguity. The authors then show this extension then behaves very similarly to the
original function $f$. Our associated function with $f$ on the other hand, namely $\mathrm{T}_{\mathcal{C}} f$, can be thought of as mild averaging of
the function $f$. Typically, the parameter $\zeta$ of our coupling would be $\Theta(1/\sqrt{n})$, thus the points $({\bf x},{\bf y})\sim\mathcal{C}$
are typically of Hamming distance $\Theta(\sqrt{n})$ apart, and so $\mathrm{T}_{\mathcal{C}} f(y)$ could be thought of as averaging over points $x$
in the mutli-slice that are at distance $\Theta(\sqrt{n})$. In this language, our invariance principle for low-degree functions (Lemma~\ref{lem:inv_low_deg})
asserts that this averaging barely affects the function $f$ if its degree is significantly smaller than $\sqrt{n}$, which is precisely the range
of degrees that~\cite{FKMW,FM2019} are able to handle.
We think however that our version of the invariance principle is perhaps more natural.
Two points to support that are as follows:
\begin{enumerate}
  \item Our proofs are much more ``generic'', i.e.~less domain specific than of~\cite{FKMW,FM2019}, and can probably be extended
to other $S_n$-modules and not only the multi-slice.
  \item The functions $f$ and $\mathrm{T}_{\mathcal{C}} f$ enjoy many of the same properties that we will be concerned with. For example, we have
  the following important properties:
  \begin{enumerate}
    \item if all of the noisy influences of $f$ are small (a notion that we only define in Section~\ref{sec:app_hardness}, but should be thought of as an
  appropriate analogue of the noisy influences from the Boolean cube), then all of the noisy influences of $\mathrm{T}_{\mathcal{C}} f$ are small.
    \item If $f$ has a coordinate $i$ with significant noisy influence, then the same coordinate would have significant noisy influence in $\mathrm{T}_{\mathcal{C}} f$.
    \item If $f$ is pseudo-random in the sense that fixing constantly many coordinates does not change the average of $f$ by too much, then the same holds
    for $\mathrm{T}_{\mathcal{C}} f$.
  \end{enumerate}
  These properties are very important for us in our applications, and can be used to convert any of the Gaussian bounds of~\cite{MosselGaussian} to the
  multi-slice (we do not do it explicitly in this paper). Thus, for example, one can use our techniques to derive some of the applications in~\cite{FKMW,FM2019},
  such as a version of the Majority is Stablest theorem over the slice.
\end{enumerate}

\subsection{Applications to Hardness of Approximation}
In this section, we discuss an application of our invariance principle to the field of hardness of approximation. More specifically, we
show that it enables us to lift a certain class of dictatorship tests with perfect completeness to hardness results, assuming a conjecture
of~\cite{BKM} that we discuss below.
\subsubsection{The basic PCP Theorem}
The primary goal in hardness of approximation is to
show that approximating solutions to optimization problems within a certain approximation ratio is NP-hard. For some problems, the basic PCP theorem combined
with the long-code based paradigm developed over the last 20 years, yield tight inapproximability results (such as the results in~\cite{Hastad,Hastadclique}).
This paradigm however fails to establish tight inapproximability results for many problems, such as Max-Cut and Vertex-Cover. Towards this end,
the Unique-Games Conjecture was proposed~\cite{Khot}, and is now known to imply tight inapproximability result for a wide range of optimization problems.
A prime example is Raghavendra's result~\cite{Rag} regarding constraint satisfaction problems, who showed a generic SDP algorithm for the class of CSPs,
and proved that it is essentially optimal, assuming the Unique-Games Conjecture. We remark that despite recent progress establishing a weaker variant of
the Unique-Games Conjecture, called the $2$-to-$1$ Games Conjecture~\cite{KMS,DKKMS-1,DKKMS-2,KMS2,BKS} (which can be used to prove a wide range
of weaker inapproximability results, see~\cite{BKhalf}), proving Unique-Games Conjecture or some its consequences independently remains
an open challenge.

To present the Unique-Games Conjecture, we first define the $2$-Prover-$1$-Round problem.
\begin{definition}
  A 2P1R Games instance $\Psi = (L\cup R, E, \Sigma_L,\Sigma_R, \Phi)$ consists of a regular, bipartite graph
  $(L\cup R,E)$, the alphabet $\Sigma_L$ for the vertex set $L$, the
  alphabet $\Sigma_R$ for the vertex set $R$, and a
  set of constraints $\Phi = \set{\phi_e}_{e\in E}$, one for each edge.
  Each vertex is supposed to receive a label from the respective alphabet.
  The constraint $\phi_{e}$ for an edge $e = (u,v)\in E, u \in L, v \in R$
  is defined by a relation $\phi_{e}\subseteq \Sigma_L\times \Sigma_R$, thought of as the set of label-pairs to
  the vertices $u$ and $v$ that satisfy the constraint.
\end{definition}
For $0<s<c\leq 1$, we denote by {\sf Gap-$2$P$1$R}$[c,~s]$ the promise problem, where given an instance $\Psi$
of 2P1R Games, one has to distinguish between the case that there is an assignment satisfying at least $c$
fraction of the constraints in $\Psi$ and the case that no assignment satisfies more than $s$ fraction
of the constraints in $\Psi$. Combining the basic PCP Theorem~\cite{FGLSS,AS,ALMSS} with the Parallel Repetition Theorem~\cite{Raz},
one can show that for all $s>0$, the problem {\sf Gap-$2$P$1$R}$[1,~s]$ is NP-hard on alphabets of size $n = {\sf poly}(1/\delta)$.
This result serves as the starting point of many hardness of approximation results. However, for many other results, one needs
a stronger structure from the constraints.

\subsubsection{The Unique-Games Conjecture and Imperfect Completeness}
\begin{definition}
  A $d$-to-$1$ Games instance is a 2P1R Games instance $\Psi = (L\cup R, E, \Sigma_L,\Sigma_R, \Phi)$, in which
  $\card{\Sigma_L} = d\card{\Sigma_R}$ and each constraint $\phi_{u,v}\colon \Sigma_L\to \Sigma_R$ is a $d$-to-$1$ map.
\end{definition}
Khot conjectured that for $d\geq 2$, the problem {\sf Gap-$d$-to-$1$Games}$[1,~s]$ remains NP-hard provided the alphabets
are sufficiently large constant (depending on $s$). For $d=1$ however, i.e.~for {\sf $1$-to-$1$ Games} problem often called the Unique-Games problem,
one cannot hope for such a strong hardness result: it is easy to see that given a satisfiable instance of Unique-Games, one can efficiently find
a satisfying assignment. For that reason, to deal with Unique-Games, one has to give up the perfect completeness, and indeed the Unique-Games
Conjecture reads:
\begin{rconj}\label{conj:UG}
  For all $\eps,s>0$, there is $n\in\mathbb{N}$, such that given a Unique-Games instance $\Psi$ with alphabet size $n$, it is NP-hard
  to distinguish between the following two cases:

  {\bf YES case}: there is an assignment to $\Psi$ satisfying at least $1-\eps$ fraction of the constraints.

  {\bf NO case}: there is no assignment to $\Psi$ satisfying more than $s$ fraction of the constraints.
\end{rconj}

It is thus clear that when working with the Unique-Games Conjecture, one cannot hope to prove hardness for satisfiable instances of problems,
and indeed all of the results based on the Unique-Games Conjecture (such as Raghavendra's result~\cite{Rag}) have imperfect completeness. We
stress here that even though the difference between perfect and imperfect completeness is often thought of as minor, it is critical for some
problems and affects their complexity dramatically: the problem of solving systems of linear equations of over finite fields is such problem~\cite{Hastad},
and the Unique-Games problem is likely to be such problem, to name a few.

This raises the following issue: how can one prove general hardness results for satisfiable instances, assuming some feasible complexity assumption?
One option is to work with $2$-to-$1$ Games for which the perfect completeness version is conjectured to be hard (and the imperfect completeness version
was already proved to be hard~\cite{KMS,DKKMS-1,DKKMS-2,KMS2}). Alas, the structure of the constraints in $2$-to-$1$ Games is often not strong enough to use the powerful reductions and
tools used in the context of Unique-Games.

\subsubsection{The Rich $2$-to-$1$ Games Conjecture}
Recently, a stronger variant of the $2$-to-$1$ Games Conjecture, called the Rich $2$-to-$1$ Games conjecture was proposed~\cite{BKM} in the hope
to shed more light on Unique-Games and the complexity of satisfiable instances, and we present this conjecture below.

Let $\Psi = (L\cup R, E, \Sigma_L,\Sigma_R, \Phi)$ be a $2$-to-$1$-Game, with
$|\Sigma_L|=n$ and $|\Sigma_R| = n/2$.
Fix a vertex $u\in L$. Let $e=(u,v)\in E$ be an edge incident on $u$ and let $\pi_e$ be the $2$-to-$1$ projection defining that constraint. The map defines a partition of
$\Sigma_L$ as
 $$\Sigma_L = \bigcup_{\rho \in \Sigma_R} \pi_e^{-1}(\rho) $$
into disjoint sets of size $2$. Let us denote by $\mathcal{P}(u)$ the distribution over partitions of $\Sigma_L$ into sets of
size $2$, given by first sampling a uniformly random edge $e=(u,v)$ incident on $u$ and then outputting the
partition of $\Sigma_L$ as above.
\begin{definition}
  An instance of Rich~$2$-to-$1$ Games is an instance of  ~$2$-to-$1$ Games with the additional property that
  for every vertex $u\in L$, the distribution
  $\mathcal{P}(u)$ is uniform over all partitions of $\Sigma_L$ into sets of size $2$.
\end{definition}
For $0<s<c\leq 1$, we denote by {\sf Gap-Rich-$2$-to-$1$}$_n[c,~s]$ the promise problem, where given an instance $\Psi$
of Rich~$2$-to-$1$ Games, one has to distinguish between the case that there is an assignment satisfying at least $c$
fraction of the constraints in $\Psi$, or all assignments satisfy at most $s$ fraction of the constraints in $\Psi$.
The following conjecture was made in~\cite{BKM}:
\begin{rconj}\label{conj:rich}
\noindent For all $\delta > 0$, there is a sufficiently large even $n\in\mathbb{N}$ such that {\sf Gap-Rich-$2$-to-$1$}$_n[1,~\delta]$ is NP-hard.
\end{rconj}
A few remarks are in order. First, one may consider the imperfect completeness variant of Conjecture~\ref{conj:rich},
i.e.~the question of whether {\sf Gap-Rich-$2$-to-$1$}$_n[1-\eps,~\delta]$ is NP-hard for all $\eps,\delta>0$ provided $n(\eps,\delta)$
is large enough. It turns out that this problem is equivalent to the Unique-Games Conjecture~\cite{BKM}, and thus Conjecture~\ref{conj:rich}
is strictly stronger than the Unique-Games Conjecture, and thus can be thought of as a variant of it with perfect completeness. It is thus
natural to ask whether one can base on it hardness of satisfiable instances of problems, the prime example being constraint satisfaction
problems.

\subsubsection{Our application: hardness from dictatorship tests with perfect completeness}
Using our invariance principle, we are able to convert certain dictatorship tests
to NP-hardness results, assuming Conjecture~\ref{conj:rich}. This result generalizes a theme from the context of Unique-Games Conjecture based hardness results,
for which one has an automatic machinery to convert dictatorship tests with imperfect completeness to hardness of approximation results (a theme that has started
with in~\cite{KKMO} and plays an important role in Raghavendra's work~\cite{Rag}). Thus, the task of proving hardness of approximation result for a problem with
imperfect completeness reduces to the design of a certain dictatorship test.

Our result extends this theme to the realm of perfect completeness, however we are not able to handle dictatorship tests in full generality and require a
certain technical condition that is nevertheless natural in existing dictatorship tests. We hope that as more general dictatorship tests emerge in the literature,
the ideas in the work herein will be useful in converting these more general tests into hardness results.

A precise statement of our result (given in Theorem~\ref{thm:dictator_to_hard} and slightly
extended in Remark~\ref{remark:extend}) requires quite a bit of set-up, and is deferred to Section~\ref{sec:app_hardness}. Instead, we give two instantiations of it,
using existing dictatorship tests from the literature.

For a collection of $r$-ary predicates $\mathcal{P}_r\subseteq\set{P\colon\set{0,1}^r\to\set{0,1}}$, an instance of
{\sf CSP-$\mathcal{P}_r$} consists of a set of variables $X = \set{x_1,\ldots,x_m}$, and a collection of constraints
$E$ of the form $P(x_{i_1},\ldots,x_{i_r}) = 1$ where $P\in\mathcal{P}_r$. The goal is to find a Boolean assignment
to the variables that satisfies as many constraints as possible. We also define the gap version,
{\sf Gap-CSP-$\mathcal{P}_r$} in the natural way.
Using the dictatorship test of~\cite{BKT}, we have:
\footnote{Strictly speaking, in~\cite{BKT} only a single predicate $P_r\colon\power{r}\to\power{}$ is used.
However, they use the folding technique, and our collection $\mathcal{P}_r$ replaces this folding.
For all $\vec{a}\in\power{r}$, we define $P_{r,\vec{a}}\colon \power{r}\to\power{}$ as
$P_{r,\vec{a}}(x) = P_r(x_1+a_1,\ldots,x_r+a_r)$ (addition is done mod $2$), and our collection
$\mathcal{P}_r$ is $\sett{P_{r,\vec{a}}}{\vec{a}\in \power{r}}$.}
\begin{corollary}\label{cor:CSP}
  Assuming Conjecture~\ref{conj:rich}, for all $r\in\mathbb{N}$ of the form $2^{m} - 1$ there is a collection of
  $\mathcal{P}_r\subseteq\set{P\colon\set{0,1}^r\to\set{0,1}}$ such that for
  all $\eps>0$, {\sf Gap-CSP-$\mathcal{P}_r$}$\left[1,\frac{2r+1}{2^r}+\eps\right]$ is NP-hard.
\end{corollary}
Unconditionally, it is known that there is a predicate $P$ for which {\sf Gap-CSP-$\mathcal{P}_r$}$\left[1,\frac{2^{\tilde{O}(r^{1/3})}}{2^r}+\eps\right]$
is NP-hard~\cite{Huang}. Corollary~\ref{cor:CSP} significantly improves upon this result, and is nearly tight (however, it is conditioned on Conjecture~\ref{conj:rich}).
Indeed, the works~\cite{charikar2009near,makarychev2012approximation} show algorithms for {\sf Gap-CSP-$\mathcal{P}_r$}$\left[1-\eps,c\frac{r}{2^r}\right]$ for
some absolute constant $c>0$.

Next, we show that assuming Conjecture~\ref{conj:rich}, one can prove strong NP-hardness results for $3$-colorable graphs.
Recently, it was shown that for all $d\geq 2$, the $d$-to-$1$ Games Conjecture implies the hardness of coloring $3$-colorable
graphs with constantly many colors~\cite{GS20}. We show that the Rich $2$-to-$1$ Games Conjecture implies the same result with
a stronger soundness guarantee asserting that there are no independent sets of significant size.
More precisely, using the dictatorship test of~\cite{DMR}, we get:
\begin{corollary}\label{cor:graph_col}
  Assuming Conjecture~\ref{conj:rich}, for any $\delta>0$, given a graph
  $G$ it is NP-hard to distinguish between the cases:
    \begin{enumerate}
    \item {\bf YES case}: $G$ is $3$-colorable;
    \item {\bf NO case}: $G$ has no independent set of fractional size $\delta$.\footnote{In particular, the chromatic number of $G$ is at least $\lceil{1/\delta\rceil}$.}
  \end{enumerate}
\end{corollary}
We remark that the work~\cite{DMR} establishes Corollary~\ref{cor:graph_col} assuming a different conjecture (Conjecture 4.8 therein), as well as a variant
of this result where the completeness case is relaxed to $4$-colorable assuming the $2$-to-$1$ Games Conjecture.

\subsection{Applications to Combinatorics}
In a subsequent paper~\cite{BKLMcomb}, we show several applications of our invariance principle
in extremal combinatorics. Below, we explain the sort of problems that will be addressed in that paper.

We write $\left[n\right]=\left\{ 1,\ldots,n\right\} $ and $\binom{\left[n\right]}{k}$
for the family of all $k$-subsets of $\left[n\right].$ The following
problems are fundamental in hypergraph theory.
\begin{enumerate}
\item \emph{The hypergraph Tur\'{a}n problem: }Given $H\subseteq\binom{\left[n\right]}{k}$.
How large can a family $\mathcal{F}\subseteq\binom{\left[n\right]}{k}$
be if it does not contain a copy of $H$?
\item \emph{The Ramsey problem: }What is the largest $r$, such that every
$r$-colouring of $\binom{\left[n\right]}{k}$ contains a copy of
$H$?
\item \emph{Removal lemmas: }Is it true that every family that contains
few copies of $H$ is close to a family without any copies of $H$?
\end{enumerate}
In the subsequent paper~\cite{BKLMcomb}, we focus on the above problems in the case where
$k$ is linear in $n$. This regime was extensively studied when $H$
is an $s$-matching, i.e.~the hypergraph consisting of $s$ pairwise
disjoint edges. The methods used for tackling the problems above are
varied and include shifting~\cite{frankl2018erd}, algebraic topology~\cite{lovasz1978kneser,alon1986chromatic},
regularity~\cite{friedgut2017kneser}, sharp threshold results~\cite{ellis2016stability},
and agreement testing~\cite{DFR}.

In the companion paper we throw our invariance principal into the
mix. We are then able to extend these results from the case where
$H$ is an $s$-matching, to the case where $H$ belongs to a wide
class of hypergraphs we call $\zeta$-forests.
\begin{definition}
We say that a hypergrph $H$ is a \emph{$\zeta$-forest} if its edges
can be ordered as $A_{1},\ldots,A_{r},$ where $\left|A_{i}\setminus\bigcup_{j=1}^{i-1}A_{j}\right|>\zeta n.$
\end{definition}

We show that every family that contains few copies of a $\zeta$-forest
$H$ is close to an $H$-free family, provided that $\left|\bigcup_{A\in H}A\right|\le\left(1-\zeta\right)n$.
We also show that every $r$-colouring of $\binom{\left[n\right]}{k}$
contains a monochromatic copy of $H,$ provided that $n\geq n_{0}\left(r,\zeta\right).$
Finally, we make progress on the Tur\'{a}n problem by reducing it
to a special case where the $H$-free family $\mathcal{F}\subseteq\binom{\left[n\right]}{k}$
is a junta. This means that there exists a set $J$ of size $O\left(1\right)$,
such that the question whether $A$ is in $\mathcal{F}$, depends
only on $A\cap J.$

We remark that such a reduction was shown to be useful Ellis et al.~\cite{ellis2016stability},
who considered the $\frac{1}{4}$-forest $M_{2,t}$ consisting of
two edges that intersect in $t$-elements. The Tur\'{a}n problem
for $M_{2,t}$ corresponds to an old problem of Erd\H{o}s and S\'{o}s~\cite{erdHos1975problems}.
One of the main ingredients in~\cite{ellis2016stability} was a reduction
of the Erd\H{o}s-S\'{o}s forbidden intersection problem to the special
case where the $M_{2,t}$-free family is a junta.

%\subsection{More related works}
%\paragraph{Invariance principle over the slice.}
%The works~\cite{FKMW,FM} established an invariance principle for low degree functions over the slices, which is the simplest instance of a
%multi-slices (over the alphabet of size $2$), and raised the problem of establishing an invariance principle between multi-slices and its product versions in
%general. Strictly speaking,
%our result is incomparable to their results: given $f\colon \sett{x\in\power{n}}{\card{x} = pn}\to\mathbb{R}$ of low degree,
%they choose a unique, canonical representation for $f$ (called ``harmonic representation'') which allows them to think of $f$ as a function
%on the entire cube $\power{n}$. They then prove that the distribution of $f({\bf x})$ and $f({\bf y})$ are close in various senses, where
%${\bf x}$ is uniform over the slice, and ${\bf y}$ is sampled according to the $p$-biased product measure over $\power{n}$. Our result instead
%defines, in this case, a slightly perturbed function $g = \mathrm{T} f\colon \power{n}\to \mathbb{R}$, and shows that
%the distributions of $f({\bf x})$, $g({\bf y})$ are very close. Roughly speaking, the action of the operator $\mathrm{T}$ could be thought of as:
%to compute $\mathrm{T} g(y)$, average over $f({\bf x})$ where ${\bf x}$ is taken from the slice to be ``close'' to $y$ in Hamming distance (typically,
%this distance would be $\Theta(\sqrt{n})$).
\subsection{Our techniques}
\subsubsection{The proof of our invariance principles}
\paragraph{Invariance principle for low-degree functions.}
The proof of our invariance principle for low-degree functions
mainly involves representation-theoretic arguments, and builds on the approach of~\cite{FKLM}.
To prove Lemma~\ref{lem:inv_low_deg},
we reduce the problem into the problem of understanding eigenvalues of a certain operator associated with our couplings.
For simplicity of presentation, let us consider the case
that the couplings $\mathcal{C}$ and $\mathcal{C}'$ in Lemma~\ref{lem:inv_low_deg} are identical, in which case we study the eigenvalues
of the operator $\mathrm{S} = \mathrm{T}_{\mathcal{C}}^{*}\mathrm{T}_{\mathcal{C}}\colon L^2(\mathcal{U}_{\vec{k}})\to L^2(\mathcal{U}_{\vec{k}})$.
Since this is a symmetric operator, we can find an eigenbasis of $L^2(\mathcal{U}_{\vec{k}})$ consisting of eigenvectors, and our claim reduces to
showing that eigenvalues of eigenfunctions which are low-degree functions are close to $1$.

Using the symmetry of $\mathcal{C}$, we show that the operator $\mathrm{S}$ preserve \emph{juntas}. Namely, if $f\colon\mathcal{U}_{\vec{k}}\to\mathbb{R}$
depends only on coordinates from a set $J\subseteq[n]$, then the value of $\mathrm{S} f$ at a point $x$ also only depends on $x_J$. This allows to further
reduce the question of studying eigenvalues corresponding to low-degree functions, to studying eigenvalues corresponding to juntas. The last part is easy
to do, using fact that the probability the coupling changes a specific coordinate is small (this is the parameter $\zeta$).

\paragraph{Invariance principle for admissible distributions.}
For simplicity of presentation, let us assume we are dealing with the case that $r=2$ and that the two multi-slices are identical,
in which case we denote the functions from Theorem~\ref{thm:basic_inv_multi} by $f$ and $g$. Roughly speaking, the proof proceeds by analyzing the contribution of the high-degree parts of $f$ and $g$ and the low-degree parts separately,
say $f = f^{\leq d} + f^{>d}$ and $g = g^{\leq d} + g^{>d}$. Lemma~\ref{lem:inv_low_deg} allows us to handle the low-degree parts, and the main effort in
our proof goes into upper bounding the contribution of the high-degree parts, e.g. $\Expect{({\bf x},{\bf x}')\sim\mu}{f^{\leq d}({\bf x}) g^{>d}({\bf y})}$.
We now turn this problem again into the problem of understanding eigenvalues of a certain operator. More specifically, we consider the operator
$\mathrm{T}_{\mu}\colon L^2(\mathcal{U}_{\vec{k}})\to L^2(\mathcal{U}_{\vec{k}})$ defined as
$\mathrm{T}_{\mu} f(x') = \cExpect{({\bf x},{\bf x}')\sim\mu}{{\bf x}' = x'}{f({\bf x})}$ so that the expectation that we wish to study is
$\inner{f^{\leq d}}{\mathrm{T}_{\mu} g^{>d}}$, and using the Cauchy-Schwarz inequality it suffices to bound $\norm{\mathrm{T}_{\mu}^{*}\mathrm{T}_{\mu} g^{>d}}$.
In other words, we want to prove that the operator $\mathrm{T}_{\mu}^{*}\mathrm{T}_{\mu}$ contracts high degree functions very strongly, and to do that we study
its eigenvalues.

To study the eigenvalues of $\mathrm{T}_{\mu}^{*}\mathrm{T}_{\mu}$ we use the trace method: first, note that this operator defines a random walk on
$L^2(\mathcal{U}_{\vec{k}})$. By considering long (but constant length) walks, we show that this random walk mixes rather well -- in the sense that the
trace of the resulting operator is $(1+\eps)^n$. Intuitively, it is easy to observe that this would happen in the product analogue of $\mu$, namely
$\tilde{\mu}$, and we know by the coupling $\mathcal{C}$ that we may consider the coupling along the random walk, so that we expect the two traces to
be same. Thus, using the trace method, we are able to bound eigenvalues whose multiplicity is exponentially large, say at least $(1+2\eps)^n$, and this
way we are able to handle functions of degree $\Omega(n)$.

To handle functions of degree $o(n)$, we also use the eigenvalues approach, and in this case
show that for each eigenvalue $\theta$ of degree $d$ function, we may find a $d$-junta which is an eigenfunction with eigenvalue $\theta$, say on
a set of variables $[d]\subseteq[n]$. At this point, we essentially project the problem on a set of coordinates slightly larger than $[d]$ --
say $[3d]$ , ignoring the rest of the coordinates. That is, we consider the long random walk $\mathrm{T}_{\mu}^{*}\mathrm{T}_{\mu}$ induces
on $[m]^{3d}$, and show again the trace of this operator is small, namely $(1+\eps)^{3d}$; intuitively this happens for similar reasons
as before, except that $n$ is effectively $3d$. Thus it remains to show that the multiplicity of our eigenvalue $\theta$ is at least
$(1+2\eps)^{3d}$, and for that we again use symmetry: we note that for each $\pi\in S_{3d}$, the function $^{\pi} f\colon\mathcal{U}_{\vec{k}}\to\mathbb{R}$
defined as $^{\pi} f(x) = f(\pi(x))$ is an eigenfunction of $\mathrm{T}_{\mu}^{*}\mathrm{T}_{\mu}$ with eigenvalue $\theta$, and then show
that these functions span a subspace of dimension at least $(1+2\eps)^{3d}$.

\subsubsection{Converting dictatorship tests into hardness results}
Our proof of Theorem~\ref{thm:dictator_to_hard}, which converts a dictatorship test into a hardness result (and implies Corollaries~\ref{cor:CSP} and~\ref{cor:graph_col})
uses roughly the same reduction as Raghavendra~\cite{Rag}. We begin by outlining a simple adaptation of this reduction, which we do not know how to analyze, and then explain
our adaptations and in particular how multi-slices enter the picture.

For simplicity of presentation, let us assume that a dictatorship test for a predicate
$P\colon \Sigma^r\to\power{}$ is a distribution $\mathcal{D}$ over $\Sigma^r$ such that:
\begin{enumerate}
  \item If $f\colon \Sigma^n\to\Sigma$ is a dictatorship (i.e.~$f(x) = x_i$ for some $i$), then
  \[
  \Expect{({\bf x}(1),\ldots,{\bf x}(r))\sim\mathcal{D}^{\otimes n}}{P(f({\bf x}(1)),\ldots,f({\bf x}(r)))} = 1.
  \]
  \item If $f\colon \Sigma^n\to\Sigma$ is far from dictatorship, i.e.~has all of low-degree influences being small, then
  \[
  \Expect{({\bf x}(1),\ldots,{\bf x}(r))\sim\mathcal{D}^{\otimes n}}{P(f({\bf x}(1)),\ldots,f({\bf x}(r)))} \leq s.
  \]
\end{enumerate}

Given a Rich $2$-to-$1$ Games instance, $\Psi = (L\cup R, E, [n],[n/2], \Phi)$, we wish
to encode a label to $u\in L$ using the long-code of $u$, and encode a label to $v\in R$ using the long code of $v$. That is, we replace each $u\in L$ with the cloud
$\set{u}\times\power{n}$ and each $v\in L$ by $\set{v}\times\power{n/2}$. The intention is that a label $\sigma\in [n]$ to $u$ would be encoded by
the dictatorship assignment $f_u\colon \Sigma^{n}\to\Sigma$ defined as $f_u(x) = x_{\sigma}$. Our goal is to test, using the predicate $P$ and the dictatorship
test $\mathcal{D}$, that the assignments $f_u$'s (and $f_v$'s) are correlated with legitimate codewords.

The idea then is to sample $v\in R$, and virtually perform the dictatorship test on its cloud. Namely, sample queries $({\bf x}(1),\ldots,{\bf x}(r))$ according
to the given dictatorship test, then sample $u_1,\ldots,u_r$ neighbours of $v$ independently, lift the points ${\bf x}(i)$ to ${\bf x}(i)'$ on the long-code
of $u_i$ using the projection map $\phi_{v,u_i}$ and then test that $f_{u_1}({\bf x}(1)'),\ldots,f_{u_r}({\bf x}(r)')$ satisfies the predicate $P$.

While the completeness of this reduction is straight-forward, the soundness analysis of this reduction is more tricky. For one, the distribution of
the points ${\bf x}(i)'$ that we consider in the reduction is not very nice: ideally, we would have liked the distribution to be close to the distribution
of ${\bf y}(i)$ if we sampled $({\bf y}(1),\ldots,{\bf y}(r))$ according to the dictatorship test in the cloud of $u_i$; alas this is not true.
The distribution of ${\bf x}(i)'$ is not a product distribution, its support is not even full (for example, each symbol appears even number
of times) and in general it is not convenient to work with. This is an artifact of the fact the constraints maps $\phi_{v,u_i}$ are only $2$-to-$1$,
as opposed to $1$-to-$1$ as in Raghavendra's theorem (in which case, these maps preserve the distributions on the long-codes up to relabeling the coordinates).

To bypass this issue, we note that if instead of the full long-code of $u$, we used a multi-slice $\mathcal{U}_{\vec{k}}\subseteq \Sigma^{n}$
with an appropriate $\vec{k}$ (actually, we will use several multislices, each matching the statistics of one of ${\bf y}(1),\ldots, {\bf y}(r)$
above) and the same for $v$, our distributions would align perfectly. Indeed, we use this idea, and use our invariance principle to argue that
the soundness of the multi-slice analogue of the dictatorship test $\mathcal{D}^{\otimes n}$ from above is roughly the same as in the product version.
To finish up the proof, we must then define analogues of low-degree influences on the multi-slice (it will actually be more convenient for us to work
with noisy influence), and prove several basic properties of them -- see Section~\ref{sec:noisy_inf_multi} for more details.

%\subsection{More related work}
%To give a basic version the invariance principle, we need to define the notion of influences. Consider the Boolean cube $\power{n}$ with the
%uniform measure. For a function $f\colon \power{n}\to \mathbb{R}$ and $i\in[n]$, we define the influence of $i$ to be
%$I_i[f] = \Expect{{\bf x}\in_R \power{n}}{(f({\bf x}) - f({\bf x}\oplus e_i))^2}$.
%In a nutshell, a basic form of the invariance principle says that if $f\colon \power{n}\to \mathbb{R}$ is
%a low-degree, multilinear polynomials with small individual influences, then looking at the multi-linear extension
%$\tilde{f}\colon \mathbb{R}^n\to\mathbb{R}$ of $f$ to $\mathbb{R}^n$, one has that the distributions $f({\bf x})$ and
%$\tilde{f}({\bf g})$ are close under smooth test functions, where ${\bf x}$ is sampled uniformly from $\power{n}$, and
%each coordinate of ${\bf g}$ is an independent standard Gaussian random variable.
%More general forms of this result relax conditions of the function $f$, such as replacing the requirement that $f$ has low-degree
%with the requirement that the Fourier tail of $f$ decays exponentially, and considering more relaxed notions of influences.

\section{Preliminaries}
\paragraph{Notations.} Throughout the paper, we denote random variables by boldface letters (such as ${\bf x},{\bf y},\bm{\pi}$).
We denote $[n] = \set{1,\ldots,n}$, and let $S_n$ be the symmetric group, i.e.~the set of all permutations $\pi\colon [n]\to[n]$.
For a vector $x$ of length $n$ and $A\subseteq[n]$, we denote by $x_A$ the vector whose length is $\card{A}$ corresponding to
the entries of $x$ from $A$. For a vector $x$ of length $n$ and $\pi\in S_n$, we denote by $^{\pi} x$ the vector of length
$n$ whose $i$th entry is $x_{\pi(i)}$.

\smallskip

In this section, we present standard background regarding various decompositions of functions the multi-slice.
More precisely, we will use the degree decomposition, as well as a refinement of it given by representation theory.
Towards that end, it will be useful for us to think of the multi-slice $\mathcal{U}_{k_1,\ldots,k_m}$ as a quotient
space $S_n/(S_{k_1}\times S_{k_2}\times\ldots S_{k_m})$, which will allow us to lift the standard representation-theoretic
decomposition of functions over $S_n$ to decompositions of functions over $\mathcal{U}_{k_1,\ldots,k_m}$.

\subsection{The degree decomposition and representation-theoretic decomposition over $S_n$}
We begin by presenting the coarse degree decomposition of functions $f\colon S_n\to\mathbb{R}$.
\begin{definition}
  We say a function $f\colon S_n\to\mathbb{R}$ is a $d$-junta if there exists a set of coordinates
  $A\subseteq[n]$ of size at most $d$, such that $f(\pi) = g(\pi|_{A})$ for some function $g\colon[n]^A\to\mathbb{R}$.
  In words, letting $a_1,\ldots,a_m$ be an ordering of the elements of $A$, the value of $f$ only depends on $(\pi(a_1),\ldots,\pi(a_m))$.
\end{definition}

We can now define the space of degree $d$ functions. For that, we introduce the expectation inner product:
for $f,g\colon S_n\to\mathbb{R}$, we define $\inner{f}{g} = \Expect{\bm{\pi}}{f(\bm{\pi})g(\bm{\pi})}$.
\begin{definition}
  For $d=0,1,\ldots,n$, we denote by $V_{d}(S_n)\subseteq \set{f\colon S_n\to\mathbb{R}}$
  the span of $d$-juntas. We often refer to this space as the space of degree $d$ functions.
  We also define $V_{=d}(S_n) = V_{d}(S_n)\cap V_{d-1}(S_n)^{\perp}$.
\end{definition}
Thus, one can write the space of real-valued functions as $V_{=0}(S_n)\oplus V_{=1}(S_n)\oplus\ldots\oplus V_{=n-1}(S_n)$, and
thus write any $f\colon S_n\to\mathbb{R}$ uniquely as $f = \sum\limits_{i=0}^{n-1} f^{=i}$ where $f^{=i}\in V_{=i}(S_n)$. We next
present a refinement of this decomposition, given by representation theory of $S_n$.

\subsubsection*{Partitions and the decomposition by partitions}
A partition of $n$, often denoted as $\lambda\vdash n$, is a monotonically non-increasing
sequence of positive integers, $\lambda = (\lambda_1,\ldots,\lambda_r)$, that sum up to $n$. It is
well known that partitions index the equivalence classes of representations of $S_n$,
and thereby give rise to a decomposition of real-valued functions over $S_n$.

Given a partition $\lambda$, a $\lambda$-tabloid is a partition of $[n]$ into sets $A_1,\ldots,A_k$
such that $\card{A_i} = \lambda_i$. Thus, for $\lambda$-tabloids $A = (A_1,\ldots,A_k)$ and $B = (B_1,\ldots,B_k)$,
we define
\[
T_{A,B} = \sett{\pi\in S_n}{\pi(A_i) = B_i ~\forall i=1,\ldots,k},
\]
and refer to any such $T_{A,B}$ as a $\lambda$-coset.

\begin{definition}
  For a partition $\lambda$ of $n$, we define the space $V_{\lambda}(S_n)$ as the linear span of indicator functions of all $\lambda$-cosets.
\end{definition}
It is easily seen that if $\lambda_1 = n-k$, then $V_{\lambda}(S_n)\subseteq V_{k}(S_n)$, thus the $V_{\lambda}(S_n)$ are
a refinement of the spaces $V_{d}(S_n)$. Next, we define a natural ordering on partitions which will
allow us to further refine the space $V_{\lambda}(S_n)$, so that they will be refinements of the pure-degree spaces
$V_{=d}(S_n)$.

\begin{definition}
  Let $\lambda = (\lambda_1,\ldots,\lambda_k)$, $\mu = (\mu_1,\ldots,\mu_{s})$ be partitions
  of $[n]$. We say that $\lambda$ dominates $\mu$, and denote $\lambda \trianglerighteq \mu$, if for all
  $j=1,\ldots,k$ it holds that $\sum\limits_{i=1}^{j} \lambda_i\geq \sum\limits_{i=1}^{j} \mu_i$.
\end{definition}

\begin{definition}
  For a partition $\lambda$ of $n$, we define $V_{=\lambda}(S_n) = V_{\lambda}(S_n)\cap\bigcap_{\mu\vartriangleright \lambda} V_{\mu}(S_n)^{\perp}$.
\end{definition}

It is well known that the spaces $V_{=\lambda}(S_n)$ are orthogonal, and that this decomposition is
a refinement of the degree decomposition, namely that $V_{=\lambda}(S_n) \subseteq V_{=k}(S_n)$ where $\lambda_1 = n-k$; see for example~\cite[Theorem 7]{EFP}.

An important parameter of a partition $\lambda\vdash n$ is its dimension ${\sf dim}(\lambda)$.
Partitions $\lambda$ are in one-to-one correspondence with irreducible representations of $S_n$, and the dimension of that representation is
defined as ${\sf dim}(\lambda)$. Thus, while the isotypical component $V_{=\lambda}(S_n)$ contains several (and more precisely, ${\sf dim}(\lambda)$ many) copies of that irreducible representation,
any sub-representation of it has dimension at least ${\sf dim}(\lambda)$. We will use this fact several times
when we bound eigenvalues of certain operators using the trace method.
Towards this end we first
define the action of $S_n$ on functions in the following way. For a function
$f\colon S_n\to\mathbb{R}$ and $\pi\in S_n$, define the function $^{\pi} f(\sigma) = f(\pi\circ \sigma)$; this way, we will
think of the space of real-valued functions over $S_n$ as a left module. With this action in hand, a situation that would often arise for
us is the following: we are given an operator $\mathrm{T}\colon L^2(S_n)\to L^2(S_n)$ which is positive semi-definite,
and has a small trace; given an eigenvector $f\in V_{=\lambda}(S_n)$ of $\mathrm{T}$, we would like to argue that the eigenvalue
$\alpha_{f}$ of $f$ is small. By the trace method, $\alpha_{f}\leq {\sf Tr}(T)/m_{\alpha_f}$, where $m_{\alpha_f}$ is the multiplicity
of $\alpha_f$, and thus we would get a decent bound on $\alpha_f$ provided the multiplicity $m_{f}$ is large. Here comes
the importance of the parameter ${\sf dim}(\lambda)$: our operators of interest $\mathrm{T}$ will always commute with the action of
$S_n$, and thus each $^{\pi} f$ is also an eigenvector of $\mathrm{T}$ with eigenvalue $\alpha_{f}$.
Thus, $m_{\alpha_{f}}\geq {\sf dim}({\sf span}(\sett{^{\pi} f}{\pi\in S_n}))$, and the latter quantity is well known to be at least
${\sf dim}(\lambda)$, as the following claim asserts.
\begin{claim}\label{claim:subrep_lb_sn}
  Suppose $f\in V_{=\lambda}(S_n)$ is not identically $0$. Then
  ${\sf dim}({\sf span}(\sett{^{\pi} f}{\pi\in S_n}))\geq {\sf dim}(\lambda)$.
\end{claim}
\begin{proof}
  Let $\rho\colon S_n\to V_{=\lambda}(S_n)$ be a representation, and let $W = {\sf span}(\sett{^{\pi} f}{\pi\in S_n})$.
  Then since $W$ is closed under the action of $S_n$, we have that $W$ is a sub-representation. Since $f$ is not identically
  $0$, $W$ is a non-trivial subspace, and therefore it follows that ${\sf dim}(W)\geq {\sf dim}(\lambda)$.
\end{proof}

Thus, to get meaningful bounds on eigenvalues this way, it suffices to have effective bounds on the dimension of partitions, and for
that we use the well-known Hook formula (or rather a corollary of it).
\begin{lemma}[Claim 1, Theorem 19 in~\cite{EFP}]\label{lem:dim_lb}
  For all $c>0$, there exists $\delta>0$ such that the following holds.
  Let $\lambda\vdash n$ be given as $\lambda = (\lambda_1,\ldots,\lambda_k)$,
  and denote $d = \min(n-\lambda_1,k-1)$.
  \begin{enumerate}
    \item If $\lambda = (n)$, then ${\sf dim}(\lambda) = 1$.
    \item If $d>0$, then ${\sf dim}(\lambda)\geq \left(\frac{n}{d\cdot e}\right)^d$.
    \item If $d > c\cdot n$, then ${\sf dim}(\lambda)\geq (1+\delta)^n$.
  \end{enumerate}
\end{lemma}

\subsection{The degree decomposition and the representation-theoretic decomposition over $\mathcal{U}_{\vec{k}}$}
In this section, we refine the degree decomposition of functions on $\mathcal{U}_{\vec{k}}$ presented in the introduction
using the representation-theoretic decomposition.
%We first define the multi-slice formally.
%\begin{definition}
%  Let $n\in\mathbb{N}$ and $\Sigma = [m]$ be a finite alphabet.
%  For a tuple of non-negative integers $\vec{k} = (k_1,\ldots,k_m)$ that add up to $n$,
%  we define the multi-slice $\mathcal{U}_{\vec{k}}\subseteq [m]^n$ as the set
%  of vectors $x\in[m]^n$ that for all $i\in[m]$, have exactly $k_i$ of their
%  coordinates equal to $i$.
%\end{definition}
%
%We will mostly be concerned with roughly balanced multi-slices, i.e.~multislices $\mathcal{U}_{\vec{k}}\subseteq[m]^n$ in which
%each $k_i$ is at least linear in $n$ (where $m$ is thought of as constant).
%\begin{definition}
%  Let $c\in (0,1)$ be a constant. We say a multi-slice $\mathcal{U}_{\vec{k}}$ over the alphabet $[m]$
%  is $\alpha$-balanced if $k_i\geq \alpha\cdot n$ for all $i\in[m]$.
%\end{definition}

\subsubsection{The maximal degree on the multi-slice}
%\begin{definition}
%  For a multi-slice $\mathcal{U}_{\vec{k}}$ over alphabet $[n]$. we say a function $f\colon\mathcal{U}_{\vec{k}}\to\mathbb{R}$ is a $d$-junta if there exists a set of coordinates
%  $A\subseteq[n]$ of size at most $d$, such that $f(x) = g(x_{A})$ for some function $g\colon[m]^A\to\mathbb{R}$.
%\end{definition}
%
%We can now define the space of degree $d$ functions. First, as before, we define the expectation inner product, i.e.~for
%$f,g\colon\mathcal{U}_{\vec{k}}\to\mathbb{R}$ we define $\inner{f}{g} = \Expect{{\bf x}\in_R \mathcal{U}_{\vec{k}}}{f({\bf x})g({\bf x})}$.
%\begin{definition}
%  For $d=0,1,\ldots,n$, we denote by $V_{d}(\mathcal{U}_{\vec{k}})\subseteq \set{f\colon\mathcal{U}_{\vec{k}}\to\mathbb{R}}$
%  the span of $d$-juntas. We often refer to this space as the space of degree $d$ functions.
%
%  We also define $V_{=d}(\mathcal{U}_{\vec{k}}) = V_{d}(\mathcal{U}_{\vec{k}})\cap V_{d-1}(\mathcal{U}_{\vec{k}})^{\perp}$.
%\end{definition}

Recalling the spaces $V_0(\mathcal{U}_{\vec{k}}),\ldots V_n(\mathcal{U}_{\vec{k}})$ from the introduction, we note that it is clear that
$V_n(\mathcal{U}_{\vec{k}})$ contains all real-valued functions on $\mathcal{U}_{\vec{k}}$. The following claim shows that something stronger happens,
and one already has that $V_{n-k_1}$ contains all real valued functions on $\mathcal{U}_{\vec{k}}$.
\begin{claim}\label{claim:very_high_deg_vanish}
  For a multi-slice $\mathcal{U}_{\vec{k}}$, one has that $V_n = V_{n-k_1}$.
\end{claim}
\begin{proof}
  Consider the set of functions
  \begin{equation}
  \mathcal{A} = \sett{1_{x_A = \alpha}}{\card{A} = n-k_1, \alpha\in\set{2,\ldots,m}^A, \card{\set{i : \alpha_i = j}} = k_j ~\forall j\in\set{2,\ldots,m}}.
  \end{equation}
  We note that each one of the functions in this set is a $n-k_1$-junta, so their span is contained in $V_{n-k_1}$
  Thus, it is enough to prove that this set spans $V_n$.  Indeed, let
  $B\subseteq[n]$, and $\beta\in[m]^n$ and partition $B = B'\cup B''$ where $B'$ is the set of coordinates $i$ of $\beta$ for which
  $\beta_i = 1$, and $B''$ is the rest. We prove by induction on $\card{B'}$ that $1_{x_B = \beta}$ is in ${\sf span}(\mathcal{A})$.

  \paragraph{Base case.} For $\card{B'} = 0$, we have $\card{B''}\leq n-k_1$ and the statement is clear. For $\card{B'} = 1$, we have
  \[
  1_{x_B = \beta}
  =
  1_{x_{B''} = \beta''}
  -
  \sum\limits_{\gamma\in\set{2,\ldots,m}}1_{x_{B'} = \gamma,x_{B''} = \beta''},
  \]
  we note that if $\card{B'\cup B''} > n-k_1$, then all of the functions in the sum
  on the right hand side are $0$ and so $1_{x_B = \beta} = 1_{x_{B''} = \beta''}\in \mathcal{A}$.
  Otherwise, we clearly get $1_{x_B = \beta}\in {\sf span}(\mathcal{A})$.

  \paragraph{Inductive step.} Write $B' = \set{i}\cup B'''$ and note that
  \[
  1_{x_B = \beta}
  =
  1_{x_{B''} = \beta'', x_{B'''} = \beta'''}
  -
  \sum\limits_{\gamma\in\set{2,\ldots,m}}1_{x_{i} = \gamma,x_{B''} = \beta'', x_{B'''} = \beta'''}.
  \]
  The statement now follows as in the base case using the inductive hypothesis.
\end{proof}

Thus, letting $k = \max_i k_i$, we have that $V_\ell = \set{0}$ for all $\ell > n-k$, and
thus the space of real-valued functions over $\mathcal{U}_{\vec{k}}$ can be written as
$V_{=0}(\mathcal{U}_{\vec{k}}) \oplus V_{=1}(\mathcal{U}_{\vec{k}})\oplus\ldots\oplus V_{=n-k}(\mathcal{U}_{\vec{k}})$.

\subsubsection{The representation-theoretic decomposition over $\mathcal{U}_{\vec{k}}$}\label{sec:rep_multi}
In this section we lift the decomposition according to partitions from $S_n$ to the multi-slice. The basic
observation is that any function $f\colon \mathcal{U}_{\vec{k}}\to\mathbb{R}$ may be identified with a function
$\tilde{f}\colon S_n\to\mathbb{R}$ in the following way. Partition $[n]$ into $m$ sets, $K_1,\ldots,K_m$ where
$K_j$ has size $k_j$, and define $\tilde{f}(\pi)  = f(x)$ where $x$ is defined as
$x_{i} = j$ for all $i\in \pi(K_j)$. It is easy to note that the mapping $f\mapsto \tilde{f}$ is linear, and
sends $d$-juntas to $d$-juntas; it thus makes sense to define
\[
V_{\lambda}(\mathcal{U}_{\vec{k}}) = \sett{f}{\tilde{f}\in V_{\lambda}(S_n)},
\qquad
V_{=\lambda}(\mathcal{U}_{\vec{k}}) = \sett{f}{\tilde{f}\in V_{=\lambda}(S_n)}.
\]
This allows us to establish an analogue of Claim~\ref{claim:subrep_lb_sn} for the multi-slice, and for that we first define the action of $S_n$ on
functions over the multi-slice as $(\pi,f)\rightarrow ^{\pi} f$, where $^{\pi}f\colon\mathcal{U}_{\vec{k}}\to\mathbb{R}$ is given by $^{\pi}f(x) = f(^{\pi}x)$.
Here and throughout, for $x\in [m]^n$ and $\pi\in S_n$ we denote by $^{\pi} x$ the vector whose $i$th coordinate is $x_{\pi(i)}$.

\begin{claim}\label{claim:subrep_lb_multislice}
  Suppose $f\in V_{=\lambda}(\mathcal{U}_{\vec{k}})$ is not identically $0$. Then
  ${\sf dim}({\sf span}(\sett{^{\pi} f}{\pi\in S_n}))\geq {\sf dim}(\lambda)$.
\end{claim}
\begin{proof}
  Let $W = {\sf span}(\sett{^{\pi} f}{\pi\in S_n})$, and
  let $\tilde{W} = \sett{\tilde{f}}{f\in W}$. Noting that $\tilde{W} = \sett{^{\pi} \tilde{f}}{\pi\in S_n}$,
  we have by Claim~\ref{claim:subrep_lb_sn} that ${\sf dim}(\tilde{W})\geq {\sf dim}(\lambda)$. As ${\sf dim}(\tilde{W})\leq {\sf dim}(W)$
  the claim follows.
\end{proof}

\subsubsection{An alternative description}\label{sec:alternative}
Lastly, we need an alternative, equivalent description of the spaces $V_{\lambda}(\mathcal{U}_{\vec{k}})$
and $V_{=\lambda}(\mathcal{U}_{\vec{k}})$ in terms of cosets. For $\lambda\vdash n$,
a $\lambda$-tabloid is a partition of $[n]$ into $A = (A_1,\ldots,A_r)$, and a partition of the multiset $\set{1,\ldots,1,\ldots,m,\ldots,m}$ (where $j$ appears $k_j$ times)
into $B = (B_1,\ldots,B_r)$, where $\card{A_i} = \card{B_i} = \lambda_i$.
Given a $\lambda$-tabloid $(A,B)$, we define the corresponding coset as
\[
\tilde{T}_{A,B} = \sett{x\in\mathcal{U}_{\vec{k}}}{\sett{x_\ell}{\ell\in A_i} = B_i\qquad \forall i=1,\ldots,r}.
\]
Following the map $f\mapsto\tilde{f}$ directly, one sees that it sends $\lambda$-tabloids over $\mathcal{U}_{\vec{k}}$ into
$\lambda$-tabloids over $S_n$, and so $V_{\lambda}(\mathcal{U}_{\vec{k}})$ is the span of $1_{\tilde{T}_{A,B}}$ for all $\lambda$-tabloids
$(A,B)$. It also follows that
\[
V_{=\lambda}(\mathcal{U}_{\vec{k}})
=V_{\lambda}(\mathcal{U}_{\vec{k}})\cap \bigcap_{\mu\vartriangleright \lambda} V_{\mu}(\mathcal{U}_{\vec{k}})^{\perp}.
\]

We will sometimes want to show that the space $V_{\lambda}$ is an invariant space for some operators, and
for that it will be useful to further decompose $V_{\lambda}$.
Towards this end, for a partition $A = (A_1,\ldots,A_r)$ of $[n]$ into $r$ sets where
$\card{A_i} = \lambda_i$, we define
\[
V_{A}(\mathcal{U}_{\vec{k}}) = \sett{f\colon\mathcal{U}_{\vec{k}}\to\mathbb{R}}{^{\pi} f = f\qquad\forall \pi\in S_n \text{ such that }\pi(A_1) = A_1,\ldots,\pi(A_r) = A_r}.
\]
\begin{claim}\label{claim:decompose_into_symmetries}
For all partitions $\lambda = (\lambda_1,\ldots,\lambda_r)$ of $n$ we have
\[
V_{\lambda}(\mathcal{U}_{\vec{k}}) = \sum\limits_{\substack{A = (A_1,\ldots,A_r)\\ \card{A_i} = \lambda_i}} V_{A}(\mathcal{U}_{\vec{k}}).
\]
\end{claim}
\begin{proof}
The direction $\subseteq$ is trivial: it is enough to observe that $1_{\tilde{T}_{A,B}}$ is in $V_A(\mathcal{U}_{\vec{k}})$ for all
$\lambda$-cosets $T_{A,B}$, which is clear.

For the other direction, fix $A = (A_1,\ldots,A_r)$, let $f\in V_{A}(\mathcal{U}_{\vec{k}})$, and fix a sequence of multi-sets $B_1,\ldots,B_r\subseteq[m]$
for which $\card{B_i}=\lambda_i$. We claim that $f$ is constant on the coset $\tilde{T}_{A,B}$. Indeed, let $x,y\in\tilde{T}_{A,B}$
then for all $j$, the multi-sets $\sett{x_i}{i\in A_j}$ and $\sett{y_i}{i\in A_j}$ are equal to $B_j$ and hence the same, so we may
find a permutation $\pi_j$ that maps $A_j$ to itself and has $[n]\setminus A_j$ as fixed point, such that $y_{\pi(i)} = x_i$ for all
$i\in A_j$. Thus, letting $\pi = \pi_1\circ\pi_2\circ\ldots\circ\pi_m$ we have that $x = \pi(y)$ and $\pi(A_j) = A_j$ for all $j$,
and so $f(x) = f(\pi(y)) = ^{\pi}f(y) = f(y)$. Thus, we get that $f$ is constant on any coset $\tilde{T}_{A,B}$ for any multi-sets
$B_1,\ldots,B_r\subseteq[m]$ such that $\card{B_i} = \lambda_i$, and we denote this value by $f_{A,B}$. Since there co-sets are disjoint,
we get that
\[
f = \sum\limits_{B} f_{A,B} 1_{\tilde{T}_{A,B}} \in V_{\lambda}(\mathcal{U}_{\vec{k}}).\qedhere
\]
\end{proof}

\subsection{Hypercontractivity}
We will need the following variant of the hypercontractive inequality over general product space. For that, given
a probability measure $\mathcal{D}$ on a finite space $\Omega$, we define expectation $L^q$ norms as
$\norm{f}_{q,\mathcal{D}} = \left(\Expect{{\bf x}\sim \mathcal{D}}{\card{f({\bf x})}^{q}}\right)^{1/q}$.
We often drop the $\mathcal{D}$ subscript from the notation whenever it is clear from context.

\begin{thm}\label{thm:hypercontractivity_prod}\cite[Theorem 10.21]{Od}
  Let $(\Omega = \Omega_1\times\ldots\times\Omega_n, \nu = \nu_1\times\ldots\times\nu_n)$ be a product space.
  Suppose $\alpha\in (0,1/2)$ is such that $\nu_i(\omega) \geq \alpha$ for all $i\in [n]$ and $\omega\in\Omega_i$.
  Then for all $d\in\mathbb{N}$, $p\geq 2$ and $f\colon\Omega\to\mathbb{R}$ of degree at most $d$ we have
  $\norm{f}_q\leq \left(\frac{10 q}{\alpha}\right)^d\norm{f}_2$.
\end{thm}

We will also need the following variant of the hypercontractive inequality on the multi-slice from~\cite{FKLM}. Again,
we use here expectation norms according to the uniform measure on $\mathcal{U}_{\vec{k}}$.
\begin{thm}\label{thm:hypercontractivity_multi}
  For all $c>0$, $d\in\mathbb{N}$, and $q\in\mathbb{N}$ there is $N = N(c,d,q)>0$ and $C = C(c,d,q)>0$ such that the following holds.
  Let $n\geq N$, $\mathcal{U}_{\vec{k}}\subseteq[m]^n$ be a $c$-balanced multi-slice and
  let $f\colon\mathcal{U}_{\vec{k}}\to\mathbb{R}$ be a function of degree at most $d$.
  Then $\norm{f}_q\leq C\norm{f}_2$.
\end{thm}

\subsection{Tail bounds}
We need the following standard tail bound.
\begin{thm}\label{thm:chernoff}
  Suppose ${\bf Z}_1,\ldots,{\bf Z}_n$ are independent random bits with the same expectation $p$. Then
  \[
  \Prob{}{\card{\sum\limits_{i=1}^{n} {\bf Z}_i - pn}\geq \eps n}\leq 2 e^{-2\eps^2 n}.
  \]
\end{thm}

We also need a tail bound for negatively {\it associated random variables}, introduced in~\cite{negativeassoc}.
\begin{definition}
  We say a collection of random variables ${\bf Z}_1,\ldots,{\bf Z}_n$ over $\mathbb{R}$ is negatively associated if for each pair of disjoint sets $I, J\subset [n]$
  and any two increasing functions $f_I\colon\mathbb{R}^I\to\mathbb{R}$, $f_J\colon\mathbb{R}^J\to\mathbb{R}$ it holds that the random variables $f_I(({\bf Z}_i)_{i\in I})$
  and $f_J(({\bf Z}_j)_{j\in J})$ have non-positive covariance.
\end{definition}

\begin{remark}\label{remark:neg_assoc}
An important example of negatively associated random that we will use proceeds as follows. Suppose an urn contains $n$ balls of different colors $i=1,\ldots,n$,
and we pick $k<n$ balls from it randomly without replacement. Then, letting ${\bf Z}_i$ be the random variable indicating that the ball of color $i$ has been chosen,
we have that the collection ${\bf Z}_i$ for $i=1,\ldots,n$ is negatively associated~\cite{negativeassoc}.

We will make use of this fact throughout the paper as follows. Suppose $\mathcal{U}_{\vec{k}}$ is a multi-slice over $[m]$.
We will be concerned with distributions $(x,y)\in \mathcal{U}_{\vec{k}}\times \mathcal{U}_{\vec{k}}$
which are uniform over all pairs such that the number of coordinates $i$ such that $(x_i,y_i) = (a,b)$ is some fixed number $n_{a,b}$ for all $a,b\in [m]$.
It is thus clear that fixing $a,b$, defining for each $i\in [n]$ the random variable ${\bf Z}_i$ indicating that $(x_i,y_i) = (a,b)$, the collection
${\bf Z}_i$ is of the above form, and hence is negatively associated.
\end{remark}
The following result is~\cite[Proposition 5]{negativeassoc2}.
\begin{thm}\label{thm:chernoff_negatively}
  Suppose ${\bf Z}_1,\ldots,{\bf Z}_n$ are negatively correlated random bits with the same expectation $p$. Then
  \[
  \Prob{}{\sum\limits_{i=1}^{n} {\bf Z}_i\geq (p+\eps)n}\leq e^{-2\eps^2 n}.
  \]
\end{thm}

\section{A basic invariance principle: the bilinear case}
%\begin{definition}
%  We say a distribution $\mu$ over $([m_1]\times\ldots\times[m_r])^n$ is symmetric under the action of $S_n$
%  (or simply symmetric) if for all $\pi\in S_n$, the following distributions are the same:
%  (a) $(\pi({\bf x}(1)), \ldots, \pi({\bf x}(r)))$, where $({\bf x}(1),\ldots,{\bf x}(r))\sim \mu$,
%  (b) $\mu$.
%\end{definition}

\begin{definition}
  We say a distribution $\mu$ over $([m]\times [m'])^n$ commutes with the action of $S_n$ if the following distributions
  are the same for all $\pi\in S_n$ and $x\in [m]^n$:
  (a) ${\bf x}'$, where $({\bf x},{\bf x'})\sim \mu$ conditioned on ${\bf x} = \pi(x)$,
  (b) $\pi({\bf x'})$, where $({\bf x},{\bf x'})\sim\mu$ conditioned on ${\bf x} = x$.
\end{definition}

\begin{claim}\label{claim:sym_implies_commute}
  Suppose a distribution $\mu$ over $[m]^n\times[m']^n$ is symmetric under the action of $S_n$. Then $\mu$
  commutes with the action of $S_n$.
\end{claim}
\begin{proof}
  Fix $x,y\in [m]^n$ and $\pi\in S_n$. Then
  \[
  \cProb{{({\bf x}, {\bf y})}}{{\bf x} = \pi(x)}{{\bf y} = y}
  =\frac{\mu(\pi(x),y)}{\Prob{({\bf x},{\bf y})\sim\mu}{{\bf x} = \pi(x)}},
  \qquad
  \cProb{{({\bf x}, {\bf y})}}{{\bf x} = x}{\pi({\bf y}) = y}
  =\frac{\mu(x,\pi^{-1}(y))}{\Prob{({\bf x},{\bf y})\sim\mu}{{\bf x} = x}}.
  \]
  By the symmetry of $\mu$, we have that
  \[
  \mu(\pi(x),y)
  =\Prob{({\bf x},{\bf y})\sim\mu}{{\bf x} = \pi(x), {\bf y} = y}
  =\Prob{({\bf x},{\bf y})\sim\mu}{\pi^{-1}({\bf x}) = x, \pi^{-1}({\bf y}) = \pi^{-1}(y)}
  =\mu(x,\pi^{-1}(y)),
  \]
  which shows that the numerators of the two fractions are equal; summing this over $y$ also shows that
  the denominators of the two fractions are equal.
\end{proof}

\subsection{Couplings between measures on $[m]^n$}
Consider the multi-slice $\mathcal{U}_{\vec{k}}$ over alphabet $[m]$, and consider the
distribution over $[m]$ defined by $\nu_{\vec{k}}(i) = k_i/n$. We will show that in several senses, the multi-slice
$\mathcal{U}_{\vec{k}}$ behaves very similarly to the product domain $([m]^n,\nu_{\vec{k}}^{\otimes n})$, and for
that we will use couplings, which we defined earlier in the introduction and recall for convenience below.
%We remark that the definition below is more general so as to capture more
%cases that will arise in our arguments.
\begin{definition}
  Let $\alpha,\zeta>0$, $m\in\mathbb{N}$.  For symmetric distributions $\nu_1,\nu_2$ over $[m]^n$,
  a $(\alpha,\zeta)$-coupling between $([m]^n,\nu_1)$ and
  $([m]^n,\nu_2)$ is a jointly distributed $\mathcal{C} = ({\bf x}, {\bf y})$ satisfying the following properties.
  \begin{enumerate}
    \item The marginal distribution of ${\bf x}$ is $\nu_1$, and the marginal distribution of ${\bf y}$ is $\nu_2$.
    \item The distribution of $\mathcal{C}$ is symmetric under the action of $S_n$.
    \item For all $i\in [n]$, $\Prob{({\bf x},{\bf y})\sim\mathcal{C}}{{\bf x}_i\neq {\bf y}_i}\leq \zeta$.
    \item Tail bounds: for all $\eps>0$, it holds that $\Prob{({\bf x},{\bf y})\sim\mathcal{C}}{\card{\sett{i}{x_i\neq y_i}}\geq \eps n}\leq
    \frac{1}{\alpha} e^{-\alpha\eps^2 n}$.
%   For each $i\in [n]$, it holds that $\Prob{(x,y)\sim \mathcal{C}}{x_i \neq y_i}\leq \delta$.
%    \item The coupling commutes with the action of $S_n$: namely, for each $\pi\in S_n$, the distribution
%    of $(\pi(x),\pi(y))$ is identical to the distribution of $(x',y')$ where we sample $(a,b)\sim \mathcal{C}$,
%    set $x' = \pi(a)$, and then sample $(a',y')\sim\mathcal{C}$ conditioned on $a' = x'$.
  \end{enumerate}
\end{definition}
We also recall that given a coupling $\mathcal{C}$ between $(\mathcal{U}_{\vec{k}},{\sf Uniform})$ and $([m]^n,\nu_{\vec{k}}^{\otimes n})$,
we may define lifting operators on functions between the two spaces, namely
$\mathrm{T}_{\mathcal{C}}\colon L^2(\mathcal{U}_{\vec{k}})\to L^2([m]^{n},\nu_{\vec{k}}^{\otimes n})$ and
its adjoint operator $\mathrm{T}_{\mathcal{C}}^{*}\colon L^2([m]^{n},\nu_{\vec{k}}^{\otimes n})\to L^2(\mathcal{U}_{\vec{k}})$, as:
\[
\mathrm{T}_{\mathcal{C}} f(y) = \cExpect{({\bf x}, {\bf y})\sim \mathcal{C}}{{\bf y} = y}{f({\bf x})},
\qquad
\mathrm{T}_{\mathcal{C}}^{*} g(x) = \cExpect{({\bf x},{\bf y})\sim \mathcal{C}}{{\bf x} = x}{f({\bf y})}.
\]
%Given a coupling $\mathcal{C}$ between $\mathcal{U}_{\vec{k}}$ with the uniform measure and $([m]^n,\nu_{\vec{k}}^{\otimes n})$,
%we may use it as a tool to lift functions from $\mathcal{U}_{\vec{k}}$ to functions over
%$([m]^{n},\nu_{\vec{k}}^{\otimes n})$, and vice versa. Towards this end, we introduce the operator
%$\mathrm{T}_{\mathcal{C}}\colon L^2(\mathcal{U}_{\vec{k}})\to L^2([m]^{n},\nu_{\vec{k}}^{\otimes n})$ and
%its adjoint operator $\mathrm{T}_{\mathcal{C}}^{*}\colon L^2([m]^{n},\nu_{\vec{k}}^{\otimes n})\to L^2(\mathcal{U}_{\vec{k}})$, defined
%as
%\[
%\mathrm{T}_{\mathcal{C}} f(y) = \cExpect{({\bf x}, {\bf y})\sim \mathcal{C}}{{\bf y} = y}{f({\bf x})},
%\qquad
%\mathrm{T}_{\mathcal{C}}^{*} g(x) = \cExpect{({\bf x},{\bf y})\sim \mathcal{C}}{{\bf x} = x}{f({\bf y})}.
%\]

\subsection{Invariance for low-degree functions: Proof of Lemma~\ref{lem:inv_low_deg}}
A key property of couplings is that at least for low-degree functions $f\colon \mathcal{U}_{\vec{k}}\to\mathbb{R}$,
we have that $f({\bf x})\approx \mathrm{T}_{\mathcal{C}}f({\bf y})$ where
$({\bf x},{\bf y})\sim \mathcal{C}$. This is essentially the content of Lemma~\ref{lem:inv_low_deg}, restated below.
We remark that due to technical reasons, we will use two different couplings
$\mathcal{C}$ and $\mathcal{C}'$. Here $\mathcal{C}$, will be used to lift functions from the multi-slice to the product space,
and $\mathcal{C}'$ will be used to sample the inputs $({\bf x},{\bf y})$.
\begin{replemma}{lem:inv_low_deg}
Let $d,m,n\in\mathbb{N}$, $\alpha, \zeta>0$, and let $\mathcal{U}_{\vec{k}}$ be multi-slice over alphabet $[m]$.
Suppose $f\colon \mathcal{U}_{\vec{k}}\to\mathbb{R}$ is a function of degree at most $d$, and $\mathcal{C},\mathcal{C}'$ are $(\alpha,\zeta)$-coupling
between $(\mathcal{U}_{\vec{k}})$ and $([m]^n, \nu_{\vec{k}}^{\otimes n})$. Then
\[
\Expect{({\bf x},{\bf y})\sim \mathcal{C}'}
{(f({\bf x}) - \mathrm{T}_{\mathcal{C}} f({\bf y}))^2} \leq  8 \sqrt{d\zeta}\norm{f}_2^2.
\]
\end{replemma}

The rest of this section is devoted to the proof of Lemma~\ref{lem:inv_low_deg}.

\subsubsection{Proof overview}
It will be more convenient for us to consider operators that map a space to itself, and more specifically
we will consider
$\mathcal{S}\colon L^{2}(\mathcal{U}_{\vec{k}})\to L^{2}(\mathcal{U}_{\vec{k}})$ defined as
$\mathcal{S} = \mathrm{T}_{\mathcal{C}'}^{*}\mathrm{T}_{\mathcal{C}}$.
With these definitions, expanding the difference in~Lemma~\ref{lem:inv_low_deg}
we see that
\begin{align}\label{eq:1}
\Expect{({\bf x},{\bf y})\sim \mathcal{C}'}
{(f({\bf x}) - \mathrm{T}_{\mathcal{C}} f({\bf y}))^2}
=
\norm{f}_2^2+\norm{\mathrm{T}_{\mathcal{C}} f}_2^2
-2\inner{f}{\mathcal{S} f}
\leq 2(\norm{f}_2^2 - \inner{f}{\mathcal{S} f})
=2\inner{f}{(\mathrm{I}-\mathcal{S}) f}.
\end{align}
We would like to reason about the eigenvalues of $\mathrm{I}-\mathcal{S}$ and say they are close to $0$ for low-degree functions,
however this may be problematic since $\mathcal{S}$ may not be diagonizable.
We thus use a standard trick by Cauchy-Schwarz to bound
\begin{align}
~\eqref{eq:1}
\leq 2\norm{f}_2\norm{(\mathrm{I}-\mathcal{S}) f}_2
=2\norm{f}_2\sqrt{\inner{(\mathrm{I}-\mathcal{S}) f}{(\mathrm{I}-\mathcal{S}) f}}
&=2\norm{f}_2\sqrt{\inner{f}{(\mathrm{I}-\mathcal{S})^{*}(\mathrm{I}-\mathcal{S}) f}} \notag\\\label{eq:22}
&\leq 2\norm{f}_2^{3/2}\norm{{(\mathrm{I}-\mathcal{S})^{*}(\mathrm{I}-\mathcal{S}) f}}_2^{1/2}.
\end{align}
Thus, we reduced our question to understanding the operator $\mathcal{S}^{\star}  = (\mathrm{I}-\mathcal{S})^{*}(\mathrm{I}-\mathcal{S})$,
for which we may study eigenvalues since it is self-adjoint.
Indeed, we show that it strongly contracts low degree functions, and for that we follow the following plan:
\begin{enumerate}
  \item We argue that $\mathcal{S}$ and its adjoint preserves juntas.  Therefore $\mathcal{S}^{\star}$ preserves juntas.
  \item We deduce that for each eigenvalue $\theta$ of $\mathcal{S}^{\star}$ corresponding to a low-degree eigenfunction $f$,
  there is a junta $g$ which is also an eigenfunction of $\mathcal{S}^{\star}$ with eigenvalue $\theta$.
  \item Finally, we note that the coupling operator is unlikely to touch a small set of coordinates
  (given that $n$ is sufficiently large in comparison to $d$), so we must have that any eigenvalue $\theta$
  corresponding to a junta must be close to $0$.
  \item Combining the above two items with Parseval's equality establishes Lemma~\ref{lem:inv_low_deg}.
\end{enumerate}
We begin with the formal proof of Lemma~\ref{lem:inv_low_deg} following the above outline.

\subsubsection{$\mathcal{S}$ preserves juntas}
For a set $J$, we denote by $V_J(\mathcal{U}_{\vec{k}})$ the set of $J$-juntas. Note that we may write
$V_{\leq d}(\mathcal{U}_{\vec{k}})$ as the sum of all $V_J(\mathcal{U}_{\vec{k}})$.
Recall the action of $S_n$ on functions over the multi-slice is defined as:
for $\pi\in S_n$, $f\colon\mathcal{U}_{\vec{k}}\to\mathbb{R}$ define $^{\pi} f = f(\pi(x))$.
To prove that $\mathcal{S}$ preserves juntas, it will be more convenient for us to use the following equivalent
criteria.
\begin{fact}\label{fact:junta_trivial}
  A function $f\colon\mathcal{U}_{\vec{k}}\to\mathrm{R}$ is a $J$-junta if and only if
  $^{\pi} f = f$ for all $\pi$ that fix $J$.
\end{fact}
\begin{proof}
The $\Rightarrow$ direction is clear, and we prove the $\Leftarrow$ direction. Suppose $^{\pi} f = f$ for all $\pi$ that
fix $J$, and define $g\colon [m]^J\to\mathbb{R}$ as $g(\alpha) = \Expect{\substack{x\in\mathcal{U}_{\vec{k}}\\ x_J = \alpha}}{f(x)}$; it
suffices to argue that $f(x) = g(x_J)$ for all $x$. Indeed, fix $x$, and denote $\alpha_J$. If $x'\in\mathcal{U}_{\vec{k}}$ is a point
such that $x'_J = \alpha$, then we have that for all $\ell = 1,\ldots,m$, the sets
\[
X_\ell = \sett{i\in[n]\setminus J}{x_i = \ell},
\qquad
X'_{\ell} = \sett{i\in[n]\setminus J}{x'_i = \ell}
\]
have the same size, and so there is $\pi\in S_{[n]\setminus J}$ that maps each $X_{\ell}$ to $X'_{\ell}$. Thus,
there is $\pi$ fixing $J$ such that $x' = \pi(x)$, and so $f(x') = ^{\pi} f(x) = f(x)$.
\end{proof}
\begin{claim}\label{claim:commute}
  For all $\pi\in S_n$ and $f\colon\mathcal{U}_{\vec{k}}\to\mathbb{R}$, we have that
  $\mathcal{S}(^{\pi} f) = ^{\pi}(\mathcal{S} f)$, as well as $\mathcal{S}^{*}(^{\pi} f) = ^{\pi}(\mathcal{S}^{*} f)$.
\end{claim}
\begin{proof}
  We prove the first assertion -- the proof of the second assertion is analogous.
  By Claim~\ref{claim:sym_implies_commute}, we get that $\mathcal{C},\mathcal{C}'$ commute with the action of $S_n$.
  Fix $x\in\mathcal{U}_{\vec{k}}$ and sample ${\bf y}$ as $({\bf x},{\bf y})\sim \mathcal{C}$
  conditioned on ${\bf x} = x$, and then $({\bf x}',{\bf y}')\sim \mathcal{C}'$ conditioned on ${\bf y}' = {\bf y}$.
  Then by definition of $\mathcal{S}$ and the definition on the adjoint operator we have
  \[
  \mathcal{S}(^{\pi} f)(x)
  = \Expect{{\bf x}'\sim \mathcal{S} x}{^{\pi} f({\bf x}')}
  = \Expect{{\bf x}'\sim \mathcal{S} x}{f(\pi({\bf x}'))}.
  \]
  Consider the points $x''=\pi(x)$, ${\bf y}''=\pi({\bf y}')$.
  Since $\mathcal{C}$ commutes with $S_n$,
  the distribution of ${\bf y}''$ is the same as of ${\bf b}$ where
  $({\bf a},{\bf b})\sim \mathcal{C}$ conditioned on ${\bf a} = x''$;
  also, the distribution of $\pi({\bf x}')$ the same as of ${\bf a}'$ where
  $({\bf a}',{\bf b}')\sim\mathcal{C}'$ conditioned on ${\bf b}' = {\bf b}$. Together,
  these two facts imply that the distribution of $\pi({\bf x}')$ is $\mathcal{S} \pi(x)$, and
  the above average is equal to $\mathcal{S}f(\pi(x)) = ^{\pi}(\mathcal{S} f)(x)$, as required.
\end{proof}

\begin{claim}\label{claim:preserve_junta}
  Suppose $\mathcal{R}$ is an operator that commutes with the action of $S_n$ on functions over the multi-slice.
  Then for each $J\subseteq[n]$, we have that $\mathcal{R} \left(V_J(\mathcal{U}_{\vec{k}})\right)\subseteq V_J(\mathcal{U}_{\vec{k}})$.
  Subsequently, $\mathcal{R}$ preserves $V_{d}(\mathcal{U}_{\vec{k}})$.
\end{claim}
\begin{proof}
  Let $\pi\in S_n$ be a permutation that fixes $J$, and let $f$ be a $J$-junta. Then,
  by Claim~\ref{claim:commute} we have
  $^{\pi}(\mathcal{S} f) = \mathcal{S}(^{\pi} f) = \mathcal{S}(f)$, where the last transition is since $f$ is a $J$-junta and
  $\pi$ fixes $J$. Since this holds for all $\pi\in S_n$ that fixes $J$, we get that $\mathcal{S}(f)$ is a $J$-junta.

  The ``subsequently'' part follows as $V_{d}(\mathcal{U}_{\vec{k}}) = \sum_{\card{J}\leq d} V_J(\mathcal{U}_{\vec{k}})$.
\end{proof}
From Claims~\ref{claim:commute},~\ref{claim:preserve_junta} it follows that
$V_J(\mathcal{U}_{\vec{k}})$ is invariant under $\mathcal{S}^{\star}$, and
since $\mathcal{S}^{\star}$ is self-adjoint we may decompose this space as sum of eigenspaces.
Let us denote by $V_J^{\theta}(\mathcal{U}_{\vec{k}})\subseteq V_J(\mathcal{U}_{\vec{k}})$ the subspace
of eigenvectors with eigenvalue $\theta$, and write $V_{\leq d}(\theta) = \sum_{\card{J}\leq d}V_J^{\theta}(\mathcal{U}_{\vec{k}})$.
\begin{claim}\label{claim:decompose_low_deg_ev}
  $V_{\leq d}(\mathcal{U}_{\vec{k}}) = \bigoplus_{\theta} V_{\leq d}(\theta)$.
\end{claim}
\begin{proof}
  To see that the space on the left hand side is the sum of the spaces on the right hand side, note that
  \[
    \sum_{\theta} V_{\leq d}(\theta) =
    \sum_{\theta} \sum_{\card{J}\leq d}V_J^{\theta}(\mathcal{U}_{\vec{k}})=
    \sum_{\card{J}\leq d} \sum_{\theta} V_J^{\theta}(\mathcal{U}_{\vec{k}})=
    \sum_{\card{J}\leq d} V_J(\mathcal{U}_{\vec{k}})=
    V_{\leq d}(\mathcal{U}_{\vec{k}}).
  \]
  The statement now follows as the spaces $V_{\leq d}(\theta)$ are orthogonal for different $\theta$'s.
\end{proof}

\subsubsection{Bounding the eigenvalues of $\mathcal{S}^{\star}$}
Next, we will estimate the eigenvalues of $\mathcal{S}^{\star}$ corresponding to low-degree functions. We begin
by the following easy claim.
\begin{claim}\label{claim:coup_small_noise}
  Let $\mathcal{U}_{\vec{k}}$ be a $c$-balanced multi-slice over alphabet $[m]$, let $J\subseteq[n]$ be of size
  at most $d$, and let $x\in\mathcal{U}$. Then
  \[
  \Prob{{\bf x}'\sim \mathcal{S}x}{{\bf x}'_J\neq x_J}\leq
  2d\zeta
  %2d\left(\delta+\frac{1}{(1-e^{-c})\sqrt{n}}\right).
  \]
\end{claim}
\begin{proof}
The claim is immediate by the third property in Definition~\ref{def:coupling} and the union bound.
\end{proof}

\begin{claim}\label{claim:ev_lb_ld}
  Let $\theta$ be such that $V_{\leq d}(\theta)\neq\set{0}$. Then $\theta\leq 16d\zeta$.
\end{claim}
\begin{proof}
  By assumption, $V_{\leq d}(\theta)\neq \set{0}$ so by definition there is $J$ of size at most $d$ such that
  $V_J^{\theta}(\mathcal{U}_{\vec{k}})\neq\set{0}$, and we take $g\in V_J^{\theta}(\mathcal{U}_{\vec{k}})$.
  Let $x\in\mathcal{U}_{\vec{k}}$ be a point achieving the maximum of $\card{g(x)}$ (note that it is necessarily positive);
  we may assume without loss of generality that $g(x) > 0$, otherwise we work with the function $-g$.
  Let ${\bf x}'\sim \mathcal{S} x$, then
  \begin{align*}
  \mathcal{S} g(x)
  = \Expect{{\bf x}'}{g({\bf x}')}
  &\geq  \Prob{{\bf x}'}{{\bf x}'_J = x_J} \cExpect{{\bf x}'}{{\bf x}'_J = x_J}{g({\bf x}')}
  - (1-\Prob{{\bf x}'}{{\bf x}'_J = x_J})g(x)\\
  &= (2\Prob{{\bf x}'}{{\bf x}'_J = x_J}-1)g(x),
  \end{align*}
  which by Claim~\ref{claim:coup_small_noise} is at least $(1-4d\zeta)g(x)$. Since clearly
  $\mathcal{S} g(x)\leq g(x)$, it follows that $\card{\mathcal{S} g(x) - g(x)}\leq 4d\zeta$.
  A similar argument shows that $\card{\mathcal{S}^{*} g(x) - g(x)}\leq 4d\zeta g(x)$, and
  also that $\card{\mathcal{S}^{*}\mathcal{S} g(x) - g(x)}\leq 8d\zeta g(x)$.
  We thus get that
  \[
   \theta g(x)
   =\card{\mathcal{S}^{\star} g(x)}
   =
   \card{g(x) - \mathcal{S} g(x) - \mathcal{S}^{*} g(x)
   +\mathcal{S}^{*}\mathcal{S} g(x)}
   \leq  16d\zeta g(x),
  \]
  and dividing by $g(x)$ finishes the proof.
\end{proof}

\subsubsection{Proof of Lemma~\ref{lem:inv_low_deg}}
  Let $f\in V_{\leq d}(\mathcal{U}_{\vec{k}})$.
  From~\eqref{eq:1} and~\eqref{eq:22} we get that
  \[
  \Expect{({\bf x},{\bf y})\sim \mathcal{C}'}
{(f({\bf x}) - \mathrm{T}_{\mathcal{C}} f({\bf y}))^2}\leq 2\norm{f}_2^{3/2}\norm{{\mathcal{S}^{\star} f}}_2^{1/2},
  \]
  and we upper bound the last norm.  By Claim~\ref{claim:decompose_low_deg_ev} we may
  write $f = \sum\limits_{\theta} f^{=\theta}$ where $f^{=\lambda}\in V_{\leq d}(\theta)$, and so
  \[
  \norm{{\mathcal{S}^{\star} f}}_2^2
  =\sum\limits_{\theta} \norm{\mathcal{S}^{\star} f^{=\theta}}_2^2
  =\sum\limits_{\theta}\theta^2 \norm{f^{=\theta}}_2^2
  \leq (16d\zeta)^2\sum\limits_{\theta}\norm{f^{=\theta}}_2^2
  =(16d\zeta)^2\norm{f}_2^2
  \]
  where we used Claim~\ref{claim:ev_lb_ld}. Plugging this into the inequality above finishes the proof.

\subsection{Invariance for admissible distributions: the bilinear case}
Next, we state and prove an invariance principle for admissible distributions as in Definition~\ref{def:admissible}. Let
$\mu$ be an admissible distribution on $\mathcal{U}_{\vec{k}}\times\mathcal{U}_{\vec{k}'}$ where the multi-slices are
over alphabets $m,m'$ respectively, and let $\tilde{\mu}$ be the product version of $\mu$ as per Definition~\ref{def:product_ver}.
In this section, we will prove Theorem~\ref{thm:basic_inv_multi} for the case that $r=2$, which captures the main ideas of
the argument, and in the subsequent section we generalize it for all $r\in\mathbb{N}$. Throughout, we will denote samples
of $\mu$ by $({\bf x},{\bf x}')\sim\mu$, and samples of $\tilde{\mu}$ by $({\bf y},{\bf y}')\sim\tilde{\mu}$.
We will be interested in couplings between the distribution $\mu$ and $\tilde{\mu}$, as in Definition~\ref{def:coupling} --
namely, a jointly distributed $\mathcal{C} = ((x,x'),(y,y'))$ satisfying the properties in Definition~\ref{def:coupling}.

%Note that by definition, we get that the distribution of $(x_i,x'_i)$ where $(x,x')\sim \mu$
%is the same for all $i$. This motivates the definition of the product analogue of $\mu$ over $[m]^n\times [m']^{\otimes}$.
%\begin{definition}\label{def:product_ver}
%  Let $\mathcal{U}_{\vec{k}(1)},\ldots,\mathcal{U}_{\vec{k}(r)}$ be multislices over alphabets $[m_1],\ldots,[m_r]$,
%  and let $\mu$ be a distribution over $\mathcal{U}_{\vec{k}(1)}\times\ldots\times\mathcal{U}_{\vec{k}(r)}$.
%  For each $i\in[n]$, define the distribution $\tilde{\mu}_i$ over $[m_1]\times\ldots\times[m_r]$ as
%  \[
%  \tilde{\mu}_i(\vec{a}) = \Prob{({\bf x}(1),\ldots{\bf x}(r))\sim\mu}{{\bf x}(1)_i = a_1,\ldots,{\bf x}(r)_i = a_r}.
%  \]
%  Then, the product analogue of $\mu$ is defined as $\tilde{\mu} = \prod\limits_{i=1}^{n}\tilde{\mu}_i$.
%\end{definition}
%
%In this section, we will only be concerned with the case that $r=2$, and denote samples according to
%$\tilde{\mu}$ by $({\bf y},{\bf y'})$. As will be the case in our applications, we will be interested
%in couplings between the distribution $\mu$ and $\tilde{\mu}$, as in Definition~\ref{def:coupling} -- that is, a jointly
%distributed $\mathcal{C} = ((x,x'),(y,y'))$ satisfying the properties in Definition~\ref{def:coupling}.

\begin{thm}\label{thm:basic_inv_bipartite}
  For all $\alpha\in (0,1)$, $m,m'\in\mathbb{N}$, $\eps>0$ there is $\zeta>0$ such that the following holds.
  Suppose $\mathcal{U}_{\vec{k}}$ and $\mathcal{U}_{\vec{k}'}$ are $\alpha$-balanced multi-slices over alphabets
  $[m]$ and $[m']$ respectively, and $\mu$ is a connected, $\alpha$-admissible distribution over $\mathcal{U}_{\vec{k}}\times \mathcal{U}_{\vec{k}'}$.
  Suppose $\mathcal{C}$ is a $(\alpha,\zeta)$-coupling between $\mathcal{U}_{\vec{k}}$ and $\nu_{\vec{k}}^{\otimes n}$,
  $\mathcal{C}'$ is a $(\alpha,\zeta)$-coupling between $\mathcal{U}_{\vec{k}'}$ and $\nu_{\vec{k}'}^{\otimes n}$,
  and that there is a $(\alpha,\zeta)$-coupling $\mathcal{C}''$ between $\mu$ and $\tilde{\mu}$.

  Then for all $f\colon \mathcal{U}_{\vec{k}}\to \mathbb{R}$ and $g\colon\mathcal{U}_{\vec{k}'}\to \mathbb{R}$  it holds that
  \[
  \card{
  \Expect{(\bf{x},\bf{x}')\sim \mu}{f({\bf x})g({\bf x}')} -
  \Expect{(\bf{y},\bf{y}')\sim \tilde{\mu}}{\mathrm{T}_{\mathcal{C}}f({\bf y})\mathrm{T}_{\mathcal{C}'}g({\bf y}')}}\leq
  \eps \norm{f}_2\norm{g}_2.
  \]
\end{thm}
The rest of this section is devoted to the proof of Theorem~\ref{thm:basic_inv_bipartite}, and we begin by outlining
the idea of the proof.
\subsubsection{Proof overview}\label{sec:basic_inv_pf_overview}
We begin by noting that Lemma~\ref{lem:inv_low_deg} implies Theorem~\ref{thm:basic_inv_bipartite}
for low degree functions. Indeed, by adding and subtracting
$\Expect{\left(\substack{{\bf x},{\bf x}'\\{\bf y},{\bf y}'}\right)\sim\mathcal{C}''}{f({\bf x})\mathrm{T}_{\mathcal{C}'}g({\bf y}')}$,
the left hand side in Theorem~\ref{thm:basic_inv_bipartite} may be upper bounded by
\begin{align*}
&\card{\Expect{\left(\substack{{\bf x},{\bf x}'\\{\bf y},{\bf y}'}\right)\sim\mathcal{C}''}{f({\bf x})(g({\bf x}') - \mathrm{T}_{\mathcal{C}'}g({\bf y}'))}}
+\card{\Expect{\left(\substack{{\bf x},{\bf x}'\\{\bf y},{\bf y}'}\right)\sim\mathcal{C}''}{(f({\bf x})-\mathrm{T}_{\mathcal{C}}f({\bf y}))\mathrm{T}_{\mathcal{C}'}g({\bf y}')}}\\
&\leq
\norm{f}_2\sqrt{\Expect{\left(\substack{{\bf x},{\bf x}'\\{\bf y},{\bf y}'}\right)\sim\mathcal{C}''}{(g({\bf x}') - \mathrm{T}_{\mathcal{C}'}g({\bf y}'))^2}}
+\norm{\mathrm{T}_{\mathcal{C}'}g}_2\sqrt{\Expect{\left(\substack{{\bf x},{\bf x}'\\{\bf y},{\bf y}'}\right)\sim\mathcal{C}''}{(f({\bf x}) - \mathrm{T}_{\mathcal{C}}f({\bf y}))^2}},
\end{align*}
where we used Cauchy-Schwarz. By Jensen's inequality we have that $\norm{\mathrm{T}_{\mathcal{C}'}g}_2\leq \norm{g}_2$,
and by Lemma~\ref{lem:inv_low_deg} the expectations are
at most $8\sqrt{d\zeta}\norm{g}^2_2$ and $8\sqrt{d\zeta}\norm{f}^2_2$ respectively, yielding the bound $8(d\zeta)^{1/4}\norm{f}_2\norm{g}_2$.

This observation suggests that we should consider the contribution from the high-degree parts and the low-degree parts
in Theorem~\ref{thm:basic_inv_bipartite} separately, and show that the contribution from the high-degree parts is small.
More precisely, we will decompose $f$ and $g$ into their high degree and low-degree parts as
$f = f^{\leq d} + f^{>d}$ and $g = g^{\leq d} + g^{>d}$, and analyze the terms corresponding to
$f^{\leq d} g^{\leq d}$, $f^{> d} g^{\leq d}$ and $f g^{> d}$ separately.
We show that only the contribution from the first term is meaningful, i.e.~of $f^{\leq d}g^{\leq d}$, which by Lemma~\ref{lem:inv_low_deg} is very
close to $\Expect{}{\mathrm{T}_{\mathcal{C}}(f^{\leq d})({\bf y})\mathrm{T}_{\mathcal{C}'}(g^{\leq d})({\bf y}')}$.

To complete the proof, we have to show that the last quantity is very close to
$\Expect{}{\mathrm{T}_{\mathcal{C}}f({\bf y})\mathrm{T}_{\mathcal{C}'}g({\bf y})}$. The first step in this
argument is to observe yet another important property of $\mathrm{T}_{\mathcal{C}}, \mathrm{T}_{\mathcal{C}'}$, namely that
they nearly commute with degree truncations. This comes in the form of Lemma~\ref{lem:coupling_preserve_deg},
asserting that $\mathrm{T}_{\mathcal{C}}(f^{\leq d})$ and $(\mathrm{T}_{\mathcal{C}}f)^{\leq d}$ are close in $2$-norm, provided that
$\zeta\ll d^{-7}$ (and similarly for $\mathrm{T}_{\mathcal{C}}(g^{\leq d})$ is close to
$(\mathrm{T}_{\mathcal{C}'}g)^{\leq d}$). Thus, $\Expect{}{\mathrm{T}_{\mathcal{C}}(f^{\leq d})({\bf y})\mathrm{T}_{\mathcal{C}'}(g^{\leq d})({\bf y}')}$
is close to $\Expect{}{(\mathrm{T}_{\mathcal{C}} f)^{\leq d}({\bf y})(\mathrm{T}_{\mathcal{C}'}g)^{\leq d}({\bf y}')}$,
and it remains to show that the contribution from the high-degree parts is small. That is, we must show that the expectation of each one of the terms
$\mathrm{T}_{\mathcal{C}}(f)(\mathrm{T}_{\mathcal{C}'}g)^{> d}$ and
$(\mathrm{T}_{\mathcal{C}}f)^{> d}(\mathrm{T}_{\mathcal{C}}g)^{\leq d}$ is small in magnitude.
Such bounds have been established in previous works, e.g. in~\cite{MosselGaussian}, and we reproduce the argument in the
appendix for completeness.

Finally, we elaborate on how we show that the expectations corresponding to products involving a high-degree functions over the multi-slice are small. To do so,
it will be convenient to us to view the distribution $\mu$ as an operator $\mathrm{T}_{\mu}\colon L^2(\mathcal{U}_{\vec{k}})\to L^2(\mathcal{U}_{\vec{k}'})$
defined as $\mathrm{T}_{\mu} f (y) = \cExpect{({\bf x},{\bf y})\sim \mu}{{\bf y}=y}{f({\bf x})}$. Thus, we are interested in studying expressions of the form
$\inner{\mathrm{T}_{\mu}  f^{>d}}{g^{\leq d}}$, which by Cauchy-Schwarz is at most
$\norm{\mathrm{T}_{\mu}  f^{>d}}_2\norm{g^{\leq d}}$. Again, it will be more convenient for us to work with
operators from a space to itself, which we may do again by defining $\mathcal{S}_{\mu} = \mathrm{T}_{\mu}^{*} \mathrm{T}_{\mu}$
and using Cauchy-Schwarz to bound
\[
\norm{\mathrm{T}_{\mu} f^{>d}}_2^2
=\inner{f^{>d}}{\mathcal{S}_{\mu} f^{>d}}
\leq \norm{f^{>d}}_2\norm{\mathcal{S}_{\mu} f^{>d}}_2.
\]
Thus, it suffices to show that the eigenvalues of $\mathcal{S}_{\mu}$ corresponding to high-degree functions are small.
We prove the latter statement using the trace method: we show that on the one hand, the multiplicity of each one of them is exponentially large in $n$,
say at least $(1+\delta)^n$, and that the sum of some constant power of them is small. For the latter part, we consider a large enough
(but constant) power $h$, and show that the operator $\mathcal{S}_{\mu}^{h}$ mixes well enough so that its trace is at most $(1+\xi(h))^n$, where
$\lim_{h\rightarrow \infty} \xi(h) = 0$.

The above argument works for $d\geq \Omega(n)$, and does not immediately apply in the case that $d = o(n)$. The reason is that the upper bound
we have on the trace, $(1+o(1))^n$, is too weak in comparison to the multiplicity we expect an eigenvalue corresponding to degree $d$
functions to have (which we expect to be $2^{\Theta(d)}$).
Fortunately, there is an elegant fix. We show that we may assume the eigenvalue we are interested in bounding, say $\theta$,
corresponds to a $d$-junta (for the same reasons as in the proof of Lemma~\ref{lem:inv_low_deg}), and thus we may ``restrict our view''
to a set of $3d$ coordinates that contains the $d$-coordinates of the junta. Morally, this view allows us to assume that
$n=3d$, in which case the previous argument appears to be applicable. Due to technical reasons, however, we use a different and more
direct argument once we have ``focused our view'' on the set of $3d$ relevant coordinates.

\subsection{Strong contraction for high-degrees functions}
Throughout this section, we analyze operators such as $\mathcal{S}_{\mu}$ defined above, and prove that
they contract high degree functions very strongly.

\begin{lemma}\label{lem:upperbound_ev}
  For all $\alpha>0$, $m\in\mathbb{N}$ there exist $\delta>0$ and $C>0$
  such that the following holds. Let $\mathcal{U}_{\vec{k}}\subseteq [m]^n$ be
  $\alpha$-balanced, and $\mu$ be a $\alpha$-admissible, connected distribution over
  $\mathcal{U}_{\vec{k}}\times \mathcal{U}_{\vec{k}}$, and suppose
  there is a $(\alpha,\zeta)$-coupling $\mathcal{C}$ between $\mu$ and $\tilde{\mu}$.

  Then for all $d\in\mathbb{N}$, if $f\in V_{> d}(\mathcal{U}_{\vec{k}'})$, we have
  $\norm{\mathrm{T}_\mu f}_2\leq C(1+\delta)^{-d}\norm{f}_2$.
\end{lemma}

The proof of Lemma~\ref{lem:upperbound_ev} proceeds differently for $d$'s that are at least linear in $n$, and $d$'s that are sub-linear.
We begin by presenting the argument for $d$'s that are at least $\gamma n$, where $\gamma>0$ is to be thought of as a small constant
(depending on $m,m'$ and $\alpha$).

\begin{remark}\label{remark:self_adjoint}
We remark that we may, and will, assume that $\mathrm{T}_{\mu}$ is self-adjoint (i.e.~symmetric as a matrix). We can do that without
loss of generality as one has, by a standard Cauchy-Schwarz argument as before, that
$\norm{\mathrm{T}_\mu f}_2^2\leq \norm{f}_2^{3/2}\norm{\mathrm{T}_\mu^{*}\mathrm{T}_\mu f}_2^{1/2}$.
The operator $\mathrm{T}_\mu^{*}\mathrm{T}_\mu$ is self-adjoint and equal to $\mathrm{T}_{\mu'}$ for a distribution $\mu'$ which
is sampled by taking $({\bf x},{\bf y})\sim \mu$, then $({\bf x'},{\bf y'})\sim \mu$ conditioned on ${\bf y'} ={\bf y}$
and outputting $({\bf x},{\bf x'})$. Clearly, if $\mu$ is $\alpha$-admissible, then $\mu'$ is $\alpha^2$-admissible,
if $\mu$ is connected then $\mu'$ is connected, and a $(\alpha,\zeta)$-coupling between $\mu$ and $\tilde{\mu}$ naturally
induces a $(\alpha/2,2\zeta)$-coupling between $\mu'$ and $\tilde{\mu'}$.
\end{remark}
\subsubsection{The case that $d\geq \gamma n$}
\begin{claim}\label{claim:upperbound_ev_large_d}
  For all $m\in\mathbb{N}$ and $\alpha,\gamma>0$, there are $\delta>0$, $C>0$ such that the following holds.
  Let $\mathcal{U}_{\vec{k}}\subseteq [m]^n$ be
  $\alpha$-balanced, and $\mu$ be a $\alpha$-admissible, connected distribution over
  $\mathcal{U}_{\vec{k}}\times \mathcal{U}_{\vec{k}}$, such that $\mathrm{T}_{\mu}$ is self-adjoint.

  Then for all $d\geq \gamma n$, if $f\in V_{=d}(\mathcal{U}_{\vec{k}})$, then
  $\norm{\mathrm{T}_{\mu} f}_2\leq C(1+\delta)^{-d}\norm{f}_2$.
\end{claim}
In the rest of this section, we will fix $m\in\mathbb{N}$, $\alpha,\gamma > 0$ and $\mu$ as in the above claim.
Consider the graph $H$, whose vertex set is $[m]$ and its edges are weighted as follows.
For $a,b\in[m]$, the weight of the edge $(a,b)$ is
\[
w(a,b)
=
\cProb{\substack{({\bf x},{\bf x'}) \sim\mu\\ i\in [n]}}{{\bf x}_i = a}{{\bf x'}_i = b}.
\]
We note that as $\mu$ is connected, it follows that $H$ is connected, and from the third property in Definition~\ref{def:admissible}
it follows that for each $a,b$, we either have $w(a,b) = 0$ or $w(a,b)\geq \alpha$. Therefore, by Claim~\ref{claim:cheeger}
it follows that $H$ has constant spectral gap, i.e.~$\lambda_2(H)\leq 1 - \Omega_{\alpha}(1)$.
Note that the stationary distribution of $H$ is precisely $\nu_{\vec{k}}$.
It follows from a standard spectral argument that random walks of length
$O_{\alpha,\eps}(1)$ from any vertex reach distribution that is $\eps$-close to $\nu_{\vec{k}}$:
\begin{claim}\label{claim:basic_graph_mix}
  For all $\eps>0$, there is $h = h(m,\alpha,\eps)\in\mathbb{N}$, such that for all $v_1\in V(H)$,
  \[
  \card{\Prob{\substack{v_2,\ldots,v_h\\ \text{ random walk on $H$ from $v_1$}}}{v_h = a} - \nu_{\vec{k}}(a)}\leq \eps.
  \]
\end{claim}
Let $\eps = \eps(m,\alpha)>0$ to be determined later, and take $h = h(m,\alpha,\eps)$ from Claim~\ref{claim:basic_graph_mix}.
We now consider $\mathrm{T}_{\mu}^{h}$, and want to argue it is almost mixing in the following sense.
\begin{claim}\label{claim:graph_mix}
For all $a,b\in[m]$, it holds that
\[
\Prob{
\substack{{\bf x}\in \mathcal{U}_{\vec{k}}
\\ {\bf x}'\sim\mathrm{T}_{\mu}^h {\bf x}}}{\card{\#\sett{i}{{\bf x}_i = a, {\bf x'}_i = b} - n\nu_{\vec{k}}(a)\nu_{\vec{k}}(b)}\geq 3\eps n}\leq
4\alpha^{-1} e^{-\alpha\eps^2 n}.
\]
\end{claim}
\begin{proof}
Let ${\bf x}(0)\in_R \mathcal{U}_{\vec{k}}$, and sample $x(1)$ as ${\bf x'}$ where $({\bf x},{\bf x'})\sim \mu$ conditioned on
${\bf x} = {\bf x}(0)$. Inductively for $j\geq 0$, once
${\bf x}(j)$ have been defined, sample ${\bf x}(j+1)$ as ${\bf x'}$ where $({\bf x},{\bf x'})\sim \mu$ conditioned on
${\bf x}  = {\bf x}(j)$.
Then the distribution of $({\bf x}(0), {\bf x}(h))$ is
the same as $({\bf x}, {\bf x'})$ in the statement of the claim, and we wish to study the number of coordinates on which
it is equal to $(a,b)$. This will be much more convenient to do over product domains, and indeed we next use our coupling
$\mathcal{C}''$ to move to that scenario.

More precisely, let ${\bf y}(0), {\bf y}(1)$ be distributed as $({\bf y},{\bf y'})$ where $(({\bf x},{\bf x'}),({\bf y},{\bf y'}))\sim \mathcal{C}''$ conditioned
on $({\bf x},{\bf x'}) = ({\bf x}(0),{\bf x}(1))$. Then inductively, for all $j>0$, once ${\bf y}(\ell-1)$ has been drawn, sample
${\bf y}(j)$ as ${\bf y'}$ where $(({\bf x},{\bf x'}),({\bf y},{\bf y'}))\sim \mathcal{C}''$ conditioned on
$({\bf x},{\bf x'},{\bf y}) = ({\bf x}(j-1),{\bf x}(j), {\bf y}(j-1))$.

Consider the distribution over $({\bf y}(0),\ldots,{\bf y}(h))$. An equivalent way to sample it,
is by first taking ${\bf y}(0)\sim \nu_{\vec{k}}^{\otimes n}$, then taking ${\bf y}(1)$ as ${\bf y'}$ where
$({\bf y},{\bf y'})\sim \tilde{\mu}$ conditioned on ${\bf y} = {\bf y}(0)$, and continuing inductively. This shows that
the random variables $({\bf y}(0)_i, {\bf y}(h)_i)$ are independent. Thus, letting ${\bf Z}_i$ be the indicator of the event that
$({\bf y}(0)_i, {\bf y}(h)_i) = (a,b)$, we get that ${\bf Z}_i$ are independent. Also, from Claim~\ref{claim:basic_graph_mix}
we have that
\[
\card{\Expect{}{{\bf Z}_i} - \nu_{\vec{k}}(a)\nu_{\vec{k}}(b)}\leq \eps,
\]
so by Chernoff bound (Theorem~\ref{thm:chernoff}) we have
\[
\Prob{}{\card{\sum\limits_{i=1}^{n}{{\bf Z}_i} - n\nu_{\vec{k}}(a)\nu_{\vec{k}}(b)}\geq \eps n}
\leq 2 e^{-2\eps^2 n}.
\]

Finally, we relate the statistics of $({\bf y}(0), {\bf y}(h))$ to that of
$({\bf x}(0),{\bf x}(h))$.
For each $j$, the distribution of $({\bf x}(j-1),{\bf x}(j),{\bf y}(j-1),{\bf y}(j))$ is $\mathcal{C}''$.
Thus, using the tail bound property, we see that the probability that ${\bf x}(0)$ and ${\bf y}(0)$ differ in at least $\eps n$ coordinates,
is at most $\alpha^{-1}e^{-\alpha\eps^2 n}$. Also, ${\bf x}(h),{\bf y}(h)$ differ in at least $\eps n$ coordinates with probability
at most $\alpha^{-1} e^{-\alpha\eps^2 n}$. We thus get
\begin{align*}
&\Prob{}{\card{\#\sett{i}{{\bf x}(0)_i = a, {\bf x}(h)_i = b'} - n\nu_{\vec{k}}(a)\nu_{\vec{k}}(b)}\geq 3\eps n}\\
&\qquad\leq 2\alpha^{-1}e^{-\alpha\eps^2 n} + \Prob{}{\card{\sum\limits_{i=1}^{n}{{\bf Z}_i} - n\nu_{\vec{k}}(a)\nu_{\vec{k}}(b)}\geq \eps n}\\
&\qquad\leq 4\alpha^{-1}e^{-\alpha\eps^2 n}.
\end{align*}
\end{proof}

Claim~\ref{claim:graph_mix} motivates us to write the operator $\mathrm{T}_{\mu}^{h}$ as a convex combination of operators
according to the statistics of $({\bf x},{\bf y})$ where ${\bf x}\in_R \mathcal{U}_{\vec{k}'}$ and ${\bf y}\sim\mathrm{T}_{\mu}^{h} {\bf x}$.
That is, for each list of non-negative integers $\vec{r} = (r_{a,b})_{a,b\in [m]}$ that sum to $n$, let
$\mathcal{R}_{\vec{r}}$ be the operator corresponding to the distribution of
$({\bf x},{\bf y})$ where ${\bf x}\in\mathcal{U}_{\vec{k}}$ is uniform and ${\bf y}\sim\mathrm{T}_{\mu}^{h} {\bf x}$,
conditioned on the statistics of $({\bf x},{\bf y})$ being $\vec{r}$. Then we may
write $\mathrm{T}_{\mu}^{h} = \sum\limits_{\vec{r}} p_{\vec{r}} \mathcal{R}_{\vec{r}}$ where $p_{\vec{r}}$ is the probability
that the statistics $\vec{r}$ is achieved by $({\bf x},\mathrm{T}_{\mu}^h {\bf x})$.

\begin{definition}
  We say $\vec{r}$ is $\eps$-reasonable if
  $\card{r_{a,b} - n\nu_{\vec{k}}(a)\nu_{\vec{k}}(b)}\leq 3\eps n$
  for all $a',b'\in[m']$.
\end{definition}
By Claim~\ref{claim:graph_mix} and the union bound, we have $\sum\limits_{\vec{r}\text{ unreasonable}} p_{\vec{r}}\leq 4m^2 \alpha^{-1}e^{-\alpha\eps^2 n}$, so
it will be enough for us to show that the operators $\mathcal{R}_{\vec{r}}$ strongly contract high degree functions for reasonable $\vec{r}$.

By the symmetry of $\mu$, it follows that for each $\vec{r}$, the distribution of $({\bf x},{\bf y})$,
where ${\bf x}\in\mathcal{U}_{\vec{k}}, {\bf y}\sim\mathcal{R}_{\vec{r}}{\bf x}$ is also symmetric, and hence
uniform over vectors $(x,y)\in \mathcal{U}_{\vec{k}}\times\mathcal{U}_{\vec{k}}$ achieving the statistics $\vec{r}$.
Also, from Claim~\ref{claim:sym_implies_commute} we get that the distribution
$({\bf x}, \mathcal{R}_{\vec{r}} {\bf x})$ commutes with the action of $S_n$, and so
the operator $\mathcal{R}_{\vec{r}}$ commutes with the action of $S_n$ on functions.
Thus, the following claim
finds invariant subspaces of $\mathcal{R}_{\vec{r}}$.

\begin{claim}\label{claim:preserve_rt}
Suppose $\mathcal{R}\colon L^2(\mathcal{U}_{\vec{k}})\to L^2(\mathcal{U}_{\vec{k}})$ is an operator such that
both $\mathcal{R}$ and $\mathcal{R}^{*}$ commute with the action of $S_n$. Then
\begin{enumerate}
  \item for all $\lambda\vdash n$,
the space $V_{=\lambda}(\mathcal{U}_{\vec{k}})$ is an invariant space of $\mathcal{R}$;
  \item consequently, for all $d$ the space $V_{=d}(\mathcal{U}_{\vec{k}})$ is an invariant space of $\mathcal{R}$.
\end{enumerate}
\end{claim}
\begin{proof}
  Let $\lambda\vdash n$, and $A = (A_1,\ldots,A_r)$ be a partition of $[n]$ such that $\card{A_i} = \lambda_i$.
  Recall the spaces $V_{A}(\mathcal{U}_{\vec{k}})$ from Section~\ref{sec:alternative}; then for all $f\in V_{A}(\mathcal{U}_{\vec{k}})$
  and $\pi\in S_n$ such that $\pi(A_i) = A_i$ it holds that
  \[
  ^{\pi}(\mathcal{R} f)
  =\mathcal{R}(^{\pi} f)
  =\mathcal{R} f,
  \]
  and so $\mathcal{R} f \in V_{A}(\mathcal{U}_{\vec{k}})$. Thus, each $V_{A}(\mathcal{U}_{\vec{k}})$ is an invariant
  space of $\mathcal{R}$, and by Claim~\ref{claim:decompose_into_symmetries} it follows that
  $V_{\lambda}(\mathcal{U}_{\vec{k}'})$ is an invariant space of $\mathcal{R}$; the same argument applies to
  $\mathcal{R}^{*}$.

  Now, letting $f\in V_{=\lambda}(\mathcal{U}_{\vec{k}})$, $\lambda'\vartriangleright \lambda$ and
  $g\in V_{=\lambda'}(\mathcal{U}_{\vec{k}})$, we have that $\mathcal{R}^{*} g\in V_{=\lambda'}(\mathcal{U}_{\vec{k}})$ and so
  \[
  \inner{\mathcal{R} f}{g}
  =\inner{f}{\mathcal{R}^{*} g}
  =0,
  \]
  so by the alternative definition from Section~\ref{sec:alternative} we get that $\mathcal{R} f\in V_{=\lambda}(\mathcal{U}_{\vec{k}})$.

  The second item follows by summing the first item over $\lambda\vdash n$ such that $\lambda_1 = n-d$.
\end{proof}

\begin{claim}\label{claim:estimate_reasonable}
  For all $\alpha,\gamma>0$, $m\in\mathbb{N}$ there exist $\delta>0$ and $\eps>0$ such that the following holds.
  Suppose $\vec{r}$ is $\eps$-reasonable. Then for all $d\geq \gamma n$ and
  $f\in V_{=d}(\mathcal{U}_{\vec{k}'})$ we have that
  $\norm{\mathcal{R}_{\vec{r}} f}_2\leq (1+\delta_2)^{-d} \norm{f}_2$.
\end{claim}
\begin{proof}
  We prove that for all $\lambda\vdash n$ such that $\lambda_1 = n-d$, and $f\in V_{=\lambda}(\mathcal{U}_{\vec{k}'})$ it holds that
  $\norm{\mathcal{R}_{\vec{r}} f}_2\leq (1+\delta)^{-d}\norm{f}_2$, from which the statement follows as
  $V_{=d}(\mathcal{U}_{\vec{k}'}) = \bigoplus_{\lambda\vdash n, \lambda_1 = n-d} V_{=\lambda}(\mathcal{U}_{\vec{k}'})$.

  Write $\mathcal{R}_{\vec{r}}^{\star} = \mathcal{R}_{\vec{r}}^{*}\mathcal{R}_{\vec{r}}$.
  Fix $\lambda\vdash n$ such that $\lambda_1 = n-d$ and $V_{=\lambda}(\mathcal{U}_{\vec{k}})\neq \set{0}$.
  By Claim~\ref{claim:preserve_rt}, the space $V_{=\lambda}(\mathcal{U}_{\vec{k}})$ is invariant under $\mathcal{R}_{\vec{r}}, \mathcal{R}_{\vec{r}}^*$,
  so we may decompose $V_{=\lambda}(\mathcal{U}_{\vec{k}'})$ as $\bigoplus_{\theta} V_{=\lambda}^{\theta}(\mathcal{U}_{\vec{k}'})$ where each
  $V_{=\lambda}^{\theta}(\mathcal{U}_{\vec{k}'})$ is an eigenspace of $\mathcal{R}_{\vec{r}}^{\star}$ with eigenvalue $\theta$.
  Therefore, to establish our claim it suffices to show that for any $\theta$ such that
  $V_{=\lambda}^{\theta}(\mathcal{U}_{\vec{k}'})\neq \set{0}$, it holds that $\card{\theta}\leq (1-\delta_2)^{n}$ for
  $\delta_2 = \delta_2(\alpha,\gamma)>0$.

  Fix such $\theta$, take a non-zero $f\in V_{=\lambda}^{\theta}(\mathcal{U}_{\vec{k}'})$, and let $m_f$ be the multiplicity of $\theta$. Then since the trace of an operator is the sum of
  its eigenvalues, we have that
  \begin{equation}\label{eq:trace_main}
  {\sf Tr}(\mathcal{R}_{\vec{r}}^{\star})\geq m_f \theta,
  \end{equation}
  and as $m_f\geq {\sf dim}({\sf span}(\sett{^{\pi} f}{\pi\in S_n}))$ (since each $^{\pi} f$ is an eigenvector of $\mathcal{R}_{\vec{r}}$
  with eigenvalue $\alpha$) we get by Claim~\ref{claim:subrep_lb_multislice} that $m_f\geq {\sf dim}(\lambda)$. By Lemma~\ref{lem:dim_lb}, the latter is at least
  $(1+\delta_1)^n$ for some $\delta_1 = \delta_1(\alpha,\gamma)>0$, where we used $d\geq \gamma n$ (which we assumed to hold) and $\lambda_1\leq (1-\alpha)n$
  (which holds by Claim~\ref{claim:very_high_deg_vanish}). Thus, it
  is enough to prove that ${\sf Tr}(\mathcal{R}_{\vec{r}}^{\star})\leq (1+\delta_1/2)^n$, and this is done by a direct computation. Let
  ${\bf x}\in\mathcal{U}_{\vec{k}}$, ${\bf y}\sim \mathcal{R}_{\vec{r}} {\bf x}$ and ${\bf z}\sim \mathcal{R}_{\vec{r}}^{*}{\bf y}$,
  then by symmetry the probability that
  ${\bf z}={\bf x}$ is
  $\frac{1}{\prod\limits_{a\in[m]}{k_{a} \choose r_{a,1},\ldots,r_{a,m}}}$,
  so by Fact~\ref{Fact:stirling_cor}
  \begin{align}
  {\sf Tr}(\mathcal{R}_{\vec{r}}^{\star})
  = \frac{{n\choose k_{1},\ldots,k_{m}}}{\prod\limits_{a\in[m]}{k_{a} \choose r_{a,1},\ldots,r_{a,m}}}
  \leq n^{m^2} \frac{2^{H\left(\frac{k_{1}}{n},\ldots,\frac{k_{m}}{n}\right)n}}
  {\prod\limits_{a\in[m]}2^{H\left(\frac{r_{a,1}}{k_{a}},\ldots,\frac{r_{a,m}}{k_{a}}\right)k_{a}}}\label{eq:3}
  = n^{m^2}\frac{2^{H(\nu_{\vec{k}}(1),\ldots,\nu_{\vec{k}}(m))n}}
  {\prod\limits_{a\in[m]}2^{H\left(\frac{r_{a,1}}{k_{a}},\ldots,\frac{r_{a,m}}{k_{a}}\right)k_{a}}}.
  \end{align}
  Since $\vec{r}$ is reasonable, for all $a,b\in[m]$ we have
  \[
  \card{\frac{r_{a,b}}{k_{a}} - \nu_{\vec{k}}(b)}
  = \frac{\card{r_{a,b}-n\nu_{\vec{k}}(a)\nu_{\vec{k}}(b)}}{k_{a}}
  \leq \frac{3\eps n}{k_{a}}
  \leq \frac{3}{\alpha} \eps.
  \]
  It follows from Fact~\ref{fact:entropy_close} that
  \[
  \card{H(\nu_{\vec{k}'}(1),\ldots,\nu_{\vec{k}'}(m')) - H\left(\frac{r_{a,1}}{k_{a}},\ldots,\frac{r_{a,m}}{k_{a}}\right)} =
  \tilde{O}_{m,\alpha}(\eps),
  \]
  and plugging this into~\eqref{eq:3} yields that
  ${\sf Tr}(\mathcal{R}_{\vec{r}}^{\star})\leq n^{m^2} 2^{\tilde{O}_{m,\alpha}(\eps)n}\leq (1+\delta_1/2)^n$ provided $\eps$ is small enough
  with respect to $m$ and $\alpha$.
  Plugging everything into~\eqref{eq:trace_main} yields that $(1+\delta_1/2)^n\geq (1+\delta_1)^n \theta$, and so
  $0\leq \theta\leq (1-\delta_2)^n$ for $\delta_2(\alpha,\gamma)>0$.
\end{proof}

We are now ready to prove Claim~\ref{claim:upperbound_ev_large_d}.
\begin{proof}[Proof of Claim~\ref{claim:upperbound_ev_large_d}]
Fix $d\geq \gamma n$, pick $\delta(\alpha,\gamma,m), \eps(\alpha,\gamma,m)>0$
from Claim~\ref{claim:estimate_reasonable} and then $h = h(\alpha,\eps) >0$ from Claim~\ref{claim:basic_graph_mix}.
By the symmetry of $\mu$, it follows from Claims~\ref{claim:sym_implies_commute},~\ref{claim:preserve_rt} that $V_{=d}(\mathcal{U}_{\vec{k}})$ is
invariant under $\mathrm{T}_{\mu}$. We may therefore decompose it as a sum of
eigenspaces of $\mathrm{T}_{\mu}$, and so by Parseval it suffices to prove the claim for each one of these eigenspaces.

Let $f\in V_{=d}(\mathcal{U}_{\vec{k}})$ be a non-zero eigenvector of $\mathrm{T}_{\mu}$ with eigenvalue $\theta$, then
\[
\card{\theta}^h\norm{f}_2
=
\norm{\mathrm{T}_{\mu}^{h} f}_2
\leq \sum\limits_{\vec{r}} p_{\vec{r}}\norm{\mathcal{R}_{\vec{r}} f}_2
=
\sum\limits_{\substack{\vec{r}\\ \text{reasonable}}} p_{\vec{r}}\norm{\mathcal{R}_{\vec{r}} f}_2
+\sum\limits_{\substack{\vec{r}\\ \text{unreasonable}}} p_{\vec{r}}\norm{\mathcal{R}_{\vec{r}} f}_2.
\]
For the first sum, we bound $\norm{\mathcal{R}_{\vec{r}} f}_2\leq (1+\delta_2)^{-n}\norm{f}_2$ by
Claim~\ref{claim:estimate_reasonable}, and the sum of the $p_{\vec{r}}$'s by $1$. For the second sum,
we bound $\norm{\mathcal{R}_{\vec{r}} f}_2\leq \norm{f}_2$ by Jensen's inequality and the sum of
$p_{\vec{r}}$'s by $4m^2 \alpha^{-1} e^{-\alpha\eps^2 n}$. Overall, we get that
\[
\card{\theta}^h\norm{f}_2\leq
(4m^2 \alpha^{-1} e^{-\alpha\eps^2 n} + (1+\delta_2)^{-n})\norm{f}_2,
\]
and dividing by $\norm{f}_2$, simplifying and taking $h$-th root finishes the proof.
\end{proof}

\subsubsection{The case that $d < \gamma n$}\label{sec:med_degs}
In this section, we prove the following claim.
\begin{claim}\label{claim:upperbound_ev_small_d}
  For all $\alpha > 0$, $m\in\mathbb{N}$ there are $\gamma>0$, $\delta>0$
  and $C>0$ such that the following holds.
  Let $\mathcal{U}_{\vec{k}}\subseteq [m]^n$ be
  $\alpha$-balanced, and $\mu$ be a $\alpha$-admissible, connected distribution over
  $\mathcal{U}_{\vec{k}}\times \mathcal{U}_{\vec{k}}$, such that $\mathrm{T}_{\mu}$
  is self-adjoint.

  For all $d\leq \gamma n$, if $f\in V_{=d}(\mathcal{U}_{\vec{k}})$, then
  $\norm{\mathrm{T}_{\mu} f}_2\leq C(1+\delta)^{-d}\norm{f}_2$.
\end{claim}
For the rest of this section, we fix $\alpha>0$, $m\in\mathbb{N}$ and $\mu$ as in the above claim.
The proof follows the same sequence of claims as in the previous section, except for Claim~\ref{claim:estimate_reasonable} whose
formulation and proof is adjusted to this case.
\begin{claim}\label{claim:estimate_reasonable2}
  For all $\alpha>0$, $m\in\mathbb{N}$ there are $\eps>0$, $\gamma>0$, $\delta>0$ such that the following holds.
  Suppose $\vec{r}$ be $\eps$-reasonable. Then for all $d\leq \gamma n$ and
  $f\in V_{=d}(\mathcal{U}_{\vec{k}})$ we have that
  $\norm{\mathcal{R}_{\vec{r}} f}_2\leq (1+\delta)^{-d} \norm{f}_2$.
\end{claim}

\paragraph{Proof overview.}
We begin with an outline of the argument. Let $\eps>0$ be sufficiently small to be determined later,
and take $\gamma =\min(\alpha/4, \eps)$. Then the $\alpha$-balancedness of
our multi-slices implies that for any $S\subseteq [n]$ of size $\gamma n$, the support of $x_S$ where $x\in\mathcal{U}_{\vec{k}}$ is
full, i.e.~$[m]^{S}$.

As before, we will reduce Claim~\ref{claim:estimate_reasonable2} to understanding the eigenvalues of $\mathcal{R}_{\vec{r}}^{\star} = \mathcal{R}_{\vec{r}}^{*}\mathcal{R}_{\vec{r}}$ on the space
$V_{=d}(\mathcal{U}_{\vec{k}})$. Here however, we will show that for any eigenvalue $\theta\neq 0$ of $\mathcal{R}_{\vec{r}}^{\star}$ in this space,
there is a $d$-junta $f\in V_{=d}(\mathcal{U}_{\vec{k}})$ for which $\mathcal{R}_{\vec{r}}^{\star} f = \theta f$. Assume without loss of generality that
the junta is $S = \set{1,\ldots,d}$. Roughly speaking, the idea then is to take a set of $N = 3d$ coordinates containing the junta coordinates,
then ``project everything'' onto these coordinates, reducing us to case very similar to the $d\geq \gamma n$ case.

More precisely, we will consider the induced measure $\nu$ on $[m]^N$,
and the natural projection of the operator $\mathcal{R}_{\vec{r}}^{\star}$ on this space, i.e.~$\mathcal{R}_{\vec{r},N}^{\star}\colon L^2([m]^N,\nu)\to L^2([m]^N,\nu)$. We will do so while keeping $f$ as an eigenvector of
$\mathcal{R}_{\vec{r},N}^{\star}$ with eigenvalue $\theta$.
We will then study the operator $\mathcal{R}_{\vec{r},N}^{\star}$, and in particular show (similarly to before)
that: (a) ${\sf Tr}((\mathcal{R}_{\vec{r},N}^{\star})^2)\leq (1+\xi(h))^d$ where $\lim_{h\rightarrow \infty} \xi(h) = 0$,
and (b) the multiplicity of the eigenvalue $\theta$ is large, namely at least $(1+\Omega(1))^d$.
Combining these two facts yields, as before, that $\card{\theta}\leq (1-\delta')^d$ for some $\delta' = \delta(\alpha,\gamma,m)>0$.

We remark that the proof of (a) is again by a direct computation (as can perhaps be expected). As for (b), here we use the fact that
the multiplicity of $f$ is at least ${\sf dim}({\sf span}(\sett{^{\pi} f}{\pi\in S_N}))$ (as each $^{\pi} f$ will automatically be an
eigenvector of $\mathrm{R}_{\vec{r},N}$), so our task reduces to showing that this is a space of large dimension. To evaluate this
dimension, we construct an invertible linear map $\psi\colon L^{2}([m]^N,\nu)\to L^2([m]^N, {\sf Uniform})$ that commutes with the
action of $S_N$, and show that the space $\psi({\sf span}(\sett{^{\pi} f}{\pi\in S_N}))$ has a large dimension by orthogonality
considerations (which are much easier to carry out in a product space).

\skipi

We now proceed to the formal proof, and we fix an $\eps$-reasonable statistics $\vec{r}$ for the rest of the proof.
By Claims~\ref{claim:sym_implies_commute},~\ref{claim:preserve_rt} the space
$V_{=d}(\mathcal{U}_{\vec{k}})$ is invariant under $\mathcal{R}_{\vec{r}}, \mathcal{R}_{\vec{r}}^{*}$. Thus, we may decompose
it as $V_{=d}(\mathcal{U}_{\vec{k}}) = \bigoplus_{\theta} V_{=d}^{\theta}(\mathcal{U}_{\vec{k}})$ where
$V_{=d}^{\theta}(\mathcal{U}_{\vec{k}})$ is an eigenspace of $\mathcal{R}_{\vec{r}}^{\star}$ with eigenvalue $\theta$.
Thus, by Parseval, to prove Claim~\ref{claim:sym_implies_commute} it is enough to show that if
$V_{=d}^{\theta}(\mathcal{U}_{\vec{k}})\neq\set{0}$, then $\theta\leq C(1-\delta)^d$ ($\theta$ is clearly non-negative).

Recall the junta spaces $V_{J}(\mathcal{U}_{\vec{k}})$. We further refine them by considering
$V_{=J}(\mathcal{U}_{\vec{k}}) = V_{J}(\mathcal{U}_{\vec{k}})\cap V_{d-1}(\mathcal{U}_{\vec{k}})^{\perp}$
for all $J$ of size $d$.
\begin{claim}\label{claim:high_deg_junta}
  Let $\theta$ be such that $V_{=d}^{\theta}(\mathcal{U}_{\vec{k}})\neq \set{0}$. Then there is
  a non-zero $d$-junta $f\in V_{=d}^{\theta}(\mathcal{U}_{\vec{k}})$.
\end{claim}
\begin{proof}
  By Claim~\ref{claim:preserve_junta} the operators $\mathcal{R}_{\vec{r}}, \mathcal{R}_{\vec{r}}^{*}$ preserve the spaces $V_{J}(\mathcal{U}_{\vec{k}})$ as
  well as $V_{d-1}(\mathcal{U}_{\vec{k}})$. Using that and Claim~\ref{claim:preserve_rt} it follows that $\mathcal{R}_{\vec{r}}, \mathcal{R}_{\vec{r}}^{*}$ preserve
  $V_{=J}(\mathcal{U}_{\vec{k}})$. Thus, we may write each such
  space as $V_{=J}(\mathcal{U}_{\vec{k}}) = \bigoplus_{\theta} V_{=J}^{\theta}(\mathcal{U}_{\vec{k}})$ where
  $V_{=J}^{\theta}(\mathcal{U}_{\vec{k}})$ is an eigenspace of $\mathcal{R}_{\vec{r}}^{\star}$ with eigenvalue $\theta$.
  Thus, we get that
  \[
  \bigoplus_{\theta} V_{=d}^{\theta}(\mathcal{U}_{\vec{k}})
  =V_{=d}(\mathcal{U}_{\vec{k}})
  =\bigoplus_{J} V_{=J}(\mathcal{U}_{\vec{k}})
  =\bigoplus_{J}\bigoplus_{\theta} V_{=J}^{\theta}(\mathcal{U}_{\vec{k}})
  =\bigoplus_{\theta}\bigoplus_{J} V_{=J}^{\theta}(\mathcal{U}_{\vec{k}}).
  \]
  By uniqueness of the decomposition into eigenvalues, we get that for all $\theta$,
  $V_{=d}^{\theta}(\mathcal{U}_{\vec{k}}) = \bigoplus_{J} V_{=J}^{\theta}(\mathcal{U}_{\vec{k}})$,
  and the claim follows.
\end{proof}

Fix $\theta$ such that $V_{=d}^{\theta}(\mathcal{U}_{\vec{k}})\neq \set{0}$, and pick $f^{\star}\neq 0$ a $J$-junta
from Claim~\ref{claim:high_deg_junta} for $\card{J} = d$. Without loss of generality, assume that $J = [d]$, and set $N = 3d$.
Let $\mathcal{D}$ be the uniform measure over $\mathcal{U}_{\vec{k}}$, and let
$\mathcal{D}_\downarrow$ be the marginal distribution of $x_{[N]}$ where $x\sim \mathcal{D}$.
Given a function $g\colon [m]^N\to\mathbb{R}$, we may lift it to $\tilde{g}\colon\mathcal{U}_{\vec{k}}\to\mathbb{R}$
as $\tilde{g}(z) = g(z_{[N]})$. Define the operator $\mathcal{R}_{\vec{r},N}: L^2([m]^n, \mathcal{D}_\downarrow) \to L^2([m]^n, \mathcal{D}_\downarrow)$ as follows:
\[
(\mathcal{R}_{\vec{r},N}^{\star} g) (x) =
\sum\limits_{y\in [m]^{n-N}}
\frac{\mathcal{D}(x,y)}{\mathcal{D}_\downarrow (x)}
(\mathcal{R}_{\vec{r}}^{\star} \tilde{g})(x,y).
\]
We note that $\mathcal{R}_{\vec{r},N}^{\star}$ is self adjoint and positive semi-definite.
Indeed, to see $\mathcal{R}_{\vec{r},N}^{\star}$ is self adjoint note that
if $g,h\colon [m]^n\to\mathbb{R}$ are functions then
\[
\inner{\mathcal{R}_{\vec{r},N}^{\star} g}{h}_{\mathcal{D}_\downarrow}
=\sum\limits_{x\in [m]^N}\mathcal{D}_\downarrow (x)
\sum\limits_{y\in [m]^{n-N}}\frac{\mathcal{D}(x,y)}{\mathcal{D}_\downarrow (x)}(\mathcal{R}_{\vec{r}}^{\star}\tilde{g})(x,y)h(x)
=\inner{\mathcal{R}_{\vec{r}}^{\star}\tilde{g}}{\tilde{h}}_{\mathcal{D}}.
\]
As $\mathcal{R}_{\vec{r}}^{\star}$ is self adjoint the last inner product is equal to $\inner{\tilde{g}}{\mathcal{R}_{\vec{r}}^{\star}\tilde{h}}$,
which by the argument above is equal to $\inner{g}{\mathcal{R}_{\vec{r},N}^{\star} h}$.
To see $\mathcal{R}_{\vec{r},N}^{\star}$ is positive semi-definite we plug in $g=h$ above and use the fact
$\mathcal{R}_{\vec{r}}^{\star}$ is positive semi-definite.

Abusing notation, since $f^{\star}$ is a $[d]$-junta and $[d]\subseteq[N]$,
we may think of $f^{\star}$ as a function over $[m]^N$. Let
$F = {\sf span}(\sett{^{\pi} f^{\star}}{\pi\in S_N})$.
\begin{claim}
  For all $f\in F$, we have
  $\mathcal{R}_{\vec{r},N}^{\star} f = \theta f$.
\end{claim}
\begin{proof}
Clearly, it is enough to prove that each $^{\pi} f^{\star}$
is an eigenvector. The proof is virtually identical for all
$\pi$ (replacing the role of $[d]$ with $\pi([d])$), and we
thus assume that $\pi$ is the identity permutation.
By definition,
  \[
  \mathcal{R}_{\vec{r},N}^{\star} f^{\star}(x)
  =\sum\limits_{y\in [m]^{n-N}}
    \frac{\mathcal{D}(x,y)}{\mathcal{D}_\downarrow (x)}
    (\mathcal{R}_{\vec{r}}^{\star} \tilde{f^{\star}})(x,y)
  =\sum\limits_{y\in [m]^{n-N}}
    \frac{\mathcal{D}(x,y)}{\mathcal{D}_\downarrow (x)}
    \theta \tilde{f^{\star}}(x,y)
  =\theta\sum\limits_{y\in [m]^{n-N}}
    \frac{\mathcal{D}(x,y)}{\mathcal{D}_\downarrow (x)}
    f^{\star}(x),
  \]
  which is equal to $\theta f^{\star}(x)$.
\end{proof}
Thus, using the trace method (along with the positive semi definiteness) we see that
${\sf Tr}(\mathcal{R}_{\vec{r},N}^{\star})\geq \theta {\sf dim}(F)$, i.e.~that
\begin{equation}\label{eq:2}
0\leq \theta\leq \frac{{\sf Tr}(\mathcal{R}_{\vec{r},N}^{\star})}{{\sf dim}(F)},
\end{equation}
and in the rest of the proof we prove an upper bound on the above trace and
lower bound on the dimension of $F$.

\begin{claim}\label{claim:dim_lb_med}
  ${\sf dim}(F)\geq {2d \choose d}$.
\end{claim}
\begin{proof}
  Using the expectation inner product according to $\mathcal{D}_{\downarrow}$, we may define the spaces
  $V_{=j}([m]^N,\mathcal{D}_{\downarrow})$ for $j=0,1\ldots,N$. It is easily seen then that
  $F\subseteq V_{=d}([m]^N,\mathcal{D}_{\downarrow})$.

  Consider the linear map $\Psi\colon V_{=d}([m]^N,\mathcal{D}_{\downarrow})\to V_{=d}([m]^N, {\sf Uniform})$
  defined as follows. Given a function $g\in V_{=d}([m]^N,\mathcal{D}_{\downarrow})$, we think
  of it as a function over $[m]^N$ with the uniform measure, and then take its projection $g_d$ onto
  $V_{=d}([m]^N, {\sf Uniform})$; i.e.~$\Psi(g)  = g_d$.

  We claim that $\Psi$ commutes with the action of $S_n$. Indeed, let $\pi\in S_n$, and write
  $^{\pi} g$ as $h_0+\ldots+h_d$ where $h_j$ is the projection of $^{\pi} g$
  onto $V_{=d}([m]^N, {\sf Uniform})$, so that $\Psi(^{\pi}(g)) = h_d$. Also, we
  have that $g(x) = ^{\pi^{-1}}h_0 + \ldots+^{\pi^{-1}}h_d$, and since
  the spaces $V_{=d}([m]^N, {\sf Uniform})$ are invariant under the action of $S_n$,
  we get that $^{\pi^{-1}}h_d$ is the projection of $g$ onto $V_{=d}([m]^N, {\sf Uniform})$.
  We get that $^{\pi}\Psi(g) = ^{\pi}(^{\pi^{-1}}h_d) = h_d$, and
  $\Psi(^{\pi} g) = h_d$, and so $\Psi$ commutes with $\pi$.

  We claim that $\Psi$ is invertible. Indeed, we show that if $\Psi(g) = 0$, then $g=0$.
  As $g\in V_{=d}([m]^N,\mathcal{D}_{\downarrow})$, we get that $g$ is a linear combination
  of $d$-juntas so $g\in V_{d}([m]^N,{\sf Uniform})$, but as $\Psi(g) = 0$ we have that
  its projection onto $V_{=d}([m]^N,{\sf Uniform})$ is $0$, so $g\in V_{d-1}([m]^N,{\sf Uniform})$.
  Thus, $g$ is a linear combination of $(d-1)$-juntas, and so
  $g\in V_{d-1}([m]^N,\mathcal{D}_{\downarrow})$; it follows that $g = 0$.
  We conclude that ${\sf dim}(F) = {\sf dim}(\Psi(F))$, and in the rest
  of the proof we lower bound the latter quantity.

  We note that as $f^{\star}$ is a $[d]$-junta, $g^{\star} = \Psi(f^{\star})$ is also
  a $[d]$-junta. Thus, since $\Psi$ commutes with the action of $S_n$, for
  all $\pi\in S_n$, the function $\Psi(^{\pi} f^{\star}) = {}^{\pi} g^{\star}$ is a $\pi([d])$-junta.
  We now consider $B = \sett{^{\pi} g^{\star}}{\pi\in S_N}$,
  and we show that $B$ contains a large orthogonal set.

  We claim that if $\pi,\pi'$ are permutations such that $\pi([d])\neq \pi'([d])$,
  then $^{\pi} g^{\star}$ and $^{\pi'} g^{\star}$ are orthogonal.
  This is equivalent to showing that if $\pi^{-1}([d])\neq [d]$, then
  $^{\pi} g^{\star}$ and $g^{\star}$ are orthogonal. To see that,
  let $I = \pi^{-1}([d])$; then we have that
  \[
  \inner{^{\pi} g^{\star}}{g^{\star}}
  =\Expect{x\in[m']^I, y\in[m']^{[N]\setminus I}}
  {g^{\star}(\pi(x,y)) g^{\star}(x,y)}
  =\Expect{x\in[m']^I}{g^{\star}(\pi(x,0))
  \Expect{y}{g^{\star}(x,y)}}.
  \]
  Letting $I'$ be the intersection of $[d]$ and $I$, we see that $I'$ has size at most $d-1$ and
  $\Expect{y}{g^{\star}(x,y)} = \Expect{z\in[m]^{[N]\setminus I'}}{g^{\star}(x_{I'},z)}=0$,
  so $\inner{^{\pi} g^{\star}}{g^{\star}} = 0$.

  Consider the collection $\tilde{L}$ of injections $\tilde{\pi}\colon [d]\to [N]\setminus [d]$,
  and extend each $\tilde{\pi}\in L$ to a permutation $\pi\in S_n$ arbitrarily, to form
  a collection of permutations $L$. We get that any two permutations from $L$ disagree on $[d]$, and
  by the above paragraph it follows that
  \[
  {\sf dim}(\Psi(F))
  \geq
  \card{L}
  =\card{\tilde{L}}
  =\frac{(2d)!}{d!^2}
  ={2d\choose d}.\qedhere
  \]
\end{proof}

\begin{claim}\label{claim:tr_up_med}
  ${\sf Tr}(\mathcal{R}_{\vec{r},N}^{\star})\leq (1+O_{\alpha}(\eps))^d$.
\end{claim}
\begin{proof}

  Let $x\in [m]^N$, and let $1_x\in \set{0,1}^{m^N}$ be the indicator vector of $x$.
  Denote the diagonal entry of $(\mathcal{R}_{\vec{r},N}^{\star})^2$ corresponding
  to $x$ by $a_{x,x}$; then $a_{x,x} =  1_x^{t}\cdot (\mathcal{R}_{\vec{r},N}^{\star})^2 1_x$.
  Expanding the definition of $\mathcal{R}_{\vec{r},N}^{\star}$, the vector
  $(\mathcal{R}_{\vec{r},N}^{\star})^2 1_x$ is a probability vector whose $x'$'s entry
  is equal to
  \[
  \cProb{\substack{{\bf w}\sim \mathcal{D}\\ {\bf w'}\sim \mathcal{R}_{\vec{r}}^{*}\mathcal{R}_{\vec{r}} {\bf w}}}{{\bf w}_{[N]} = x}{{\bf w'}_{[N]} = x'},
  \text{ so}
  \qquad\qquad
  a_{x,x} =
  \cProb{\substack{{\bf w}\sim \mathcal{D}\\ {\bf w'}\sim \mathcal{R}_{\vec{r}}^*\mathcal{R}_{\vec{r}} {\bf w}}}{{\bf w}_{[N]} = x}{{\bf w'}_{[N]} = x}.
  \]
  Fix any $w$ such that $w_{[N]} = x$. In the rest of the proof, we show that
  $\Prob{{\bf w'}\sim \mathcal{R}_{\vec{r}}^*\mathcal{R}_{\vec{r}} w}{{\bf w'}_{[N]}=x}\leq \mathcal{D}_{\downarrow}(x) (1+O_{\alpha}(\eps))^N$,
  from which the claim quickly follows: it implies that $a_{x,x}\leq \mathcal{D}_{\downarrow}(x) (1+O_{\alpha}(\eps))^N$,
  which by summing over $x$ and using $N = 3d$ gives us
  ${\sf Tr}(\mathcal{R}_{\vec{r},N}^{\star})\leq (1+O_{\alpha}(\eps))^d$.

  Indeed, using conditional probabilities we have that
  \[
  \Prob{{\bf w'}\sim \mathcal{R}_{\vec{r}}^{*}\mathcal{R}_{\vec{r}} w}{{\bf w'}_{[N]}=x}
  =\prod\limits_{i=1}^{N} \cProb{{\bf w'}\sim \mathcal{R}_{\vec{r}}^*\mathcal{R}_{\vec{r}} w}{{\bf w'}_{<i} = x_{<i}}{{\bf w'}_i=x_i}.
  \]
  By definition of the operator $\mathcal{R}_{\vec{r}}$, we have for $i=1$ that
  \[
  \Prob{{\bf w'}\sim \mathcal{R}_{\vec{r}}^{*}\mathcal{R}_{\vec{r}} w}{{\bf w'}_1=x_1} =
  \sum\limits_{a}\frac{r_{x_1,a}r_{a,x_1}}{n^2},
  \]
  and
  for all $i>1$
  \begin{align*}
  \cProb{{\bf w'}\sim \mathcal{R}_{\vec{r}}^{*}\mathcal{R}_{\vec{r}} w}{{\bf w'}_{<i} = x_{<i}}{{\bf w'}_i=x_i}
  \leq
  \sum\limits_{a}\frac{r_{x_i,a}r_{a,x_i}}{(k_{x_i}-i)(k_a-i)}
  &\leq \sum\limits_{a}\frac{r_{x_i,a}r_{a,x_i}}{(k_{x_i}-N)(k_{a}-N)}\\
  &\leq \sum\limits_{a}\frac{r_{x_i,a}r_{a,x_i}}{k_{x_i}k_{a}}
  \max\left(\frac{k_{x_i}}{k_{x_i}-N}, \frac{k_{a}}{k_{a}-N}\right)^2,
  \end{align*}
  as $N\leq \gamma n\leq \eps n, \alpha n/2$ and $k_{x_i}, k_{\alpha_j}\geq \alpha n$,
  the last expression is at most $p_{x_i}(1+O_{\alpha}(\eps))$,
  where $p_{x_i} = \sum\limits_{a}\frac{r_{x_i,a}r_{a,x_i}}{k_{x_i}k_{a}}$. Thus, we get that
  \[
  \Prob{{\bf w'}\sim \mathcal{R}_{\vec{r}}^{*}\mathcal{R}_{\vec{r}} w}{{\bf w'}_{[N]}=x}
  \leq (1+O_{\alpha}(\eps))^{N} \prod\limits_{i=1}^{N}p_{x_i}.
  \]
  Next, note that
  \[
  \mathcal{D}_{\downarrow}(x)
  \geq \prod\limits_{i=1}^{N} \frac{k_{x_i} - i}{n}
  \geq (1-\eps)^N\prod\limits_{i=1}^{N} \frac{k_{x_i}}{n}
  =(1-\eps)^N\prod\limits_{i=1}^{N} \nu_{\vec{k}}(x_i),
  \]
  so we get that
  \begin{equation}\label{eq:4}
  \frac{\Prob{{\bf w'}\sim \mathcal{R}_{\vec{r}}^{*}\mathcal{R}_{\vec{r}} w}{{\bf w'}_{[N]}=x}}{\mathcal{D}_{\downarrow}(x)}
  \leq (1+O_{\alpha}(\eps))^N \prod\limits_{i=1}^{N}\frac{p_{x_i}}{\nu_{\vec{k}}(x_i)}.
  \end{equation}
  Since $\vec{r}$ is $\eps$-reasonable and $k_{\alpha}\geq \alpha n$ for all $\alpha$, we get
  \[
  p_{x_i} \leq \sum\limits_{a}
  \frac{n^2\nu_{\vec{k}}(x_i)^2\nu_{\vec{k}}(a)^2}
  {n^2\nu_{\vec{k}}(x_i)\nu_{\vec{k}}(a)}
  +O_{\alpha}(\eps)
  =\nu_{\vec{k}}(x_i) +O_{\alpha}(\eps).
  \]
  Plugging this into~\eqref{eq:4} shows that
  $\frac{\Prob{{\bf w'}\sim (\mathcal{R}_{\vec{r}}^{*}\mathcal{R}_{\vec{r}})^2 w}{{\bf w'}_{[N]}=x}}{\mathcal{D}_{\downarrow}(x)}\leq (1+O_{\alpha}(\eps))^N$,
  finishing the proof.
\end{proof}
Plugging Claims~\ref{claim:dim_lb_med},~\ref{claim:tr_up_med} into~\eqref{eq:2} bounds all eigenvalues of
$\mathcal{R}_{\vec{r}}$ for reasonable $\vec{r}$, and the rest of the proof of Claim~\ref{claim:estimate_reasonable2}
is identical to the proof of Claim~\ref{claim:upperbound_ev_large_d} and we do not repeat it.

\subsubsection{Proof of Lemma~\ref{lem:upperbound_ev}}
By Remark~\ref{remark:self_adjoint}, we assume $\mathrm{T}_{\mu}$ is self-adjoint.
Fix $\alpha>0$ and $m\in\mathbb{N}$. Take $\gamma_1, \delta_1>0$ and $C_1>0$ from Claim~\ref{claim:upperbound_ev_small_d}.
Now take $\delta_2, C_2>0$ from Claim~\ref{claim:upperbound_ev_large_d} for $c,m$ and $\gamma_1$; we show that the lemma holds for
$C = \max(C_1,C_2)$ and $\delta = \min(\delta_1,\delta_2)$.

Let $f\in V_{>d}(\mathcal{U}_{\vec{k}})$, and write $f = \sum\limits_{j>d+1} f^{=j}$ where
$f^{=j}\in V_{=j}(\mathcal{U}_{\vec{k}})$. By Claims~\ref{claim:sym_implies_commute},~\ref{claim:preserve_rt},
the spaces $V_{=j}(\mathcal{U}_{\vec{k}})$ are invariant under $\mathrm{T}_{\mu}$, so we get that
$(\mathrm{T}_{\mu} f)^{=j} = \mathrm{T}_{\mu}(f^{=j})$. Thus by Parseval
$\norm{\mathrm{T}_{\mu} f}_2^2 = \sum\limits_{j>d}\norm{\mathrm{T}_{\mu}(f^{=j})}_2^2$.
By Claims~\ref{claim:upperbound_ev_small_d},~\ref{claim:upperbound_ev_large_d}
we have that $\norm{\mathrm{T}_{\mu} f^{=j}}_2\leq C (1-\delta)^{j}\norm{f^{=j}}_2$, so summing and
using Parseval again gives us the statement of the lemma.\qed

\subsection{Proof of Theorem~\ref{thm:basic_inv_bipartite}}
Let $d\in\mathbb{N}$ to be chosen later, and write $f = f^{\leq d} + f^{>d}$,
$g = g^{\leq d} + g^{>d}$ Then we
have that
\begin{equation}\label{eq:5}
\Expect{(\bf{x},\bf{x}')\sim \mu}{f({\bf x})g({\bf x'})}
=
\Expect{(\bf{x},\bf{x}')\sim \mu}{f^{\leq d}({\bf x})g^{\leq d}({\bf x'})}
+
\Expect{(\bf{x},\bf{x}')\sim \mu}{f^{\leq d}({\bf x})g^{>d}({\bf x'})}
+
\Expect{(\bf{x},\bf{x}')\sim \mu}{f^{>d}({\bf x})g({\bf x'})},
\end{equation}
and we first show that the third expectation is small in absolute value.
Indeed,
writing it as an inner product we have that it is equal to
\[
\card{\inner{\mathrm{T}_{\mu} f^{> d}}{g}}
\leq\norm{\mathrm{T}_{\mu} f^{>d}}_2\norm{g}_2
=\inner{\mathrm{T}_{\mu}^{*} \mathrm{T}_{\mu}  f^{>d}}{f^{>d}}^{1/2} \norm{g}_2
\leq \norm{\mathrm{T}_{\mu}^{*}\mathrm{T}_{\mu}  f^{>d}}_2^{1/2}\norm{f^{>d}}_2^{1/2} \norm{g}_2.
\]
Consider the distribution $\nu$ defined as: pick $({\bf x},{\bf x'})\sim\mu$, pick
$({\bf x''},{\bf x'''})$ conditioned on ${\bf x'''} = {\bf x'}$, and output $({\bf x},{\bf x''})$.
Note that as $\mu$ is connected, $\nu$ is connected, as $\mu$ is $\alpha$-admissible, $\nu$
is $\alpha^2$-admissible, and that the coupling between $\mu$ and $\tilde{\mu}$ naturally induces
a coupling between $\nu$ and $\tilde{\nu}$ with similar parameters. Finally, note that $\mathrm{T}_{\nu} = \mathrm{T}_{\mu}^{*}\mathrm{T}_{\mu}$.

Write $f^{>d} = \sum\limits_{\ell>d} f^{=\ell}$ for $f^{=\ell}\in V_{=\ell}(\mathcal{U}_{\vec{k}})$.
By Claim~\ref{claim:preserve_rt}, the operator $\mathrm{T}_{\nu}$ preserves the spaces
$V_{=\ell}(\mathcal{U}_{\vec{k}})$, so by Parseval we have
$\norm{\mathrm{T}_{\nu} f^{>d}}_2^2 = \sum\limits_{\ell >d}\norm{\mathrm{T}_{\nu} f^{=\ell}}_2^2$.
Using Lemma~\ref{lem:upperbound_ev}, we get that
\[
\norm{\mathrm{T}_{\nu}f^{>d}}_2^2
\leq
\sum\limits_{\ell >d}C^2 (1+\delta)^{-2\ell}\norm{f^{=\ell}}_2^2
\leq C^2 (1+\delta)^{-2d} \norm{f}_2^2,
\]
so the third expectation in~\eqref{eq:5} is at most
$\sqrt{C}(1+\delta)^{-d/2} \norm{f}_2\norm{g}_2$. The same argument applies to show that
the absolute value of the second expectation is also upper bounded by that. We thus get that
\begin{equation}\label{eq:6}
\card{
\Expect{({\bf x},{\bf x'})\sim \mu}{f({\bf x})g({\bf x'})}
-
\Expect{({\bf x},{\bf x'})\sim \mu}{f^{\leq d}({\bf x})g^{\leq d}({\bf x'})}
}\leq
2\sqrt{C}(1+\delta)^{-d/2} \norm{f}_2\norm{g}_2.
\end{equation}

We now argue that
\begin{equation}\label{eq:7}
\card{
\Expect{({\bf y},{\bf y'})\sim \tilde{\mu}}{\mathrm{T}_{\mathcal{C}} f({\bf y}) \mathrm{T}_{\mathcal{C}'} g({\bf y'})}
-
\Expect{({\bf y},{\bf y'})\sim \tilde{\mu}}{(\mathrm{T}_{\mathcal{C}}f)^{\leq d}({\bf y}) (\mathrm{T}_{\mathcal{C}'} g)^{\leq d}({\bf y'})}
}\leq
2\sqrt{C'}(1+\delta')^{-d/2} \norm{f}_2\norm{g}_2.
\end{equation}
Indeed, we show that by upper bounding the expectation of each one of the functions
$(\mathrm{T}_{\mathcal{C}}f)^{> d}({\bf y}) (\mathrm{T}_{\mathcal{C}'} g)({\bf y'})$
and $(\mathrm{T}_{\mathcal{C}}f)^{\leq d}({\bf y}) (\mathrm{T}_{\mathcal{C}'} g)^{> d}({\bf y'})$.
The arguments are identical, and we demonstrate on the first one.
Writing it as an inner product and using Cauchy-Schwarz we get that
\begin{align*}
\Expect{({\bf y},{\bf y'})\sim \tilde{\mu}}{
(\mathrm{T}_{\mathcal{C}}f)^{> d}({\bf y}) (\mathrm{T}_{\mathcal{C}'} g)({\bf y'})}
&=\inner{\mathrm{T}_{\tilde{\mu}}^{*} (\mathrm{T}_{\mathcal{C}}f)^{> d}}{\mathrm{T}_{\mathcal{C}'} g}
\leq
\norm{\mathrm{T}_{\tilde{\mu}} (\mathrm{T}_{\mathcal{C}}f)^{> d}}_2
\norm{\mathrm{T}_{\mathcal{C}'} g}_2\\
&\leq
\norm{\mathrm{T}_{\tilde{\mu}}^{*}\mathrm{T}_{\tilde{\mu}} (\mathrm{T}_{\mathcal{C}}f)^{> d}}_2^{1/2}
\norm{(\mathrm{T}_{\mathcal{C}}f)^{> d}}_2^{1/2}
\norm{\mathrm{T}_{\mathcal{C}'} g}_2.
\end{align*}
Consider the distribution $\mu'$ on $[m]\times [m]$ corresponding to picking
$({\bf a},{\bf b})\sim \tilde{\mu}_1$, and then $({\bf a'},{\bf b'})\sim \tilde{\mu}_1$ conditioned on ${\bf b'} = {\bf b}$,
outputting ${\bf a}, {\bf a'}$. By properties of $\mu$ we get that $\mu'$ is connected and the probability
of each atom is at least $\alpha^2$. Also, $\mathrm{T}_{\mu'} = \mathrm{T}_{\tilde{\mu}}^{*}\mathrm{T}_{\tilde{\mu}}$,
so from Lemma~\ref{lem:high_deg_dies_prod_connected}
\[
\norm{\mathrm{T}_{\tilde{\mu}}^{*}\mathrm{T}_{\tilde{\mu}} (\mathrm{T}_{\mathcal{C}}f)^{> d}}_2
\leq (1-\Omega_{\alpha}(1))^d\norm{(\mathrm{T}_{\mathcal{C}}f)^{> d}}_2
\leq (1-\Omega_{\alpha}(1))^d\norm{f}_2.
\]
This completes the proof of~\eqref{eq:7}.

By Lemma~\ref{lem:coupling_preserve_deg} below,
\[
\norm{(\mathrm{T}_{\mathcal{C}}f)^{\leq d} - \mathrm{T}_{\mathcal{C}}(f^{\leq d})}_2 = O_{\alpha,d}(\zeta^{1/4})\norm{f}_2,
\qquad
\norm{(\mathrm{T}_{\mathcal{C}'}g)^{\leq d} - \mathrm{T}_{\mathcal{C}'}(g^{\leq d})}\leq O_{\alpha,d}(\zeta^{1/4})\norm{g}_2,
\]
so by another hybrid argument and Cauchy-Schwarz
\begin{equation}\label{eq:18}
\card{
\Expect{({\bf y},{\bf y'})\sim \tilde{\mu}}{(\mathrm{T}_{\mathcal{C}}f)^{\leq d}({\bf y}) (\mathrm{T}_{\mathcal{C}'} g)^{\leq d}({\bf y'})}
-\hspace{-2ex}
\Expect{({\bf y},{\bf y'})\sim \tilde{\mu}}{\mathrm{T}_{\mathcal{C}}(f^{\leq d})({\bf y}) \mathrm{T}_{\mathcal{C}'}(g^{\leq d})({\bf y'})}
}
\leq
O_{\alpha,d}(\zeta^{1/4})\norm{f}_2\norm{g}_2.
\end{equation}

Using Lemma~\ref{lem:inv_low_deg} the same way we did as in Section~\ref{sec:basic_inv_pf_overview}, we get that
\begin{align}
\card{
\Expect{({\bf x},{\bf x'})\sim \mu}{f^{\leq d}({\bf x})g^{\leq d}({\bf x'})}
-
\Expect{({\bf y},{\bf y'})\sim \tilde{\mu}}{\mathrm{T}_{\mathcal{C}}(f^{\leq d})({\bf y}) \mathrm{T}_{\mathcal{C}'}(g^{\leq d})({\bf y'})}
}
&\leq O_d(\zeta^{1/4})\norm{f^{\leq d}}_2\norm{g^{\leq d}}_2 \notag\\\label{eq:8}
&\leq O_d(\zeta^{1/4})\norm{f}_2\norm{g}_2.
\end{align}
Combining~\eqref{eq:6},~\eqref{eq:7},~\eqref{eq:18} and ~\eqref{eq:8} with the triangle inequality, and choosing
$d$ large enough so that $2\sqrt{C'}(1+\delta')^{-d/2},2\sqrt{C}(1+\delta)^{-d/2}\leq \eps/4$, and then
$\zeta$ small enough so that $O_d(\sqrt{\zeta}),O_{\alpha,d}(\zeta^{1/4})\leq \eps/4$, finishes the proof.\qed

\subsection{$\mathrm{T}_{\mathcal{C}}$ almost commutes with degree truncations}
In this section, we prove the following lemma, asserting that if $\mathcal{C}$ is a coupling with good enough
parameters in comparison to $d$, then $(\mathrm{T}_{\mathcal{C}} f)^{\leq d}\approx \mathrm{T}_{\mathcal{C}}(f^{\leq d})$.
More precisely:
\begin{lemma}\label{lem:coupling_preserve_deg}
  Suppose $\mathcal{C}$ is a $(\alpha,\zeta)$-coupling between $(\mathcal{U}_{\vec{k}},{\sf Uniform})$ and $([m]^n,\nu_{\vec{k}}^{\otimes n})$
  as in Definition~\ref{def:coupling}. If $f\colon \mathcal{U}_{\vec{k}}\to\mathbb{R}$ is a function, and $d\in\mathbb{N}$, then
  $\norm{(\mathrm{T}_{\mathcal{C}} f)^{\leq d} - \mathrm{T}_{\mathcal{C}}(f^{\leq d})}_2\leq O\left(d^{7/4} \alpha^{-1/4} \zeta^{1/4} \norm{f}_2\right)$.
\end{lemma}
  The rest of this section is devoted to the proof of Lemma~\ref{lem:coupling_preserve_deg}.

  Let $\mathrm{T}_{\rho}$ be the standard noise operator on $L^{2}([m]^n,\nu_{\vec{k}})$ from Section~\ref{sec:std_noise}.
  Our main goal is to show that for all $\rho>0$ and $f\in V_{=\ell}(\mathcal{U}_{\vec{k}})$, it holds that
  \begin{equation}\label{eq:16}
    \norm{\mathrm{T}_{\rho} \mathrm{T}_{\mathcal{C}} f - \rho^{\ell} \mathrm{T}_{\mathcal{C}} f}_2^2\leq 3\sqrt{\frac{4\ell}{\alpha}\zeta}\norm{f}_2^2.
  \end{equation}
  Intuitively, this would say that $\mathrm{T}_{\mathcal{C}} f$ is nearly of purely degree $\ell$,
  and indeed once established, we will quickly deduce Lemma~\ref{lem:coupling_preserve_deg} from~\eqref{eq:16}.
  \begin{claim}\label{claim:move_star}
  Suppose that for every $\rho>0$, $\ell\in \mathbb{N}$ and $g\in V_{=\ell}(\mathcal{U}_{\vec{k}})$ we have that
  \begin{equation}\label{eq:17}
  \norm{(\mathrm{T}_{\mathcal{C}}^{*}\mathrm{T}_{\rho} \mathrm{T}_{\mathcal{C}} - \rho^{\ell} \mathrm{I}) f}_2^2\leq \frac{4\ell}{\alpha}\zeta\norm{f}_2^2.
  \end{equation}
  Then~\eqref{eq:16} holds for all $\rho>0$ and $\ell\in\mathbb{N}$.
  \end{claim}
  \begin{proof}
  \begin{align*}
  \norm{\mathrm{T}_{\rho} \mathrm{T}_{\mathcal{C}} f - \rho^{\ell} \mathrm{T}_{\mathcal{C}} f}_2^2
  &=
  \norm{\mathrm{T}_{\rho} \mathrm{T}_{\mathcal{C}} f}_2^2 + \rho^{2\ell}\norm{\mathrm{T}_{\mathcal{C}} f}_2^2
  -2\rho^{\ell}\inner{\mathrm{T}_{\rho} \mathrm{T}_{\mathcal{C}} f}{\mathrm{T}_{\mathcal{C}} f}\\
  &=
  \inner{\mathrm{T}_{\mathcal{C}}^{*}\mathrm{T}_{\rho^2} \mathrm{T}_{\mathcal{C}} f}{f} +
  \rho^{2\ell}\inner{\mathrm{T}_{\mathcal{C}}^{*}\mathrm{T}_{\mathcal{C}} f}{f}
  -2\rho^{\ell}\inner{\mathrm{T}_{\mathcal{C}}^{*}\mathrm{T}_{\rho} \mathrm{T}_{\mathcal{C}} f}{ f}\\
  &=
  \inner{(\mathrm{T}_{\mathcal{C}}^{*}\mathrm{T}_{\rho^2} \mathrm{T}_{\mathcal{C}}-\rho^{2\ell} \mathrm{I})f}{f}
  +\rho^{2\ell}\inner{(\mathrm{T}_{\mathcal{C}}^{*}\mathrm{T}_{\mathcal{C}}-\mathrm{I})f}{f}
  -2\rho^{\ell}\inner{(\mathrm{T}_{\mathcal{C}}^{*}\mathrm{T}_{\rho} \mathrm{T}_{\mathcal{C}}-\rho^{\ell} \mathrm{I})f}{f}\\
  &\leq 3\sqrt{\frac{4\ell}{\alpha}\zeta}\norm{f}_2^2,
  \end{align*}
  where in the last transition we used Cauchy-Schwarz and our assumption.
  \end{proof}

  Let $\mathrm{S}_{\rho} = \mathrm{T}_{\mathcal{C}}^{*}\mathrm{T}_{\rho} \mathrm{T}_{\mathcal{C}} - \rho^{\ell} \mathrm{I}$,
  and think of $\mathrm{S}_{\rho}: L^2(\mathcal{U}_{\vec{k}})\to L^2(\mathcal{U}_{\vec{k}})$.
  By symmetry of $\mathcal{C}$, it follows that $\mathrm{T}_{\mathcal{C}}, \mathrm{T}_{\mathcal{C}}^{*}$ commute with the action
  of $S_n$ on functions, and so $\mathrm{S}_{\rho}$ commutes with the action of $S_n$ on functions. Thus, by Claim~\ref{claim:preserve_junta}
  it preserves juntas. Clearly, $\mathrm{S}_{\rho}$ is self-adjoint.

  By Claim~\ref{claim:preserve_rt}, $\mathrm{S}_{\rho}$ preserves the spaces $V_{=\ell}(\mathcal{U}_{\vec{k}})$, hence
  we may decompose it as a sum of eigenspaces $\bigoplus_{\theta}V_{=\ell}^{\theta}(\mathcal{U}_{\vec{k}})$. Thus, to establish~\eqref{eq:17},
  it suffices to show that if $V_{=\ell}^{\theta}(\mathcal{U}_{\vec{k}})\neq \set{0}$, then $\card{\theta}\leq \frac{2\ell}{\alpha}\zeta$. We fix
  $\theta^{\star}$ such that $V_{=\ell}^{\theta^{\star}}(\mathcal{U}_{\vec{k}})\neq \set{0}$ henceforth.
  Recall the spaces $V_{=J}(\mathcal{U}_{\vec{k}})$ from Section~\ref{sec:med_degs}; it follows that these spaces are invariant
  under $\mathrm{S}_{\rho}$, so we may decompose each one of them as a sum of eigenspaces
  $\bigoplus_{\theta} V_{=J}^{\theta}(\mathcal{U}_{\vec{k}})$. Denote
  $V_{\ell}(\theta) = \bigoplus_{\card{J}=\ell} V_{=J}^{\theta}(\mathcal{U}_{\vec{k}})$, then
  \[
  \bigoplus_{\theta} V_{\ell}(\theta)
  =\bigoplus_{\theta}\bigoplus_{\card{J}=\ell} V_{=J}^{\theta}(\mathcal{U}_{\vec{k}})
  =\bigoplus_{\card{J}=\ell} \bigoplus_{\theta}V_{=J}^{\theta}(\mathcal{U}_{\vec{k}})
  =\bigoplus_{\card{J}=\ell} \bigoplus_{\theta}V_{=J}(\mathcal{U}_{\vec{k}})
  =V_{=\ell}(\mathcal{U}_{\vec{k}}),
  \]
  so $V_{\ell}(\theta)$ are the eigenspaces of $\mathrm{S}_{\rho}$. Therefore, as $V_{\ell}^{\theta^{\star}}(\mathcal{U}_{\vec{k}})\neq\set{0}$,
  there is $\card{J} = \ell$ such that $V_{=J}^{\theta^{\star}}(\mathcal{U}_{\vec{k}})\neq \set{0}$. We thus find a non-zero
  $f^{\star}\in V_{=J}^{\theta^{\star}}(\mathcal{U}_{\vec{k}})$.

  Next, define an auxiliary operator $R_{\rho,J}$. For $x\in\mathcal{U}_{\vec{k}}$, we choose a set
  ${\bf S}\subseteq J$ randomly by including each $i\in [n]$ in ${\bf S}$ with probability $\rho$. We then pick
  ${\bf y}\in_R \mathcal{U}_{\vec{k}}$ conditioned on ${\bf y}_{\bf S} = x_{\bf S}$. We finish the proof with the following two claims.

  \begin{claim}
    $R_{\rho,J}f^{\star} = \rho^{\ell}f^{\star}$.
  \end{claim}
  \begin{proof}
    Write $R_{\rho,J} = \sum\limits_{T\subseteq J}{ p_T R_{\rho,J,T}}$, where $p_T$ is the probability that $S=T$, and
    $R_{\rho,J,T}$ is the action of the operator conditioned on ${\bf S}=T$. Node that
    $(R_{\rho,J,T} f^{\star})(y)$ is just the average of $f^{\star}({\bf x})$ over ${\bf x}\in\mathcal{U}_{\vec{k}}$ conditioned on ${\bf x}_T = y_T$,
    so it is proportional to the inner product $\inner{f^{\star}}{1_{x_T = y_T}}$.
    If $\card{T}\leq \ell-1$, then the last inner product is $0$ as $1_{x_T = y_T}\in V_{\ell-1}(\mathcal{U}_{\vec{k}})$. Thus,
    $R_{\rho,J} f^{\star} = p_J \cdot R_{\rho,J} f^{\star} = \rho^{\ell} f^{\star}$, where the last transition is since $f^{\star}$ is a $J$ junta,
    and $p_J = \rho^{\card{J}} = \rho^{\ell}$.
  \end{proof}

  \begin{claim}
    $\card{\theta^{\star}}\leq \frac{4\ell}{\alpha}\zeta$.
  \end{claim}
  \begin{proof}
    By the previous claim,
    $\mathrm{S}_{\rho} f^{\star} = (\mathrm{T}_{\mathcal{C}}^{*}\mathrm{T}_{\rho} \mathrm{T}_{\mathcal{C}} - R_{\rho,J}) f^{\star}$.
    Take $x$ that maximizes $\card{f^{\star}(x)}$.
    We now sample ${\bf y}$ according to $\mathrm{T}_{\mathcal{C}}^{*}\mathrm{T}_{\rho} \mathrm{T}_{\mathcal{C}}x$,
    and ${\bf y'}$ according to $R_{\rho,J} x$ in a coupled way so that $\Prob{}{{\bf y}'_J\neq {\bf y}_J}\leq \frac{2\card{J}}{\alpha}\zeta$.

    To do that, first sample ${\bf T}\subseteq [n]$ by including each $i\in[n]$ in ${\bf T}$ with probability $\rho$ independently.
    Sample ${\bf y}(0)\sim \mathrm{T}_{\mathcal{C}}x$, and sample ${\bf y}(1)$ by taking ${\bf y}(1)_{\bf T} = {\bf y}(0)_{\bf T}$ and
    resampling ${\bf y}(1)_i\sim\nu_{\vec{k}}$ for each $i\not\in {\bf T}$ independently, and finally sample
    ${\bf y}\sim \mathrm{T}_{\mathcal{C}}^{*}{\bf y}(1)$. To sample ${\bf y'}$, run the sampling procedure of $R_{\rho,J} x$ with $S = {\bf T}\cap J$.

    We note that if none of the coupling operators $\mathrm{T}_{\mathcal{C}}, \mathrm{T}_{\mathcal{C}}^{*}$ change the coordinates in $J$, then
    we indeed have ${\bf y}'_J= {\bf y}_J$. Since $\mathcal{C}$ is a $(\alpha,\zeta)$-coupling, it follows from the union bound that
    this probability is at least $1-\frac{2\ell\zeta}{\alpha}$. \footnote{Strictly speaking, the second property in Definition~\ref{def:coupling} asserts that
    this is true when we sample $({\bf x},{\bf y})\sim\mathcal{C}$. However the symmetry of $\mathcal{C}$ and the balancedness of our multi-slices
    imply that for all $i$ and $x$, when we sample $({\bf x},{\bf y})\sim\mathcal{C}$ conditioned on ${\bf x} = x$, we have that
    $\Prob{}{{\bf y}_i\neq x_i}\leq \zeta/\alpha$.}
    It follows that
    \[
    \card{\theta^{\star}}\card{f^{\star}(x)} =
    \card{\mathrm{S}_{\rho} f^{\star}(x)} =
    \card{\Expect{{\bf y},{\bf y'}}{f^{\star}({\bf y})-f^{\star}({\bf y'})}}
    \leq \Prob{}{{\bf y}'_J\neq {\bf y}_J}2\card{f^{\star}(x)}
    \leq \frac{4\ell\zeta}{\alpha} \card{f^{\star}(x)},
    \]
    and dividing by $\card{f^{\star}(x)}$ finishes the proof.
  \end{proof}

  It thus follows that~\eqref{eq:17} holds, and by Claim~\ref{claim:move_star} we have that~\eqref{eq:16} holds. Finally, to establish
  Lemma~\ref{lem:coupling_preserve_deg}, we require the following claim, asserting that $\mathrm{T}_{\mathcal{C}}$ cannot decrease degrees.
  \begin{claim}\label{claim:only_higher_degs}
    $f\in V_{d}(\mathcal{U}_{\vec{k}})^{\perp}$, then $\mathrm{T}_{\mathcal{C}} f \in V_{d}([m]^n,\nu_{\vec{k}}^{\otimes n})^{\perp}$.
  \end{claim}
  \begin{proof}
  To show that, it is enough to show that if
  $g\in V_{d}([m]^n,\nu_{\vec{k}}^{\otimes n})$ then $\mathrm{T}_{\mathcal{C}}^{*} g\in V_{d}(\mathcal{U}_{\vec{k}})$,
  as that would say that
  \[
  \inner{\mathrm{T}_{\mathcal{C}} f}{g}
  =\inner{f}{\mathrm{T}_{\mathcal{C}}^{*} g}
  =0,
  \]
  implying $\mathrm{T}_{\mathcal{C}} f \in V_{d}([m]^n,\nu_{\vec{k}}^{\otimes n})^{\perp}$.

  By linearity of the operator $\mathrm{T}_{\mathcal{C}}^{*}$, it suffices to show that if $\card{A}\leq d$,
  and $g$ is a $A$-junta, then $\mathrm{T}_{\mathcal{C}}^{*} g$ is also a $A$ junta. Fix $A$; then as $\mathcal{C}$ is symmetric it
  follows that $\mathrm{T}_{\mathcal{C}}^{*}$ commutes with the action of $S_n$ on functions, so for all $\pi\in S_n$ that fixes the coordinates
  of $A$ we have that $^{\pi}\left(\mathrm{T}_{\mathcal{C}}^{*} g\right) = \mathrm{T}_{\mathcal{C}}^{*} {}^{\pi}g = \mathrm{T}_{\mathcal{C}}^{*} g$,
  so by Fact~\ref{fact:junta_trivial} we get that $\mathrm{T}_{\mathcal{C}}^{*} g$ is indeed a $A$-junta.
  \end{proof}

  We are now ready to prove Lemma~\ref{lem:coupling_preserve_deg}.
  \begin{proof}[Proof of Lemma~\ref{lem:coupling_preserve_deg}]
  Let $\rho>0$ to be determined later.  By Claim~\ref{claim:only_higher_degs} we get that $(\mathrm{T}_{\mathcal{C}} f)^{\leq d} = (\mathrm{T}_{\mathcal{C}} (f^{\leq d}))^{\leq d}$.
  Let $g_j = \mathrm{T}_{\mathcal{C}} (f^{=j})$, and write $g_j = \sum\limits_{i=0}^{n} g_j^{=i}$. Then by Fact~\ref{fact:noise_operator_ev} it follows that
  $\mathrm{T}_{\rho} g_j^{=i} = \rho^{i}g_j^{=i}$, and so by linearity $\mathrm{T}_{\rho} g_j= \sum\limits_{i=0}^{n} \rho^{i} g_j^{=i}$.
  On the other hand, by~\eqref{eq:16} $\norm{\mathrm{T}_{\rho} g_j - \rho^j g_j}\leq 6\sqrt{\frac{j}{\alpha}\zeta}\norm{f^{=j}}_2^2$,
  and plugging in the expansion of $g_j$ we get that
  \[
  \sum\limits_{i=0}^{n}(\rho^j - \rho^{i})^2\norm{g_j^{=i}}_2^2
  \leq 6\sqrt{\frac{j}{\alpha}\zeta}\norm{f^{=j}}_2^2.
  \]
  It follows that $\sum\limits_{i\neq j}\norm{g_j^{=i}}_2^2\leq 6\frac{\sqrt{j\zeta}}{\rho^{2j}(1-\rho)^2\sqrt{\alpha}}\norm{f^{=j}}_2^2$,
  and therefore by Parseval
  \[
  \norm{\sum\limits_{i \neq j }g_j^{=i}}_2
  \leq
  \frac{\sqrt{6}(j\zeta)^{1/4}}{\rho^j(1-\rho)\alpha^{1/4}}\norm{f^{=j}}_2,
  \]
   Summarizing, we get that
  \begin{align*}
    \norm{(\mathrm{T}_{\mathcal{C}} f)^{\leq d} - \mathrm{T}_{\mathcal{C}}(f^{\leq d})}_2
    = \norm{(\mathrm{T}_{\mathcal{C}} f^{\leq d})^{\leq d} - \mathrm{T}_{\mathcal{C}}(f^{\leq d})}_2
    &\leq \sum\limits_{j\leq d} \norm{(\mathrm{T}_{\mathcal{C}} f^{ = j})^{\leq d} - \mathrm{T}_{\mathcal{C}}(f^{=j})}_2\\
    &= \sum\limits_{j\leq d}\norm{\sum\limits_{i > d} g_j^{=i}}_2\\
    &\leq \sum\limits_{j\leq d}\norm{\sum\limits_{i \neq j} g_j^{=i}}_2\\
    &\leq \sum\limits_{j\leq d} \frac{\sqrt{6}(j\zeta)^{1/4}}{\rho^j(1-\rho)\alpha^{1/4}}\norm{f^{=j}}_2\\
    &\leq  \frac{\sqrt{6}(d\zeta)^{1/4}}{(1-\rho)\alpha^{1/4}}
    \sqrt{\sum\limits_{j\leq d} \rho^{-j}}\sqrt{\sum\limits_{j\leq d} \norm{f^{=j}}_2^2}\\
    &\leq \frac{\sqrt{6}(d\zeta)^{1/4}}{(1-\rho)\alpha^{1/4}}\sqrt{\frac{\rho^{-(d+1)} - 1}{\rho^{-1} - 1}}\norm{f}_2\\
    &=\frac{\sqrt{6}(d\zeta)^{1/4}}{(1-\rho)^{3/2}\alpha^{1/4}}\rho^{-d/2}\norm{f}_2.
  \end{align*}
  Choosing $\rho = 1-\frac{1}{2d}>0$ finishes the proof.
  \end{proof}

\section{Invariance principle in the multi-linear case}
In this section, we generalize Theorem~\ref{thm:basic_inv_bipartite} to the case that the distribution $\mu$ is over $r$-tuples,
where $r>2$.
%For that, we first define what it means for a distribution over $r$-tuples to be connected.
%\begin{definition}
%Let $r,m_1,\ldots,m_r\in\mathbb{N}$, and let $\mu$ be a distribution over $[m_1]^n\times\ldots[m_r]^n$. We say $\mu$ is connected,
%if for all $i$, the graph $H_i = (V_1\cup V_2, E)$ defined as follows is connected:
%(1) $V_1 \subseteq [m_i]^n$, $V_2 = \prod\limits_{j\neq i} [m_j]^n$
%are the support of the corresponding marginal distributions of $\mu$, and
%(2) $(v_1,v_2)$ is an edge if letting $v = x(i)$,
%$v_2 = (x(1),\ldots,x(i-1),x(i+1),\ldots,x(r))$ for $x(j)\in [m_j]^n$, we have that $(x(1),\ldots,x(r))$ is in the support of
%$\mu$.
%\end{definition}
\subsection{Invariance principle for products}
In this section we prove Theorem~\ref{thm:basic_inv_multi}, restated below.
\begin{reptheorem}{thm:basic_inv_multi}
  For all $\alpha\in (0,1)$, $M, r\in\mathbb{N}$, $m_1,\ldots,m_{r}\in\mathbb{N}$, $\eps>0$
  there are $\zeta>0$, $N\in\mathbb{N}$ such that the following holds for $n\geq N$.
  Suppose $\mathcal{U}_{\vec{k}(1)},\ldots,\mathcal{U}_{\vec{k}(r)}$ are $\alpha$-balanced multi-slices over alphabets
  $[m_1],\ldots,[m_r]$ respectively, $\mu$ is a connected, $\alpha$-admissible distribution over
  $\prod\limits_{i=1}^{r}\mathcal{U}_{\vec{k}(i)}$, and let $\tilde{\mu}$ the product version of $\mu$ as in Definition~\ref{def:product_ver}.
  Suppose $\mathcal{C}_1,\ldots,\mathcal{C}_r$ are couplings such that $\mathcal{C}_i$ is a
  $(\alpha,\zeta)$-coupling between $\mathcal{U}_{\vec{k}(i)}$ and $([m_i]^n, \nu_{\vec{k}(i)}^{\otimes n})$,
  and that there is a $(\alpha,\zeta)$-coupling, say $\mathcal{C}$, between $\mu$ and $\tilde{\mu}$.

  Then for all $f_i\colon \mathcal{U}_{\vec{k}(i)}\to\mathbb{R}$ such that $\norm{f_i}_{2r}\leq M$ for all $i$, it holds that
  \[
  \card{
  \Expect{({\bf x}(1),\ldots,{\bf x}(r))\sim \mu}{\prod\limits_{i=1}^{r}f_i({\bf x}(i))} -
  \Expect{({\bf y}(1),\ldots, {\bf y}(r))\sim \tilde{\mu}}{\prod\limits_{i=1}^{r}\mathrm{T}_{\mathcal{C}_i}f_i({\bf y}(i))}
  }
  \leq \eps.
  \]
\end{reptheorem}
The rest of this section is devoted to the proof of Theorem~\ref{thm:basic_inv_multi}. We choose a
sequence of degrees $d_1\leq d_2\leq\ldots\leq d_r$, where $d_1$ is sufficiently large function of
$\alpha,r,m_1,\ldots,m_r$ and $\eps$, and for each $i$, once $d_i$ has been picked, $d_{i+1}$ is picked
to be sufficiently large. Finally, $N$ is picked to be sufficiently large, and $\zeta$ is picked to be sufficiently small.

\begin{claim}\label{claim:deg_truncate_multilin}
\begin{equation*}
  \card{
  \Expect{({\bf x}(1),\ldots,{\bf x}(r))\sim \mu}{\prod\limits_{i=1}^{r}f_i({\bf x}(i))}
  -
  \Expect{({\bf x}(1),\ldots,{\bf x}(r))\sim \mu}{\prod\limits_{i=1}^{r}f_i^{\leq d_i}({\bf x}(i))}
  }
  \leq \frac{\eps}{3}.
\end{equation*}
\end{claim}
\begin{proof}
  Let $g_i = f_i^{\leq d_i}$. We intend to prove for all $j=0,1,\ldots,r-1$ that
  \[
  \card{
  \Expect{({\bf x}(1),\ldots,{\bf x}(r))\sim \mu}{\prod\limits_{i=1}^{j}g_i({\bf x}(i))\prod\limits_{i=j+1}^{r}f_i({\bf x}(i))}
  -
  \Expect{({\bf x}(1),\ldots,{\bf x}(r))\sim \mu}{\prod\limits_{i=1}^{j+1}g_i({\bf x}(i))\prod\limits_{i=j+2}^{r}f_i({\bf x}(i))}
  }
  \leq \frac{\eps}{3r},
  \]
  and the claim follows from summing this up and using the triangle inequality. Fix $j$, then the above expression is equal to
  \begin{equation}\label{eq:10}
  \card{
  \Expect{({\bf x}(1),\ldots,{\bf x}(r))\sim \mu}{h({\bf x}(1),\ldots,{\bf x}(j),{\bf x}(j+2),\ldots,{\bf x}(r)) f_{j+1}^{>d_{j+1}}({\bf x}(j+1))}
  },
  \end{equation}
  where $h(x(1),\ldots,x(j),x(j+2),\ldots,x(r)) =
  \prod\limits_{i=1}^{j}g_i(x(i))\prod\limits_{i=j+2}^{r}f_i(x(i))$. Next, we interpret this expectation as an inner product.
  Consider the operator $\mathrm{T}_{j+1}\colon L^2(\mathcal{U}_{\vec{k}(j+1)}) \to L^2(\prod\limits_{i\neq j+1}\mathcal{U}_{\vec{k}(i)})$
  defined as
  \[
  \mathrm{T}_{j+1} f(z) =
  \cExpect{({\bf x}(1),\ldots,{\bf x}(r))\sim \mu}{({\bf x}(1),\ldots,{\bf x}(j), {\bf x}(j+2),\ldots,{\bf x}(r)) = z}
  {f({\bf x}(j+1))}.
  \]
 % and then the operator $\mathcal{S}_{j+1}\colon L^2(\mathcal{U}_{\vec{k}(j+1)})\to L^2(\mathcal{U}_{\vec{k}(j+1)})$ defined as
%  $\mathcal{S}_{j+1} = \mathrm{T}_i^{*}\mathrm{T}_i$.
  Consider the distribution $\nu_{j+1}$ over $\mathcal{U}_{\vec{k}(j+1)}\times \mathcal{U}_{\vec{k}(j+1)}$ as the distribution of
  $({\bf x}(j+1), {\bf x}(j+1)')$ where we first pick
  $({\bf x}(1),\ldots,{\bf x}(h))\sim \mu$, and then $({\bf x}(1)',\ldots,{\bf x}(h))\sim \mu$ conditioned on
  ${\bf x}(i)' = {\bf x}(i)$ for all $i\neq j+1$. Then note that $\mathrm{T}_{\nu} = \mathrm{T}_{j+1}^{*}\mathrm{T}_{j+1}$,
  and so
  \begin{align}
  \eqref{eq:10}
  = \inner{h}{\mathrm{T}_{j+1} f_{j+1}^{>d_{j+1}}}
  \leq \norm{h}_2\norm{\mathrm{T}_{j+1}  f_{j+1}^{>d_{j+1}}}_2
  &= \norm{h}_2\sqrt{\inner{\mathrm{T}_{j+1}  f_{j+1}^{>d_{j+1}}}{\mathrm{T}_{j+1}  f_{j+1}^{>d_{j+1}}}} \notag\\\label{eq:11}
  &\leq \norm{h}_2 \norm{f_{j+1}^{>d_{j+1}}}_2^{1/2} \norm{\mathrm{T}_{\nu} f_{j+1}^{>d_{j+1}}}_2^{1/2}.
  \end{align}
  First, we use Lemma~\ref{lem:upperbound_ev} to bound $\norm{\mathrm{T}_{\nu} f_{j+1}^{>d_{j+1}}}_2$. Indeed, the connectedness
  of $\mu$ implies that $\nu$ is connected, and the $\alpha$-admissibility of $\mu$ implies that $\nu$ is $\alpha^2$ admissible.
  Also, using the coupling $\mathcal{C}$ one may easily construct a $(\alpha/2,2\zeta)$-coupling $\mathcal{C}'$ between $\nu$
  and $\tilde{\nu}$. We thus have by Lemma~\ref{lem:upperbound_ev} that
  \[
  \norm{\mathrm{T}_{\nu} f_{j+1}^{>d_{j+1}}}_2
  \leq
  C(m,\alpha) (1+\delta)^{-d_{j+1}}\norm{f_{j+1}^{>d_{j+1}}}_2
  \leq
  C(m,\alpha) (1+\delta)^{-d_{j+1}}\norm{f_{j+1}}_2,
  \]
  which is at most $C(m,M,\alpha) (1+\delta)^{-d_{j+1}}$ as $\norm{f_{j+1}}_2\leq \norm{f_{j+1}}_{2r}\leq M$.
  Here $\delta>0$ depends only on $\alpha$ and $m$.
  Plugging this into~\eqref{eq:11} we get that
  $\eqref{eq:10}\leq C(m,\alpha) (1+\delta)^{-d_{j+1}} \norm{h}_2$,
  and we next upper bound the norm of $h$. By H\"{o}lder's inequality we get that
  \[
  \norm{h}_2^2
  \leq \Expect{}{\prod\limits_{i=1}^{j}g_i({\bf x}(i))^2\prod\limits_{i' = j+2}^{r} f_{i'}({\bf x}(i'))^2}
  \leq \prod\limits_{i=1}^{j}\norm{g_i^2}_{r-1}\hspace{-1ex}\prod\limits_{i'=j+2}^{r}\hspace{-0.5ex}\norm{f_{i'}^2}_{r-1}
  = \prod\limits_{i=1}^{j}\norm{g_i}_{2(r-1)}^2\hspace{-1ex}\prod\limits_{i'=j+2}^{r}\hspace{-0.5ex}\norm{f_{i'}}_{2(r-1)}^2,
  \]
  and we argue that $\norm{g_i}_{2(r-1)},\norm{f_{i'}}_{2(r-1)} = O_{r,M,d_1,\ldots,d_j}(1)$. Indeed, using monotonicity of the norms and then
  hypercontractivity (Theorem~\ref{thm:hypercontractivity_multi}) we get
  \[
  \norm{g_i}_{2(r-1)}
  \leq \norm{g_i}_{2r}
  \leq O_{r,d_1,\ldots,d_j}(1) \norm{g_i}_2
  \leq O_{r,d_1,\ldots,d_j}(1) M
  \leq O_{r,M,d_1,\ldots,d_j}(1),
  \]
  where we used Parseval to bound $\norm{g_i}_2\leq \norm{f_i}_2\leq \norm{f_i}_{2r} \leq M$. For $f_{i'}$, we have
  $\norm{f_{i'}}_{2(r-1)}\leq \norm{f_{i'}}_{2r}\leq M$.

  Plugging everything
  into~\eqref{eq:11} we get that
  $\eqref{eq:10} \leq O_{\alpha,m,r,d_1,\ldots,d_j}(1) (1+\delta)^{-d_{j+1}}$, so picking $d_{j+1}$ large enough gives that
  $\eqref{eq:10}\leq \frac{\eps}{3r}$.
\end{proof}

\begin{claim}\label{claim:deg_truncate_multilin_prod}
\begin{equation*}
  \card{
  \Expect{({\bf x}(1),\ldots,{\bf x}(r))\sim \tilde{\mu}}{\prod\limits_{i=1}^{r}\mathrm{T}_{\mathcal{C}_i}f_i({\bf x}(i))}
  -
  \Expect{({\bf x}(1),\ldots,{\bf x}(r))\sim \tilde{\mu}}{\prod\limits_{i=1}^{r}(\mathrm{T}_{\mathcal{C}_i}f_i)^{\leq d_i}({\bf x}(i))}
  }
  \leq \frac{\eps}{3}.
\end{equation*}
\end{claim}
\begin{proof}
  The proof is identical to the proof of Claim~\ref{claim:deg_truncate_multilin}, except that we use Lemma~\ref{lem:high_deg_dies_prod_connected}
  instead of Lemma~\ref{lem:upperbound_ev}, and Theorem~\ref{thm:hypercontractivity_prod} instead of Theorem~\ref{thm:hypercontractivity_multi}.
\end{proof}

\begin{claim}\label{claim:low_deg_inv_multi_lin}
\begin{equation*}
  \card{
  \Expect{({\bf x}(1),\ldots,{\bf x}(r))\sim \mu}{\prod\limits_{i=1}^{r}f_i^{\leq d_i}({\bf x}(i))}
  -
  \Expect{({\bf x}(1),\ldots,{\bf x}(r))\sim \tilde{\mu}}{\prod\limits_{i=1}^{r}(\mathrm{T}_{\mathcal{C}_i}f_i)^{\leq d_i}({\bf x}(i))}
  }
  \leq \frac{\eps}{3}.
\end{equation*}
\end{claim}
\begin{proof}
 Let $\mathcal{C}$ be a $(\alpha,\zeta)$-coupling between $\mu$ and $\tilde{\mu}$ guaranteed to exist by the assumption of the theorem.
 Applying the hybrid argument again, it is enough to show that for all $j=0,\ldots,r-1$, the absolute value of
 \begin{align}\label{eq:12}
 \Expect{\left(\substack{{\bf x}(1),\ldots,{\bf x}(r)\\ {\bf y}(1),\ldots,{\bf y}(r)}\right)\sim \mathcal{C}}
 {\prod\limits_{i=1}^{j}f_i^{\leq d_i}({\bf x}(i))
 \hspace{-0.5ex}
 \left(f_{j+1}^{\leq d_{j+1}}({\bf x}(j+1)) - (\mathrm{T}_{\mathcal{C}_{j+1}} f_{j+1})^{\leq d_{j+1}}({\bf y}(j+1))\right)
 \hspace{-1.5ex}\prod\limits_{i=j+2}^{r}(\mathrm{T}_{\mathcal{C}_i} f_i)^{\leq d_i}({\bf y}(i))}
 \end{align}
 is at most $\frac{\eps}{3r}$.  By Cauchy-Schwarz, this absolute value is at most
 \[
 \norm{h}_2
 \sqrt{\Expect{\left(\substack{{\bf x}(1),\ldots,{\bf x}(r)\\ {\bf y}(1),\ldots,{\bf y}(r)}\right)\sim \mathcal{C}}
 {\left(f_{j+1}^{\leq d_{j+1}}({\bf x}(j+1)) - (\mathrm{T}_{\mathcal{C}_{j+1}} f_{j+1})^{\leq d_{j+1}}({\bf y}(j+1))\right)^2}},
 \]
 where $h = \prod\limits_{i=1}^{j}f_i^{\leq d_i}(x(i)) \prod\limits_{i=j+2}^{r}(\mathrm{T}_{\mathcal{C}_i} f_i)^{\leq d_i}(y(i))$.
 The norm of $h$ is bounded as in the previous two claims using hypercontractivity by $O_{r,M,m,\alpha,d_1,\ldots,d_r}(1)$.
 For the other expectation, note first that by Lemma~\ref{lem:coupling_preserve_deg}
 \[
 \norm{(\mathrm{T}_{\mathcal{C}_{j+1}} f_{j+1})^{\leq d_{j+1}}  - \mathrm{T}_{\mathcal{C}_{j+1}} (f_{j+1}^{\leq d_{j+1}})}_2
 \leq O_{d_{j+1},\alpha}(\zeta^{1/4}),
 \]
 so by $(a+b)^2\leq 2(a^2+b^2)$ the expectation is upper bounded by
 \[
 O_{d_{j+1},\alpha}(\sqrt{\zeta})
 +
 2\Expect{\left(\substack{{\bf x}(1),\ldots,{\bf x}(r)\\ {\bf y}(1),\ldots,{\bf y}(r)}\right)\sim \mathcal{C}}
 {\left(f_{j+1}^{\leq d_{j+1}}({\bf x}(j+1)) - \mathrm{T}_{\mathcal{C}_{j+1}} (f_{j+1}^{\leq d_{j+1}})({\bf y}(j+1))\right)^2}.
 \]
 Finally, Lemma~\ref{lem:inv_low_deg} implies that the last expectation is upper bounded by $8\sqrt{d_{j+1}\zeta}$. Thus, overall~\eqref{eq:12} is bounded
 by $O_{r,M,m,\alpha,d_1,\ldots,d_r}(\zeta^{1/4})$, so taking $\zeta$ small enough gives it is at most $\frac{\eps}{3r}$.
 Summing~\eqref{eq:12} over all $j$ and using the triangle inequality finishes the proof.
\end{proof}

Theorem~\ref{thm:basic_inv_multi} now follows from Claims~\ref{claim:deg_truncate_multilin},~\ref{claim:deg_truncate_multilin_prod},~\ref{claim:low_deg_inv_multi_lin}
and the triangle inequality.

\subsection{Invariance principle for label-assignments}\label{sec:label_assignments}
In this section, we state a version of our invariance principle, Theorem~\ref{thm:basic_inv_multi}, that will be useful in
our PCP reduction. Let $\Sigma_i = [m_i]$ be alphabets for $i=1,\ldots,r$, and let $P\colon \prod\limits_{i=1}^{r}\Sigma_i\to\power{}$ be a predicate.
Let $f_i\colon\mathcal{U}_{\vec{k}(i)}\to[m_i]$ be functions where $\mathcal{U}_{\vec{k}(i)}\subseteq [m_i]^n$ are
all $\alpha$-balanced multislices.
We will want to prove invariance for expressions of the form $\Expect{({\bf x}(1),\ldots,{\bf x}(r))\sim \mu}{P(f_1({\bf x}(1)),\ldots, f_1({\bf x}(r)))}$.
At the moment, it does not make any sense -- the function $\mathrm{T}_{\mathcal{C}_i} f_i$ is not even well defined
(as $\Sigma$ does not have a sensible additive structure). We therefore view, as is standard, the alphabets as standard simplices.

Let $\Delta_m = \{ (t_0,\ldots,t_{m-1}) | t_i \geq 0, \sum_{i=0}^{m-1} t_i =1\}$ be the standard $m$-dimensional simplex. We identify a symbol $\sigma\in \Sigma_i$ with
the standard basis element $e_{\sigma}\in \Delta_{m_i}$. This allows us to view a given function $f_i\colon\mathcal{U}_{\vec{k}(i)}\to[m_i]$ as a function
$f_i\colon\mathcal{U}_{\vec{k}(i)}\to\Delta_{m_i}$, which we will do (slightly abusing notation by denoting the two functions the same way).
Thus, now given a coupling $\mathcal{C}_i$ between $\mathcal{U}_{\vec{k}(i)}$ and $([m]^n,\nu_{\vec{k}(i)})$, we define
\[
\mathrm{T}_{\mathcal{C}_i} f_i = ( \mathrm{T}_{\mathcal{C}_i} f_{i,1},\ldots, \mathrm{T}_{\mathcal{C}_i} f_{i,m_i}),
\]
where $f_{i,j}$ is the $j$th coordinate of $f_i$. We remark that clearly, if $f_i\colon\mathcal{U}_{\vec{k}(i)}\to\Delta_{m_i}$, then
as $\mathrm{T}_{\mathcal{C}_i}$ is an averaging operator and $\Delta_{m_i}$ is convex, we get that
$\mathrm{T}_{\mathcal{C}_i} f_i\colon\mathcal{U}_{\vec{k}(i)}\to\Delta_{m_i}$

Next, we define the natural extension of our predicate $P$, denoted by $\tilde{P}\colon \prod\limits_{i=1}^{r}\Delta_{m_i}\to [0,1]$.
If for $i=1,\ldots,r$, each $y(i)\in\Delta_{m_i}$ is standard basis elements, say
$y(i) = e_{\sigma_i}$, then we define $\tilde{P}(y(1),\ldots, y(r)) = P(\sigma_1,\ldots,\sigma_r)$. For the rest
of the points, we extend $\tilde{P}$ multilinearly; let $y(i)\in\Delta_{m_i}$ for $i=1,\ldots,r$, and express each $y(i)$
using the standard basis as $y(i) = \sum\limits_{\sigma\in \Sigma_i}{y(i)_{\sigma} e_{\sigma}}$. Then we define
\[
\tilde{P}(y(1),\ldots,y(r)) =
\sum\limits_{\sigma_1\in\Sigma_1,\ldots,\sigma_r\in\Sigma_r}
P(\sigma_1,\ldots,\sigma_r)\prod\limits_{i=1}^{r} y(i)_{\sigma_i}.
\]
\begin{thm}\label{thm:inv_label_assignments}
  For all $\alpha\in (0,1)$, $r\in\mathbb{N}$, $m_1,\ldots,m_{r}\in\mathbb{N}$, $\eps>0$
  there are $\zeta>0$, $N\in\mathbb{N}$ such that the following holds for $n\geq N$.
  Suppose $\mathcal{U}_{\vec{k}(1)},\ldots,\mathcal{U}_{\vec{k}(r)}$ are $\alpha$-balanced multi-slices over alphabets
  $[m_1],\ldots,[m_r]$ respectively, and $\mu$ is a connected, $\alpha$-admissible distribution over
  $\prod\limits_{i=1}^{r}\mathcal{U}_{\vec{k}(i)}$.
  Suppose for each $i\in[r]$, $\mathcal{C}_i$ is a $(\alpha,\zeta)$-coupling between $\mathcal{U}_{\vec{k}(i)}$ and $([m_i]^n,\nu_{\vec{k}(i)}^{\otimes n})$,
  and that there is a $(\alpha,\zeta)$-coupling between $\mu$ and $\tilde{\mu}$.
  Then for all $P\colon \prod\limits_{i=1}^{r}[m_i]\to [-1,1]$ and $f_i\colon \mathcal{U}_{\vec{k}(i)}\to [m_i]$ it holds that
  \[
  \card{
  \Expect{({\bf x}(1),\ldots,{\bf x}(r))\sim \mu}{P(f_1({\bf x}(1)),\ldots, f_1({\bf x}(r)))} -
  \Expect{({\bf y}(1),\ldots, {\bf y}(r))\sim \tilde{\mu}}{\tilde{P}(\mathrm{T}_{\mathcal{C}_1}f_1({\bf y}(1)),\ldots \mathrm{T}_{\mathcal{C}_r}f_r({\bf y}(r)))}
  }
  \leq \eps.
  \]
\end{thm}
\begin{proof}
  Set $\eps' = \eps/(m_1\cdots m_r)$, and pick $\delta,N$ from Theorem~\ref{thm:basic_inv_multi} for $\alpha,r,m_1,\ldots,m_r$ and
  $\eps'$. Then the left hand side is at most
  \[
  \sum\limits_{\sigma_1\in[m_1],\ldots,\sigma_r\in[m_r]}
  \card{P(\sigma_1,\ldots,\sigma_r)}
  \card{
  \Expect{({\bf x}(1),\ldots,{\bf x}(r))\sim \mu}{\prod\limits_{i=1}^r f_{i,\sigma_i}({\bf x}(i))} -
  \Expect{({\bf y}(1),\ldots, {\bf y}(r))\sim \tilde{\mu}}{\prod\limits_{i=1}^r\mathrm{T}_{\mathcal{C}_i}f_{i,\sigma_i}({\bf y}(i))}
  }.
  \]
  By Theorem~\ref{thm:basic_inv_multi}, the second absolute value is at most $\eps'$, hence we get the bound
  \[
  \sum\limits_{\sigma_1\in[m_1],\ldots,\sigma_r\in[m_r]}
  \card{P(\sigma_1,\ldots,\sigma_r)} \eps'\leq \eps
  .\qedhere
  \]
 % the second absolute value is at most $\eps'$,
%  We may write $P(z_1,\ldots,z_r) = \sum\limits_{S\subseteq[r]} \widehat{P}(S) \prod_{i\in S} z_i$ where
%  $\widehat{P}(S) = \Expect{z\sim\set{-1,1}^r}{P(z)\prod_{i\in S} z_i}$, so the difference in the statement of the Theorem
%  is at most
%  \[
%  \sum\limits_{S\subseteq[r]}
%  \card{\widehat{P}(S)}
%  \card{
%  \Expect{({\bf x(1)},\ldots,{\bf x(r)})\sim \mu}{\prod\limits_{i\in S} f_i({\bf x(i)})} -
%  \Expect{({\bf y(1)},\ldots, {\bf y(r)})\sim \tilde{\mu}}{\prod\limits_{i\in S}\mathrm{T}_{\mathcal{C}_i}f_i({\bf y(i)})}
%  }.
%  \]
%  By Theorem~\ref{thm:basic_inv_multi} we have that for all $S\subseteq[r]$, the difference of expectations is at most $\eps'$,
%  and as $\card{\widehat{P}(S)}\leq 1$ the proof is concluded.
\end{proof}

\subsection{Using our invariance principle: a construction of a useful coupling}\label{sec:example_coupling}
In this section, we construct a useful coupling between admissible distributions over products of balanced multi-slices, and their product version.
\begin{definition}\label{def:neg_correlated}
  For each $i=1,\ldots,r$, let $\mathcal{U}_{\vec{k}(1)}$ be a multi-slice over the alphabet $[m_i]$, and
  let $\mu$ be a distribution over $\mathcal{U}_{\vec{k}(1)}\times\ldots\times \mathcal{U}_{\vec{k}(r)}$.
  We say $\mu$ is negatively correlated if for all $\vec{a}\in [m_1]\times\ldots\times[m_r]$, defining the events
  $A_i = \sett{(x(1),\ldots,x(r))}{(x(1)_i,\ldots,x(r)_i) = \vec{a}}$, the random variables $1_{A_i}$ for $i=1,\ldots,n$ are negatively associated
  (where the sampling of $(x(1),\ldots,x(r))$ is done under $\mu$).
\end{definition}

\begin{proposition}\label{prop:coupling_construction}
  For all $\alpha>0$, $r,m_1,\ldots,m_r\in \mathbb{N}$ there are $\alpha',K>0$ such that the following holds.
  Let $\mu$ be an $\alpha$-admissible distribution, negatively correlated distribution
  over $\mathcal{U}_{\vec{k}(1)}\times\ldots\times \mathcal{U}_{\vec{k}(r)}$,
  where for each $i$, the multi-slice $\mathcal{U}_{\vec{k}(i)}$ has alphabet $[m_i]$ and is $\alpha$-balanced.
  Let $\tilde{\mu}$ be the product version of $\mu$.

  Then there is a $(\alpha', K \frac{1}{\sqrt{n}})$-coupling between $\mu$ and $\tilde{\mu}$.
\end{proposition}
\begin{proof}
We define
the coupling $\mathcal{C} = (({\bf x}(1),\ldots,{\bf x}(r)),({\bf y}(1),\ldots,{\bf y}(r)))$ as follows:
\begin{enumerate}
  \item Sample a statistics of $({\bf y}(1)',\ldots,{\bf y}(r)')\sim \tilde{\mu}$.
    That is, sample $({\bf y}(1)',\ldots,{\bf y}(r)')\sim \tilde{\mu}$, and for each $\vec{a} = (a_1,\ldots,a_r)\in [m_1]\times\ldots\times[m_r]$,
    denote by $r_{\vec{a}}$ the number of coordinates $i\in[n]$ such that ${\bf y}(1)'_i = a_1,\ldots,{\bf y}(r)'_i = a_r$.
  \item Sample a statistics of $({\bf x}(1)',\ldots,{\bf x}(r)')\sim \mu$.
    That is, sample $({\bf x}(1)',\ldots,{\bf x}(r)')\sim \mu$, and for each $\vec{a} = (a_1,\ldots,a_r)\in [m_1]\times\ldots\times[m_r]$ denote by
    $k_{\vec{a}}$ the number of coordinates $i\in[n]$ such that ${\bf x}(1)'_i = a_1,\ldots, {\bf x}(r)'_i = a_r$.
  \item
  Let us consider the lexicographical ordering on $[m_1]\times\ldots\times[m_r]$.
  Sample $\bm{\pi}\in S_n$, and divide the integers $1,\ldots,n$ into consecutive segments
  $(I_{\vec{a}})_{\vec{a}\in [m_1]\times\ldots\times[m_r]}$ of lengths $k_{\vec{a}}$, according to the lexicographical ordering on
  $[m_1]\times\ldots\times[m_r]$. Also, divide the integers $1,\ldots,n$ into segments
  $(J_{\vec{a}})_{\vec{a}\in [m_1]\times\ldots\times[m_r]}$  of lengths $r_{\vec{a}}$ again according to the lexicographical ordering
  on $[m_1]\times\ldots\times[m_r]$.

  \item Define ${\bf x}(1),\ldots,{\bf x}(r)$ by setting ${\bf x}(1)_{\bm{\pi}(\ell)} = a_1,\ldots, {\bf x}(r)_{\bm{\pi}(\ell)} = a_r$
  for all $\vec{a}\in [m_1]\times\ldots\times[m_r]$ and $\ell\in I_{\vec{a}}$. Also, define ${\bf y}(1),\ldots,{\bf y}(r)$
  by setting ${\bf y}(1)_{\bm{\pi}(\ell)} = a_1,\ldots, {\bf y}(r)_{\bm{\pi}(\ell)} = a_r$ for all $\vec{a}\in [m_1]\times\ldots\times[m_r]$ and
  $\ell\in J_{\vec{a}}$.
\end{enumerate}
We now argue that $\mathcal{C}$ is a $(\Omega_{\alpha,r,m_1,\ldots,m_r}(1),O_{\alpha,r,m_1,\ldots m_r}(1/\sqrt{n}))$-coupling according to Definition~\ref{def:coupling}.
The first, second items are immediate, and we focus on the third and fourth items.

For the fourth item, let $R\subseteq [m_1]\times\ldots\times[m_r]$ be the set of symbols that occur with positive
probability in a coordinate of $\mu$, and let $\vec{a}\in R$.
Set $m = m_1\cdots m_r$, $\eps' = \frac{\eps}{4 m^{3}}$.
Fix $\vec{a} \in R$, and note that for $i=1,\ldots,n$,
the events $({\bf x}(1)'_i,\ldots,{\bf x}(r)'_i) = \vec{a}$ are negatively associated, so by Theorem~\ref{thm:chernoff_negatively} we have
that $k_{\vec{a}}\leq n\mu_{\vec{a}} + \eps' n$ with probability at least $1-e^{-\Omega_{m}(\eps^2 n)}$, so
by the union bound $k_{\vec{a}}\leq n\mu_{\vec{a}} + \eps' n$ for all $\vec{a}\in R$ with
probability at least $1-m e^{-\Omega_{m}(\eps^2 n)}$. We note that in this case, we have for all $\vec{a}\in R$ that
\[
k_{\vec{a}}
\geq n - \sum\limits_{\substack{\vec{b}\in R \\ \vec{b}\neq \vec{a}}} k_{\vec{b}}
\geq n - \sum\limits_{\substack{\vec{b}\in R \\ \vec{b}\neq \vec{a}}} (n\mu_{\vec{b}} + \eps' n),
\]
which is at least $\mu_{\vec{a}} n- m\eps' n$. We get that
$\card{k_{\vec{a}} - \mu_{\vec{a}} n} \leq m\eps' n$ for all $\vec{a}\in R$
with probability $1-m e^{-\Omega_{m}(\eps^2 n)}$. The same argument applies
for $r_{\vec{a}}$ to show that $\card{r_{\vec{a}} - \mu_{\vec{a}} n} \leq m\eps' n$ for all $\vec{a}\in R$
with probability $1-m e^{-\Omega_{m}(\eps^2 n)}$.
Therefore, with probability $1-m e^{-\Omega_{\alpha,m}(\eps^2 n)}$ we have
$\card{k_{\vec{a}} - r_{\vec{a}}} = 2m\eps' n$, and we show that in this
case $({\bf x}(1),\ldots,{\bf x}(r))$ and $({\bf y}(1),\ldots,{\bf y}(r))$ differ in at most $\eps n$ coordinates.
Indeed, note that
\[
\card{I_{\vec{a}}\triangle J_{\vec{a}}}
\leq
\sum\limits_{\vec{b}\preceq \vec{a}}2\card{k_{\vec{b}} - r_{\vec{b}}}
\leq
4m^2\eps'n,
\]
and the set $B = \bigcup_{\vec{a}\in R} I_{\vec{a}}\triangle J_{\vec{a}}$ contains all $i$'s in which
$({\bf x}(1)_i,\ldots,{\bf x}(r)_i)\neq ({\bf y}(1)_i,\ldots,{\bf y}(r)_i)$. Thus,
$\card{B}\leq 4m^3\eps'n\leq \eps n$, and the fourth item is established.

For the third item, again we see that the number of coordinates (as a random variable) is at most
$2m\sum\limits_{\vec{b}\preceq \vec{a}}\card{k_{\vec{b}} - r_{\vec{b}}}$, so its expectation is at most
\[
2m \sum\limits_{\vec{b}\preceq \vec{a}}\Expect{}{\card{k_{\vec{b}} - r_{\vec{b}}}}
\leq
2m \sum\limits_{\vec{b}\preceq \vec{a}}\sqrt{\Expect{}{\card{k_{\vec{b}} - r_{\vec{b}}}^2}}
\leq 2m \sum\limits_{\vec{b}\preceq \vec{a}} O_{m}(\sqrt{n})
=O_{m}(\sqrt{n}).
\]
The third item thus follows from symmetry.
\end{proof}

\section{Beyond connected distributions?}\label{sec:beyond?}
In this section, we remark that one can prove the invariance principle from the multi-slice to
its product version continues to holds for a class of distributions $\mu$ in which the ``contribution
of high-degree functions'' is always small. For such distributions, one only has to care about
the contribution of the low-degree parts of the functions, and for that we use Lemma~\ref{lem:inv_low_deg}.
In this language, the main content of the previous section is showing that in admissible,
connected distributions, the contribution of high-degree functions is very small, i.e.~Lemma~\ref{lem:upperbound_ev}.

\begin{definition}\label{def:ann}
  Let $r,m\in\mathbb{N}$ and let $([m]^n,\nu_{1,n})$,\ldots,$([m]^n,\nu_{r,n})$ be measure spaces for all $n\in\mathbb{N}$.
  Let $\mu = \set{\mu_n}_{n\in\mathbb{N}}$ be a sequence of distribution over $(\prod\limits_{i=1}^{r} [m])^n$
  whose marginals match $\nu_{i,n}$.  We say $\mu$ annihilates high-degree functions,
  if for all $\eps>0$, $M\in\mathbb{N}$, there are $q, d\in\mathbb{N}$ and $N\in\mathbb{N}$ such that the following holds for all $n\geq N$.

  Whenever $f_1,\ldots,f_r\colon [m]^n \to \mathbb{R}$ are functions such that:
  \begin{enumerate}
    \item for all $i=1,\ldots,r$ it holds that $\Expect{{\bf x}\sim\nu_{i,n}}{ f_i({\bf x})^{q}}\leq M$, and
    \item there is $i=1,\ldots,r$ such that $f_i\in V_{>d}([m]^n,\nu_{i,n})$,
  \end{enumerate}
  it holds that $\card{\Expect{({\bf x}(1),\ldots,{\bf x}(r))\sim\mu_n}{ \prod\limits_{i=1}^{r} f_i({\bf x}(i))}}\leq \eps$.
\end{definition}

Thus one gets the following result.
\begin{thm}\label{thm:inv_annh}
  Let $\alpha>0$, $r,m_1,\ldots,m_{r}\in\mathbb{N}$.
  For each $n\in\mathbb{N}$, let $\mathcal{U}^n_{\vec{k}(1)},\ldots,\mathcal{U}^n_{\vec{k}(r)}$ be $\alpha$-balanced multi-slices over alphabets
  $[m_1],\ldots,[m_r]$ of vectors of length $n$, respectively, let $\mu = \set{\mu_n}$ be a sequence of measures on
  $\prod\limits_{i=1}^{r}\mathcal{U}^n_{\vec{k}(i)}$ whose
  marginal on each $i$ is uniform on the respective multi-slice, and let $\tilde{\mu} = \set{\tilde{\mu_n}}_{n\in\mathbb{N}}$ be the product version of $\mu$.

  Suppose, that $\mu$ and $\tilde{\mu}$ both annihilate high-degree functions. Then for all $\eps>0$ there are $\zeta>0$, $N\in\mathbb{N}$ such that the following hold
  for $n\geq N$. If $\mathcal{C}_{1},\ldots,\mathcal{C}_{r}$ are couplings, where $\mathcal{C}_{i}$ is a
  $(\alpha,\zeta)$-coupling between $\mathcal{U}^n_{\vec{k}(i)}$ and $\nu_{\vec{k}(i)}^{\otimes n}$,
  and there is a $(\alpha,\zeta)$-coupling between $\mu_n$ and $\tilde{\mu_n}$, then
  for all $f_i\colon \mathcal{U}_{\vec{k}(i)}\to[-1,1]$ we have
  \[
  \card{
  \Expect{({\bf x}(1),\ldots,{\bf x}(r))\sim \mu_n}{\prod\limits_{i=1}^{r}f_i({\bf x}(i))} -
  \Expect{({\bf y}(1),\ldots, {\bf y}(r))\sim \tilde{\mu_n}}{\prod\limits_{i=1}^{r}\mathrm{T}_{\mathcal{C}_i}f_i({\bf y}(i))}
  }
  \leq \eps.
  \]
\end{thm}
We omit the proof since it follows easily from the argument in the previous section by replacing the use of Lemmas~\ref{lem:upperbound_ev}
and Lemma~\ref{lem:high_deg_dies_prod_connected} with the property that $\mu$ and $\tilde{\mu}$ annihilate high degree functions. We believe it
would be interesting to identify new classes of distributions that annihilate high degrees. We also remark that one immediately gets an
analogue of Theorem~\ref{thm:inv_label_assignments} for distributions $\mu$ such that both $\mu$ and $\tilde{\mu}$ annihilate high-degree functions.

\section{Applications to hardness of approximation}\label{sec:app_hardness}
In this section, we use our invariance principle in order to convert dictatorship tests
into NP-hardness results assuming the Rich $2$-to-$1$ Games Conjecture. We begin with some
definitions to formally set up terminology.
\begin{definition}
  Let $\Sigma$ be a finite alphabet, $\mathcal{P}\subseteq \set{P\colon\Sigma^r\to\power{}}$ be a collection
  of predicates.
  An instance of CSP-$\mathcal{P}$ is $(Z, E)$ whete $Z = \set{z_1,\ldots,z_n}$ is a set of variables,
  and $E$ is the set of constraints of the form $P(z_{i_1},\ldots,z_{i_r}) = 1$ where $P\in\mathcal{P}$.
\end{definition}
Given an instance $\Psi = (Z,E)$ of CSP-$\mathcal{P}$, we define the value of $\Psi$ as follows. An assignment is
a mapping $A\colon Z\to\Sigma$, and the value of the assignment ${\sf val}_{\Psi}(A)$ is the fraction of constraints that are
satisfied by it, i.e.~$\Expect{e\in E}{1_{A\text{ satisfies }e}}$. The value of the instance,
${\sf val}(\Psi)$, is defined to be $\max\limits_{A\colon Z\to\Sigma} {\sf val}_{\Psi}(A)$.

\begin{definition}
  Let $0<s < c\leq 1$, $r,m\in\mathbb{N}$ and let $\mathcal{P}\subseteq\set{P\colon[m]^r\to\power{}}$ be a collection of $r$-ary predicates
  over alphabet $[m]$. We denote by {\sf Gap-CSP-$\mathcal{P}$}$[c,s]$ the promise problem where one has to
  distinguish between the following two cases, given an instance $\Psi$ of {\sf CSP-$\mathcal{P}$}:
  \begin{enumerate}
    \item {\bf YES case}: ${\sf val}(\Psi) \geq c$;
    \item {\bf NO case}: ${\sf val}(\Psi)\leq s$.
  \end{enumerate}
\end{definition}

Next, we wish to define dictatorship tests, and for that we first define
regular functions.
\begin{definition}
  Let $\Sigma = [m]$, let $\mathcal{D} = \nu_1\times\ldots\nu_n$ be a distribution over $\Sigma^n$, and let
  $f\colon\Sigma^n\to\mathbb{R}^m$.
  The influence of variable $i\in [n]$ is defined as
  \[
  I_i[f;\mathcal{D}] =
  \Expect{{\bf x}\sim\mathcal{D}}{\norm{f({\bf x}) - \mathrm{E}_i f({\bf x})}_2^2}.
  \]
  Here, $\mathrm{E}_i$ is the averaging operator over coordinate $i$ defined as
  $\mathrm{E}_i f({\bf x}) = \Expect{{\bf y}_i\sim\nu_i}{f({\bf x}_{-i}, {\bf y}_i)}$.
\end{definition}

Note that denoting $f\colon\Sigma^n\to\mathbb{R}^m$ as $f = (f_1,\ldots,f_m)$ for real-valued $f_{\ell}$'s,
we have $I_i[f] = \sum\limits_{j\in[m]} I_i[f_j]$.

\begin{definition}
  Let $\Sigma = [m]$, let $\mathcal{D} = \nu_1\times\ldots\nu_n$ be a distribution over $\Sigma^n$,
  $d\in\mathbb{N}$, and let $f\colon\Sigma^n\to\mathbb{R}^m$.
  The degree $d$ influence of variable $i\in [n]$ is
  $I_i^{\leq d}[f;\mathcal{D}] = I_i[f^{\leq d};\mathcal{D}]$.
  Here, $f^{\leq d} = ({f_1}^{\leq d},\ldots,{f_m}^{\leq d})$.
\end{definition}

\begin{definition}
  Let $\Sigma = [m]$ and $\mathcal{D}$ be a distribution over $\Sigma^r$ with marginals $\mathcal{D}_1,\ldots,\mathcal{D}_r$,
  and let $f\colon\Sigma^n\to\Sigma$. As in Section~\ref{sec:label_assignments}, view $f$ as a function from $\Sigma^n$ to $\Delta_m$.
  We say $f$ is $(d,\tau)$-regular with respect to $\mathcal{D}$, if for all $i\in[r]$ and $j\in [n]$
  we have that $I_j^{\leq d}[f; \mathcal{D}_i^{\otimes n}]\leq \tau$.

  For a collection of distributions $\set{\mathcal{D}(t)}_{t\in\mathcal{T}}$ and a measure $p\colon\mathcal{T}\to[0,1]$,
  we say $f\colon\Sigma^n\to\Sigma$ is $(\gamma,d,\tau)$-regular with respect to $\mathcal{T}$ if choosing
  ${\bf t}\sim\mathcal{T}$, with probability at least $1-\gamma$, we have that $f$ is $(d,\tau)$-regular with respect
  to $\mathcal{D}({\bf t})$.
\end{definition}

\begin{definition}\label{def:dictator_test}
  Let $r\in\mathbb{N}$, $0<s<c\leq 1$, and $\mathcal{P}\subseteq \set{P\colon\Sigma^r\to\power{}}$ be a
  collection of $r$-ary predicates. A $(c,s)$-dictatorship test for $\mathcal{P}$ consists of a probability
  measure $w\colon\mathcal{P}\to[0,1]$, a collection of distributions $\set{\mathcal{D}(t,P)}_{t\in\mathcal{T}, P\in\mathcal{P}}$
  over $\Sigma^r$, and a measure $p\colon \mathcal{T}\to [0,1]$ such that
  \begin{enumerate}
    \item Completeness: for all $n$, if $f\colon\Sigma^n\to\Sigma$ is a dictatorship, i.e.~$f(x) = x_i$ for some $i$, then
    \[
    \Expect{\substack{{\bf t}\sim p_{\mathcal{T}}\\{\bf P}\sim w}}{\Expect{({\bf x}(1),\ldots,{\bf x}(r))\sim \mathcal{D}({\bf t},{\bf P})^{\otimes n}}{{\bf P}(f({\bf x}(1)),\ldots,f({\bf x}(r)))}}\geq c.
    \]
    \item Soundness: for all $\gamma,\eps>0$, there are $\tau>0$ and $d,N\in\mathbb{N}$ such that the following holds. If $n\geq N$, and
    $f\colon\Sigma^n\to\Sigma$ is a $(\gamma,d,\tau)$-regular function with respect to $\set{\mathcal{D}(t,P)}_{t\in\mathcal{T}, P\in \mathcal{P}}$, then
    \[
     \Expect{\substack{{\bf t}\sim p_{\mathcal{T}}\\{\bf P}\sim w}}{\Expect{({\bf x}(1),\ldots,{\bf x}(r))\sim \mathcal{D}({\bf t},{\bf P})^{\otimes n}}{{\bf P}(f({\bf x}(1)),\ldots,f({\bf x}(r)))}}\leq s+\eps.
    \]
  \end{enumerate}
\end{definition}
Raghavendra~\cite{Rag} showed that a given $(c,s)$-dictatorship test for a predicate can be converted into
a $(c-\eps,s+\eps)$ NP-hardness result for the predicate, assuming the Unique-Games Conjecture. When
$c<1$, this result is easily seen to be equivalent to a $(c,s + \eps)$ NP-hardness result, however the
situation is very different for the case of perfect completeness, i.e.~$c=1$, which we focus on henceforth.
In this case, the loss of perfect completeness in Raghavendra's result stems from two different places, each one of
which seems fundamental. First, in order to carry out his analysis, Raghavendra has to work
with connected distributions $\mu$; this is easy to achieve if one allows a small change in the completeness parameter
$c$, by mixing each one of the distributions $\mu_t$ with a small multiple of the uniform
distribution (i.e.~working with the distribution $\mu_t'(x) = (1-\eps) \mu_t(x) + \frac{\eps}{\card{\Sigma}^r}$).
Second, Raghanvedra starts off the reduction from the Unique-Games problem, which inherently has to have
imperfect completeness.

Roughly speaking, our result shows that if one is willing to assume the stronger Rich $2$-to-$1$ Games Conjecture,
then only the first of these issues persists. More precisely, we show:
\begin{thm}\label{thm:dictator_to_hard}
  Let $r\in\mathbb{N}$, $m\in\mathbb{N}$, and
  suppose $\mathcal{P}\subseteq \set{P\colon [m]^r\to\set{0,1}}$ is a collection of predicates.
  Suppose there is a $(1,s)$ dictatorship test, $(w,\set{\mathcal{D}(t)}_{t\in\mathcal{T}},p)$, for $\mathcal{P}$.
  If each $\mathcal{D}(t)$ is connected, then assuming Conjecture~\ref{conj:rich}, for all $\eps>0$, the problem
  {\sf Gap-$\mathcal{P}$}$[1,s+\eps]$ is NP-hard.
\end{thm}
The rest of this section is devoted to proving Theorem~\ref{thm:dictator_to_hard}.
Our reduction is very similar to Raghavendra's reduction; however, we do not know how to analyze the most natural version of it
using product spaces. As is natural in many PCP reductions, the constraint maps of our starting initial Rich-$2$-to-$1$ Games instance define
a projection operator from the ``large long-code'' to the ``small long-code''. The issue is that unlike in Unique-Games, these projection operators
do not behave well with respect to the measure on these spaces, and this introduces several technical difficulties.

Our idea therefore is to use multi-slices instead of product spaces, for which we show that the above issues no longer exist.
As is often the case,
the completeness of this reduction is trivial, and one only has to worry about the soundness of the test. This is where our invariance
principle enters the picture: intuitively, it asserts that the performance of our dictatorship test, when applied on multi-slices, is very
close to its performance of the original dictatorship test, which we know how to analyze. There are still technical issues that arise because
of multi-slices and the fact we are working with $2$-to-$1$ constraints (as opposed to $1$-to-$1$ constraints), mainly in the decoding phase.
For that, we define an appropriate notion of ``noisy influences'' that has similar properties to the standard noisy influences (e.g. the total
noisy influence is small), and also interacts well with $2$-to-$1$ projections and couplings.

\subsection{An appropriate notion of noisy influences}\label{sec:noisy_inf_multi}
\begin{definition}\label{def:noise_operator_S}
  Suppose $0<\beta\leq \alpha < 1$, and $n\in\mathbb{N}$.
  Suppose $\mathcal{U}_{\vec{k}}\subseteq[m]^n$ is a $\alpha$-balanced multi-slice,
  and $\beta n$ is an integer. For each $x\in \mathcal{U}_{\vec{k}}$,
  define the distribution ${\bf y}\sim \mathrm{S}_{1-\beta} x$
  over $\mathcal{U}_{\vec{k}}$ as:
  \begin{enumerate}
    \item for each $j\in[m]$, choose ${\bf A}_j\subseteq\sett{i\in [n]}{x_i = j}$ of size $\beta n$ randomly;
    \item take ${\bf y}\in \mathcal{U}_{\vec{k}}$,
    where ${\bf y}_i = x_i$ if $i\not\in\bigcup_{j\in [m]} {\bf A}_j$, ${\bf y}_i = j+1$ if $i\in {\bf A}_j$, $j\neq m$
    and ${\bf y}_i = 1$ if $i\in {\bf A}_{m}$.
  \end{enumerate}
\end{definition}

We will assume henceforth that $\beta n/2$ is an integer, and the point is so that the operator $\mathrm{S}_{1-\beta}$
is well defined on $\mathcal{U}_{\vec{k}}$ and $\mathcal{U}_{\vec{k}/2}$.
We remark that by adjusting the definition of $\mathrm{S}_{1-\beta}$ slightly, the
statements below remain true without this assumption, however this only introduces more cumbersome notations.
\begin{definition}
  Suppose $0<\beta\leq \alpha < 1$, and $n\in\mathbb{N}$.
  Suppose $\mathcal{U}_{\vec{k}}\subseteq[m]^n$ is a $\alpha$-balanced multi-slice,
  and $\beta n$ is an integer. The noise operator $\mathrm{S}_{1-\beta}\colon L^2(\mathcal{U}_{\vec{k}})\to L^2(\mathcal{U}_{\vec{k}})$
  is defined as
  \[
  (\mathrm{S}_{1-\beta} f)(x) = \Expect{{\bf y}\sim \mathrm{S}_{1-\beta} x}{f({\bf y})}.
  \]
\end{definition}

We will also define influences of functions on the multi-slice.
\begin{definition}
  Let $\mathcal{U}_{\vec{k}}\subseteq[m]^n$, $f\colon \mathcal{U}_{\vec{k}}\to\mathbb{R}$, and $i\in [n]$.
  The influence of coordinate $i$ is $I_i[f] = \Expect{{\bf j}\in [n]}{\card{f(^{\pi_{i,{\bf j}}} x) - f(x)}^2}$,
  where $\pi_{i,j}$ is the transposition permutation between $i$ and $j$.
  The total influence of $f$ is $I[f] = \sum\limits_{i=1}^{n} I_i[f]$.
\end{definition}

\begin{definition}
  Suppose $0<\beta\leq \alpha < 1$, and $n\in\mathbb{N}$.
  Suppose $\mathcal{U}_{\vec{k}}\subseteq[m]^n$ is a $\alpha$-balanced multi-slice,
  and $\beta n$ is an integer. For $f\colon \mathcal{U}_{\vec{k}}\to\mathbb{R}$,
  the $\beta$-noisy influence of a variable $i\in [n]$ is defined as
  $I^{(\beta)}_i[f] = I_i[\mathrm{S}_{1-\beta} f]$.  The total noisy influence of $f$ is $I^{(\beta)}[f] = \sum\limits_{i=1}^{n} I^{(\beta)}_i[f]$.
\end{definition}
We next state several properties of noisy influences in multi-slices, mostly analogous to properties
of noisy influences in product spaces, which will be crucial for us in the proof of Theorem~\ref{thm:dictator_to_hard}. Namely:
\begin{enumerate}
  \item Lemma~\ref{lem:sum_of_noisy_small}:
  the total $\beta$-noisy influence of a bounded function is $O_{\beta}(1)$ (so the number of coordinates with $\beta$-noisy influence at least $\tau$ is $O_{\beta}(1/\tau)$).
  \item Lemma~\ref{lem:noisy_inf_coupling}:  if $f\colon\mathcal{U}_{\vec{k}}\to[-1,1]$ is a function in which all noisy influences are small,
  and $\mathcal{C}$ is a good enough coupling between $(\mathcal{U}_{\vec{k}},{\sf Uniform})$ and $([m]^n,\nu_{\vec{k}}^{\otimes n})$, then all of the
  low-degree influences of $\mathrm{T}_{\mathcal{C}} f$ are small.
  \item Lemma~\ref{lem:projection_of_high_inf}: in Definition~\ref{def:projections} we consider the notion of projections
  of a function over the multi-slice, which plays a crucial role in our analysis. Given a vector $\vec{k} = (k_1,\ldots,k_m)$ of even integers summing up
  to $n$ and a $2$-to-$1$ map $\pi\colon [n]\to[n/2]$, we define the projection of $f\colon\mathcal{U}_{\vec{k}} \to\mathbb{R}$ along $\pi$ as
  $f|_{\pi}\colon \mathcal{U}_{\vec{k}/2}\to\mathbb{R}$, given by $f|_{\pi}(x) = f(y)$ where $y$ is the point in which $y_i = x_{\pi(i)}$.
  In Lemma~\ref{lem:projection_of_high_inf} we prove that if $f$ has a coordinate $i$ with significant noisy influence, then for a random projection $\bm{\pi}$, the coordinate
  $\bm{\pi}(i)$ has significant $\beta$-noisy influence in $f|_{\bm{\pi}}$ with significant probability.
\end{enumerate}

\subsubsection{The total noisy influence is constant}
In this section we show that the sum of noisy influences of a function over the multi-slice is constant. We begin
with the following variant of the well-known connection between noisy influences and the Fourier transform.
\begin{claim}\label{claim:trivial_inf_bd}
  For all $f\colon \mathcal{U}_{\vec{k}}\to\mathbb{R}$ we
  have that $I[f]\leq \sum\limits_{d>0}{d\norm{f^{=d}}_2^2}$.
\end{claim}
\begin{proof}
  We recall the mapping $f\to\tilde{f}$ from Section~\ref{sec:rep_multi} that given a function over the multi-slice
  produces a function $\tilde{f}\colon S_n\to\mathbb{R}$. We note that as we have already observed, this mapping is
  symmetric with respect to the action of $S_n$ on function, so it preserves degrees, i.e.~sends
  $f^{=d}$ to $\tilde{f}^{=d}$. This mapping also clearly preserves norms, so $\norm{\tilde{f}^{=d}}_2 = \norm{f^{=d}}_2$.

  Our claim will now easily follow from the results of~\cite{FOW}. To be more precise, defining the influence
  of $g\colon S_n\to\mathbb{R}$ on a transposition  $\pi_{i,j}\in [n]$ as $I_{i,j}[g] = \Expect{\bm{\pi}}{(g(\pi^{i,j} \pi) - g(\pi))^2}$,
  we see that $I_i[f] = \frac{1}{n}\sum\limits_{j=1}^{n}I_{i,j}[\tilde{f}]$. Thus, taking $I[g] = \sum\limits_{i=1}^{n} I_i[g]$,
  and using~\cite[inequality (32)]{FOW} (adapting the definition of Laplacians and $d_{\lambda}$ from there), we get that
  \[
  I[f]
  = \frac{1}{n} I[\tilde{f}]
  = \frac{{n\choose 2}}{n}\inner{\tilde{f}}{\mathrm{L}\tilde{f}}
  \leq \frac{n-1}{2} \sum\limits_{\lambda\vdash n} d_{\lambda} \norm{\tilde{f}^{=\lambda}}_2^2.
  \]
  We now note that by~\cite[Corollary 21]{FOW}, one has that
  $d_{\lambda}\leq \frac{2}{n-1}(n-\lambda_1)$, so we get that
  \[
  I[f] \leq \sum\limits_{\lambda\vdash n} (n-\lambda_1) \norm{\tilde{f}^{=\lambda}}_2^2
  =\sum\limits_{d > 0}d\sum\limits_{\substack{\lambda\vdash n \\ \lambda_1 = n-d}} \norm{\tilde{f}^{=\lambda}}_2^2
  =\sum\limits_{d > 0}d \norm{\tilde{f}^{=d}}_2^2
  =\sum\limits_{d > 0}d \norm{f^{=d}}_2^2.\qedhere
  \]
\end{proof}
\begin{lemma}\label{lem:sum_of_noisy_small}
  Suppose $m\in\mathbb{N}$, $0<\beta\leq \alpha < 1$ and $n\in\mathbb{N}$ are such that $\beta n$ is an integer.
  If $\mathcal{U}_{\vec{k}}\subseteq[m]^n$ is a $\alpha$-balanced multi-slice, then for all $f\colon \mathcal{U}_{\vec{k}}\to \mathbb{R}$ we
  have $I^{(\beta)}[f] \leq O_{m,\beta}(\norm{f}_2^2)$.
\end{lemma}
\begin{proof}
  From Claim~\ref{claim:trivial_inf_bd}
  \begin{equation}\label{eq:9}
  I^{(\beta)}[f]
  =I[\mathrm{S}_{1-\beta} f]
  \leq \sum\limits_{d>0}{d\norm{(\mathrm{S}_{1-\beta} f)^{=d}}_2^2}.
  \end{equation}
  By symmetry of $\mathrm{S}_{1-\beta}$, we by Claim~\ref{claim:preserve_rt} that
  it preserves degrees, so $(\mathrm{S}_{1-\beta} f)^{=d} = \mathrm{S}_{1-\beta} (f^{=d})$.
  Note that the operator $\mathrm{S}_{1-\beta}$ is the same as the operator $\mathrm{T}_{\mu_{\beta}}$
  where $\mu_{\beta}$ is the distribution $({\bf x}, \mathrm{S}_{1-\beta}{\bf x})$, which is
  $\beta$-admissible and connected. Also, $\mu_{\beta}$ is negatively correlated as per Definition~\ref{def:neg_correlated} (see Remark~\ref{remark:neg_assoc}),
  so by Proposition~\ref{prop:coupling_construction} there is
  a $(\Omega_{m,\beta}(1), O_{m,\beta}(1/\sqrt{n}))$-coupling between $\mu_{\beta}$ and $\tilde{\mu_{\beta}}$.
  Thus from Lemma~\ref{lem:upperbound_ev} we get
  $\norm{\mathrm{S}_{1-\beta} (f^{=d})}\leq C(m,\beta)(1+\delta(m,\beta))^{-d}\norm{f^{=d}}_2\leq C(m,\beta)(1+\delta(m,\beta))^{-d}\norm{f}_2$
  for some $C(m,\beta),\delta(m,\beta)>0$. Plugging this into~\eqref{eq:9} we get
  \[
  I^{(\beta)}[f]\leq C \norm{f}_2^2 \sum\limits_{d>0} d (1+\delta(m,\beta))^{-d}
  =O_{m,\beta}(\norm{f}_2^2).\qedhere
  \]
\end{proof}

For technical reasons, we will also need the following claim. In spirit, it is close to Lemma~\ref{lem:sum_of_noisy_small} since
$\mathrm{T}_{\mathcal{C}} \mathrm{S}_{1-\beta}$ may be thought of as a noise operator. The bounds we get are however much worse (and are very likely
not optimal), but are still good enough for our purposes.
\begin{claim}\label{claim:sum_of_noisy_coupled_small}
  Suppose $0<\beta\leq \alpha < 1$, and $n\in\mathbb{N}$ such that $\beta n$ is an integer.
  Suppose $\mathcal{U}_{\vec{k}}\subseteq[m]^n$ is a $\alpha$-balanced multi-slice,
  and $\mathcal{C}$ is a $(\alpha,\zeta)$-coupling between $\mathcal{U}_{\vec{k}}$ and $\nu_{\vec{k}}$.
  Then for all $f\colon \mathcal{U}_{\vec{k}}\to[-1,1]$ we
  have that $I[\mathrm{T}_{\mathcal{C}} \mathrm{S}_{1-\beta} f] \leq O_{m,\beta,\alpha}(\log^2 n) + O_{m,\beta,\alpha}(\zeta n\log^2 n)$.
\end{claim}
\begin{proof}
  Take $D = M(m,\beta)\log n$ for sufficiently large $M$ depending on $m,\beta$.
  We will use several basic notions from Section~\ref{sec:influences_product}.
  By Fact~\ref{fact:inf_laplacian}
  \[
  I[\mathrm{T}_{\mathcal{C}} \mathrm{S}_{1-\beta} f] =
  \inner{\mathrm{T}_{\mathcal{C}} \mathrm{S}_{1-\beta}f}{\mathrm{L}\mathrm{T}_{\mathcal{C}} \mathrm{S}_{1-\beta}f}
  =\inner{\mathrm{S}_{1-\beta}^{*}\mathrm{T}_{\mathcal{C}}^{*}\mathrm{L}\mathrm{T}_{\mathcal{C}} \mathrm{S}_{1-\beta}f}{f}.
  \]
  Each one of the operators $S_{1-\beta}, S_{1-\beta}^{*}, \mathrm{T}_{\mathcal{C}},\mathrm{T}_{\mathcal{C}}^{*}$ commutes
  with the action of $S_n$ on functions, and it is also easy to check that $\mathrm{L}$ commutes with the action of $S_n$. Thus,
  $R = \mathrm{S}_{1-\beta}^{*}\mathrm{T}_{\mathcal{C}}^{*}\mathrm{L}\mathrm{T}_{\mathcal{C}} \mathrm{S}_{1-\beta}$ commutes with the action of $S_n$,
  and therefore by Claim~\ref{claim:preserve_rt} it preserves the spaces
  $V_{=d}(\mathcal{U}_{\vec{k}})$, so we may continue the last chain of equalities as
  \begin{align*}
  I[\mathrm{T}_{\mathcal{C}} \mathrm{S}_{1-\beta} f]
  &=\sum\limits_{d=0}^{n}
  \inner{(R f)^{=d}}{f^{=d}}
  = \sum\limits_{d=0}^{n}
  \inner{R(f^{=d})}{f^{=d}}.
  \end{align*}
  We bound terms corresponding to $d\geq D$ and $d< D$ separately.

  \paragraph{Contribution from $d>D$.}
  By Cauchy-Schwarz and Jensen
  \[
  \sum\limits_{d=D}^n
  \inner{R (f^{=d})}{f^{=d}}
  \leq
  \sum\limits_{d=D}^n
  \norm{R(f^{=d})}_2\norm{f^{=d}}_2
  \leq
  \sum\limits_{d=D}^n
  \norm{\mathrm{L}\mathrm{T}_{\mathcal{C}} \mathrm{S}_{1-\beta}(f^{=d})}_2\norm{f^{=d}}_2,
  \]
  as $\norm{\mathrm{L} g}_2\leq n\norm{g}_2$ (which follows for example from Fact~\ref{fact:influence_formula}) we get that the
  last sum is upper bounded by
  $n\sum\limits_{d=D}^n \norm{\mathrm{T}_{\mathcal{C}}\mathrm{S}_{1-\beta}(f^{=d})}_2\norm{f^{=d}}_2
  \leq n\sum\limits_{d=D}^n \norm{\mathrm{S}_{1-\beta}(f^{=d})}_2\norm{f^{=d}}_2$.
  By Lemma~\ref{lem:upperbound_ev}, we have
  \[
  \norm{\mathrm{S}_{1-\beta}(f^{=d})}_2\leq O_{m,\beta}(1) (1+\Omega_{m,\beta}(1))^{d}\norm{f^{=d}}_2,
  \]
  so we get that the above sum is at most
  \[
  n\cdot O_{m,\beta}(1) (1+\Omega_{m,\beta}(1))^{D}
  \sum\limits_{d=D}^n \norm{f^{=d}}_2^2
  = n O_{m,\beta}(1) (1+\Omega_{m,\beta}(1))^{D}\norm{f}_2^2
  = O_{m,\beta}(1)
  \]
  by the choice of $D$.

  \paragraph{Contribution from $d\leq D$.}
  By definition of $R$
  \begin{equation}\label{eq:20}
  \sum\limits_{d\leq D}
  \inner{R(f^{=d})}{f^{=d}}
  =\sum\limits_{d\leq D}
  \inner{\mathrm{L}\mathrm{T}_{\mathcal{C}} \mathrm{S}_{1-\beta}(f^{=d})}{\mathrm{T}_{\mathcal{C}} \mathrm{S}_{1-\beta}(f^{=d})}
  =\sum\limits_{d\leq D}
  I[\mathrm{T}_{\mathcal{C}} \mathrm{S}_{1-\beta}(f^{=d})].
  \end{equation}
  We now bound $I[\mathrm{T}_{\mathcal{C}} \mathrm{S}_{1-\beta}(f^{=d})]$ for each $d$ separately.
  Let $g_d = \mathrm{S}_{1-\beta}(f^{=d})$, and note that $g_d\in V_d(\mathcal{U}_{\vec{k}})$ by Claim~\ref{claim:preserve_junta}
  as $\mathrm{S}_{1-\beta}$ commute with the action of $S_n$.
  Thus, by
  Claim~\ref{claim:bd_inf_coup_low_deg} we get that $I[\mathrm{T}_{\mathcal{C}}g_d]\leq \left(\frac{d^2}{\alpha^2}\zeta n + d^2\right)\norm{g_d}_2^2$,
  and plugging this into~\eqref{eq:20} gives
  \[
  ~\eqref{eq:20}
  \leq
  \left(\frac{D^2}{\alpha^2}\zeta n + D^2\right)
  \sum\limits_{d\leq D}\norm{g_d}_2^2
  =\left(\frac{D^2}{\alpha^2}\zeta n + D^2\right)
  \norm{\mathrm{S}_{1-\beta}(f^{\leq D})}_2^2
  \leq
  \left(\frac{4D^2}{\alpha^2}\zeta n + D^2\right).\qedhere
  \]
  \end{proof}

  \begin{claim}\label{claim:bd_inf_coup_low_deg}
    Let $\mathcal{U}_{\vec{k}}\subseteq[m]^n$ be a $\alpha$-balanced multi-slice, and let $\mathcal{C}$ be a $(\alpha,\zeta)$-coupling between
    $\mathcal{U}_{\vec{k}}$ and $([m]^n, \nu_{\vec{k}}^{\otimes n})$. Then for all $g\in V_{d}(\mathcal{U}_{\vec{k}})$,
    we have $I[\mathrm{T}_{\mathcal{C}} g]\leq \left(\frac{4 d^2}{\alpha^2}\zeta n + d^2\right)\norm{g}_2^2$.
  \end{claim}
  \begin{proof}
  Deferred to Section~\ref{sec:deferred_pfs}.
  \end{proof}

  %If ${\bf x}_J =  {\bf x''}_J$ then the difference is $0$; also,
%  by properties of the coupling, we have that $\Prob{{\bf x},{\bf x''}}{{\bf x}_J\neq {\bf x''}_J}\leq O(d\zeta)$, so we get that
%  \begin{align*}
%  \Expect{\substack{{\bf y}, {\bf y'}\\ {\bf x}, {\bf x''}}}
%  {\left(g({\bf x}) - g({\bf x''})\right)^2}
%  &\leq
%  O(d\zeta)
%  \cExpect{\substack{{\bf y}, {\bf y'}\\ {\bf x}, {\bf x''}}}
%  {{\bf x}_J\neq {\bf x''}_J}
%  {\left(g({\bf x}) - g({\bf x''})\right)^2}\\
%  &\leq
%  O(d\zeta)
%  \cExpect{\substack{{\bf y}, {\bf y'}\\ {\bf x}, {\bf x''}}}
%  {{\bf x}_J\neq {\bf x''}_J}
%  {g({\bf x})^2 + g({\bf x''})^2}\\
%  &\leq
%  O(d\zeta)
%  \cExpect{\substack{{\bf y}, {\bf y'}\\ {\bf x}, {\bf x''}}}
%  {{\bf x}_J\neq {\bf x''}_J}
%  {g({\bf x})^2}
%  \end{align*}
%  For each $z\in [m]^J$, consider
%  $p_{z} = \cProb{\substack{{\bf y}, {\bf y'}\\ {\bf x}, {\bf x''}}}{{\bf x}_J\neq {\bf x''}_J}{{\bf x}_J = z}$.
%
%  By Bayes' formula
%  \[
%  p_z = \frac{\cProb{\substack{{\bf y}, {\bf y'}\\ {\bf x}, {\bf x''}}}{{\bf x}_J = z}{{\bf x}_J\neq {\bf x''}_J}}
%  {\Prob{\substack{{\bf y}, {\bf y'}\\ {\bf x}, {\bf x''}}}{{\bf x}_J\neq {\bf x''}_J}}
%  \Prob{\substack{{\bf y}, {\bf y'}\\ {\bf x}, {\bf x''}}}{{\bf x}_J = z}
%  \]
%  By properties of the coupling,
%  $\cProb{\substack{{\bf y}, {\bf y'}\\ {\bf x}, {\bf x''}}}{{\bf x}_J = z}{{\bf x}_J\neq {\bf x''}_J}$
%  We note that for any two possible values $z,z\in[m]^J$ of ${\bf x}_J$, it holds that
%  \end{proof}

\subsubsection{Relating noisy influences of $f$ and  low-degree influences of $\mathrm{T}_{\mathcal{C}} f$}
The goal of this section is to show that if $\mathcal{C}$ is a coupling between our multi-slice and
the product domain, then if a function $f$ has all of its noisy influences small, then all low-degree influences of
$\mathrm{T}_{\mathcal{C}} f$ are small. To be more precise, we prove:
\begin{lemma}\label{lem:noisy_inf_coupling}
  For all $\tau>0$, $\alpha>0$, $m,d\in\mathbb{N}$ there
  are $\beta_0,\tau'>0$, such that for all $0<\beta\leq \beta_0$ there is $N\in\mathbb{N}$
  such that the following holds for all $n\geq N$, $\zeta>0$ such that $\zeta \leq \frac{1}{\log^3 n}$.

  Let $\mathcal{U}_{\vec{k}}\subseteq[m]^n$ be a $\alpha$-balanced mutlislice, and let $\mathcal{C}$ be a $(\alpha,\zeta)$-coupling between
  $\mathcal{U}_{\vec{k}}$ and $([m]^n,{\nu_{\vec{k}}}^{\otimes n})$.
  Then for all $f\colon \mathcal{U}_{\vec{k}}\to[-1,1]$, if
  $\max_{i} I_i^{(\beta)}[f]\leq \tau'$, then $\max_i I_i^{\leq d}[\mathrm{T}_{\mathcal{C}} f]\leq \tau$.
\end{lemma}

The proof of the above lemma relies on two auxiliary claims. The first asserts that if all noisy influences of $f$ are small, then
$I_i[\mathrm{T}_{\mathcal{C}} \mathrm{S}_{1-\beta}f]$ are all small.
\begin{claim}\label{claim:inf_move}
  Let $\mathcal{U}_{\vec{k}}\subseteq[m]^n$ be $\alpha$-balanced,
  $f\colon \mathcal{U}_{\vec{k}}\to\mathbb{R}$, and let $\mathcal{C}$ be a $(\alpha,\zeta)$-coupling between
  $\mathcal{U}_{\vec{k}}$ and $([m]^n,{\nu_{\vec{k}}}^{\otimes n})$.
  Then for all $i\in[n]$ we have that $I_i[\mathrm{T}_{\mathcal{C}} f]\leq \frac{2}{\alpha} I_i[f] + \frac{2}{\alpha^2 n} I[\mathrm{T}_{\mathcal{C}} f]$.
\end{claim}
\begin{proof}
  First, consider the following definition of influences over product spaces, which will be easier for us to establish the claim on.
  For $g\colon ([m]^n,\nu_{\vec{k}}^{\otimes n})\to\mathbb{R}$ define
  $Q_i(g) = \Expect{{\bf x}\sim\mathcal{D}, {\bf j}\in [n]}{(g(^{\pi_{i,{\bf j}}} {\bf x}) - g({\bf x}))^2}$.

  Sample $({\bf x},{\bf y})\sim\mathcal{C}$, and ${\bf j}\in [n]$. Note that by the symmetry of $\mathcal{C}$,
  the distribution of $^{\pi_{i,{\bf j}}} {\bf x}$ is the same as of ${\bf x}'$ where $({\bf x}',{\bf y}')\sim\mathcal{C}$
  conditioned on ${\bf y}' = ^{\pi_{i,{\bf j}}} {\bf y}$. Thus
  \[
  Q_i[\mathrm{T}_{\mathcal{C}} f]
  = \Expect{{\bf y}, {\bf j}}
  {\card{\mathrm{T}_{\mathcal{C}} f(^{\pi_{i,{\bf j}}} {\bf y}) - \mathrm{T}_{\mathcal{C}} f({\bf y})}^2}
  = \Expect{{\bf y}, {\bf j}}
  {\card{\cExpect{{\bf x}}{{\bf y},{\bf j}}{f(^{\pi_{i,{\bf j}}}{\bf x}) - f({\bf x})}}^2}
  \leq \Expect{{\bf x}}{\card{f(^{\pi_{i,{\bf j}}}{\bf x}) - f({\bf x})}^2},
  \]
  which is precisely $I_i[f]$.

  We now argue that $Q_i(g)\geq \frac{\alpha}{2} I_i[g] - \frac{1}{\alpha n}I[g]$ to finish the proof (taking
  $g = \mathrm{T}_{\mathcal{C}} f$). Sample ${\bf x}\sim\nu_{\vec{k}}^{\otimes n}$
  and ${\bf j}\in[n]$. Consider the point ${\bf z}$ which is the same as ${\bf x}$ on all coordinates except $i$, where
  it is equal to the ${\bf j}$ coordinate of ${\bf x}$. Then
  writing $(g(^{\pi_{i,{\bf j}}}{\bf x}) - g({\bf x})) = (g(^{\pi_{i,{\bf j}}}{\bf x}) - g({\bf z})) + (g({\bf z}) - g({\bf x}))$
  and using $(a+b)^2 \geq \half a^2 - b^2$, we get that
  \[
  Q_i(g)
  \geq
  \half \Expect{{\bf x},{\bf j}}{(g({\bf z}) - g({\bf x}))^2} - \Expect{{\bf x},{\bf j}}{(g(^{\pi_{i,{\bf j}}}{\bf x}) - g({\bf z}))^2}.
  \]
  For the first expectation, note that the points ${\bf z},{\bf x}$ are equal on all coordinates except the $i$th coordinate and distributed
  according to $\nu_{\vec{k}}$ on these coordinates. As $\nu_{\vec{k}}(a)\geq \alpha$ for all $a\in[m]$ it follows that
  $\Expect{{\bf x},{\bf j}}{(g({\bf z}) - g({\bf x}))^2}\geq \alpha I_i[g]$. For the second expectation, conditioning on ${\bf j} = j$
  we have by the same argument that
  $\Expect{{\bf x}}{(g(^{\pi_{i,j}}{\bf x}) - g({\bf z}))^2}\leq \frac{1}{\alpha} I_j[g]$, and we are done.
\end{proof}

The second claim directly relates $\max_{i} I_i[\mathrm{T}_{\mathcal{C}} \mathrm{S}_{1-\beta} f]$ to the low-degree influences
of $\mathrm{T}_{\mathcal{C}} f$, showing that if the former are all small, then the latter are also small.
\begin{claim}\label{claim:weird_noisy_to_noisy}
  For all $\tau>0$, $\alpha>0$, $d\in\mathbb{N}$ there are $\beta_0,\tau'>0$,
  such that for all $0<\beta\leq \beta_0$ there are $N\in\mathbb{N}$, $\zeta>0$ such that the following holds for $n\geq N$.
  Let $\mathcal{U}_{\vec{k}}\subseteq[m]^n$ be a $\alpha$-balanced mutlislice, and suppose $\mathcal{C}$ is a
  $(\alpha,\zeta)$-coupling between $(\mathcal{U}_{\vec{k}},{\sf Uniform})$, and
  $([m]^n,\nu_{\vec{k}}^{\otimes n})$. Then for all $f\colon \mathcal{U}_{\vec{k}}\to[-1,1]$,
  if $\max_{i} I_i[\mathrm{T}_{\mathcal{C}} \mathrm{S}_{1-\beta} f]\leq \tau'$, then
  $\max_i I_i^{\leq d}[\mathrm{T}_{\mathcal{C}} f]\leq \tau$.
\end{claim}
\begin{proof}
  Fix $\tau,\alpha>0$ and $d\in\mathbb{N}$. We choose $\beta_0 = \frac{\alpha}{8d}$, and $\tau'>0$ small enough.
  Fix $0<\beta\leq \beta_0$; our proof uses a parameter $D\in\mathbb{N}$ which is large enough with respect to
  $1/\beta,1/\tau,1/\alpha,d$, and then $N$ large enough with respect to $D$.

  Consider the distribution $\mu$ defined by $({\bf x}, \mathrm{S}_{1-\beta} {\bf x})$ where ${\bf x}\in_R \mathcal{U}_{\vec{k}}$,
  and consider its product version $\tilde{\mu}$. By Proposition~\ref{prop:coupling_construction} we may find a $(\Omega_{\alpha,m}(1), \zeta')$-coupling
  $\mathcal{C}' = (({\bf x}(1),{\bf x}(2)), ({\bf y}(1), {\bf y}(2)))$ between $\mu$ and $\tilde{\mu}$, where $\zeta' = O_{\alpha,\beta,m}(1/\sqrt{n})$
  (see Remark~\ref{remark:neg_assoc}). We intend to show that
  the functions $\mathrm{T}_{\mathcal{C}}\mathrm{S}_{1-\beta} f$ and $\mathrm{T}_{\tilde{\mu}} \mathrm{T}_{\mathcal{C}} f$
  are very close to each other in the $2$-norm, which will tell us that their individual influences are very close.
  From this, the statement of the claim quickly follows.

  Consider $g_1 = \mathrm{T}_{\mathcal{C}} ((\mathrm{S}_{1-\beta} f)^{\leq D})$,
  $g_2 = (\mathrm{T}_{\tilde{\mu}} \mathrm{T}_{\mathcal{C}} f)^{\leq D}$.
%  By Lemma~\ref{lem:coupling_preserve_deg},
%  \begin{equation}\label{eq:24}
%    \norm{g_1 - (\mathrm{T}_{\mathcal{C}}(\mathrm{S}_{1-\beta} f)^{\leq D})}_2
%    \leq O_{\alpha,\beta,D}(\zeta'^{1/4}),
%  \end{equation}
%  \[
%    \norm{g_2 - \mathrm{T}_{\tilde{\mu}} \mathrm{T}_{\mathcal{C}}(f^{\leq D})}_2
%    \leq \norm{(\mathrm{T}_{\mathcal{C}} f)^{\leq D} - \mathrm{T}_{\mathcal{C}}(f^{\leq D})}
%    \leq O_{\alpha,\beta,D}(\zeta'^{1/4}).
%  \]
  Note that the operators $\mathrm{T}_{\tilde{\mu}}$, $\mathrm{S}_{(1-\beta)}$ and their adjoint operators preserve juntas, and therefore preserve
  $V_{=\ell}([m]^n,\nu_{\vec{k}})$ and $V_{=\ell}(\mathcal{U}_{\vec{k}})$ respectively for all $\ell$. Thus,
  $g_2 = \mathrm{T}_{\tilde{\mu}} ((\mathrm{T}_{\mathcal{C}} f)^{\leq D})$, and $(\mathrm{S}_{1-\beta} f)^{\leq D} = \mathrm{S}_{1-\beta}(f^{\leq D})$.
  We therefore get that
    \begin{align*}
  \norm{\mathrm{T}_{\mathcal{C}}\mathrm{S}_{1-\beta} f - g_1}_2
  =\norm{\mathrm{T}_{\mathcal{C}}\mathrm{S}_{1-\beta}(f^{>D})}_2
  \leq \norm{\mathrm{S}_{1-\beta} f^{>D} }_2
  \leq
  (1+\Omega_{\beta,m}(1))^{-D}\norm{f^{>D} }_2,
   \end{align*}
%  \begin{align*}
%  \norm{\mathrm{T}_{\mathcal{C}}\mathrm{S}_{1-\beta} f - g_1}_2
%  =\norm{\mathrm{T}_{\mathcal{C}}\mathrm{S}_{1-\beta}(f^{>D})}_2 + O_{\alpha,\beta,D}(\zeta'^{1/4})
%  &\leq \norm{\mathrm{S}_{1-\beta} f^{>D} }_2+O_{\alpha,\beta,D}(\zeta'^{1/4})\\
%  &\leq
%  (1+\Omega_{\beta,m}(1))^{-D}\norm{f^{>D} }_2 + O_{\alpha,\beta,D}(\zeta'^{1/4})\\
%  &\leq (1+\Omega_{\beta,m}(1))^{-D} + O_{\alpha,\beta,D}(\zeta'^{1/4}),
%   \end{align*}
  where the third transition is by Lemma~\ref{lem:upperbound_ev};
  this is at most $(1+\Omega_{\beta,m}(1))^{-D}$ as $\norm{f^{>D} }_2\leq \norm{f}_2\leq 1$. We also get that
  \[
  \norm{\mathrm{T}_{\tilde{\mu}} \mathrm{T}_{\mathcal{C}} f
  -g_2}_2
  = \norm{\mathrm{T}_{\tilde{\mu}} (\mathrm{T}_{\mathcal{C}} f)^{>D}}_2
  \leq (1+\Omega_{\beta,m}(1))^{-D}\norm{(\mathrm{T}_{\mathcal{C}} f)^{>D}}_2
  \leq (1+\Omega_{\beta,m}(1))^{-D}\norm{\mathrm{T}_{\mathcal{C}} f}_2,
  \]
  which is at most $(1+\Omega_{\beta,m}(1))^{-D}$ as $f$ is $[-1,1]$-valued; in the second transition we used Lemma~\ref{lem:high_deg_dies_prod_connected}.
  Thus,
  \[
  \norm{\mathrm{T}_{\tilde{\mu}} \mathrm{T}_{\mathcal{C}} f - \mathrm{T}_{\mathcal{C}}\mathrm{S}_{1-\beta}}_2
  \leq \norm{g_1 - g_2}_2 + 2(1+\Omega_{\beta,m}(1))^{-D},
  \]
  and we bound $\norm{g_1 - g_2}_2\leq O_{\alpha,D}(\sqrt{\zeta+\zeta'})$.

  \begin{claim}
    $\norm{g_1 - g_2}_2\leq O_{\alpha,D}(\sqrt{\zeta+\zeta'})$.
  \end{claim}
  \begin{proof}
  Let $\mathcal{C}'_{i,j}$ denote the marginal distribution of ${\bf x}(i)$ and ${\bf y}(j)$, and
  define $g_1' = \mathrm{T}_{\mathcal{C}'_{1,2}} (f^{\leq D})$,
  $g_2' = \mathrm{T}_{\tilde{\mu}}\mathrm{T}_{\mathcal{C}} (f^{\leq D})$. Then
  \begin{equation}\label{eq:23}
  \norm{g_1 - g_2}_2\leq \norm{g_1 - g_1'}_2 + \norm{g_1' - g_2'}_2 + \norm{g_2'-g_2}_2,
  \end{equation}
  and we upper bound each one of the norms on the right hand side separately.
  For the second norm in the right hand side of~\eqref{eq:23}, note that
  \[
  g_1'(y) = \cExpect{\left(\substack{{\bf x}(1),{\bf x}(2)\\ {\bf y}(1), {\bf y}(2)}\right)\sim\mathcal{C}'}{{\bf y}(1) = y}{f^{\leq D}({\bf x}(2))},
  ~~~~
  g_2'(y) = \cExpect{\left(\substack{{\bf x}(1),{\bf x}(2)\\ {\bf y}(1), {\bf y}(2)}\right)\sim\mathcal{C}'}{{\bf y}(1) = y}{\mathrm{T}_{\mathcal{C}}(f^{\leq D})({\bf y}(2))},
  \]
  so by Cauchy-Schwarz
  \[
  \norm{g_1' - g_2'}_2^2
  \leq
  \Expect{\left(\substack{{\bf x}(1),{\bf x}(2)\\ {\bf y}(1), {\bf y}(2)}\right)\sim\mathcal{C}'}{\card{f^{\leq D}({\bf x}(2))-\mathrm{T}_{\mathcal{C}}(f^{\leq D})({\bf y}(2))}^2}
  \leq O(\sqrt{D\zeta'}),
  \]
  where the last inequality is by Lemma~\ref{lem:inv_low_deg}. To bound $\norm{g_1 - g_1'}_2$, consider the following distribution:
  sample $({\bf x}(1),{\bf x}(2), {\bf y}(1),{\bf y}(2))\sim\mathcal{C}'$, and then
  ${\bf y}$ as $({\bf x},{\bf y})\sim \mathcal{C}$ conditioned on ${\bf x} = {\bf x}(1)$.
  Then
  \[
  g_1(y) = \cExpect{\left(\substack{{\bf x}(1),{\bf x}(2)\\ {\bf y}(1), {\bf y}(2)}\right)\sim\mathcal{C}', {\bf y}}
  {{\bf y} = y}{f^{\leq D}({\bf x}(2))}.
  \]
  The above sampling procedure naturally defines $(\Omega_{\alpha,m}(1),\zeta + \zeta')$-couplings $\mathcal{C}''$ of $({\bf x}(2), {\bf y}(1))$ and
  a coupling $\mathcal{C}'''$ of $({\bf x}(2), {\bf y})$, and with this notation $g_1'(y) = \mathrm{T}_{\mathcal{C}''} f^{\leq D}(y)$ and
  $g_1(y) = \mathrm{T}_{\mathcal{C}'''} f^{\leq D}(y)$, so by Lemma~\ref{lem:inv_low_deg}
  \[
  \Expect{\mathcal{C}'''}{(g_1({\bf y}) -  f^{\leq D}({\bf x}(2)))^2}\leq O(\sqrt{D(\zeta+\zeta')}),
  \qquad
  \Expect{\mathcal{C}'''}{(g_1'({\bf y}) -  f^{\leq D}({\bf x}(2)))^2}\leq O(\sqrt{D(\zeta+\zeta')}).
  \]
  %\[
%  \Expect{\mathcal{C}''}{(g'_1({\bf y}(1)) -  f^{\leq D}({\bf x}(2)))^2}\leq O(D(\zeta+\zeta')),
%  \qquad
%  \Expect{\mathcal{C}'''}{(g'_1({\bf y}) -  f^{\leq D}({\bf x}(2)))^2}\leq O(D(\zeta+\zeta')).
%  \]
  Thus, by the triangle inequality
  \[
  \Expect{{\bf y}\sim\nu_{\vec{k}}^{\otimes n}}{(g_1({\bf y}) -  g_1'({\bf y}))^2}\leq O(\sqrt{D(\zeta+\zeta')}),
  \]
  i.e.~$\norm{g'_1 - g_1} \leq O((D(\zeta+\zeta'))^{1/4})$.

  Finally, we bound the third norm in~\eqref{eq:23}:
  \begin{align*}
  \norm{g_2 - g_2'}
  =
  \norm{(\mathrm{T}_{\tilde{\mu}} \mathrm{T}_{\mathcal{C}} f)^{\leq D} - \mathrm{T}_{\tilde{\mu}} \mathrm{T}_{\mathcal{C}}(f^{\leq D})}
  &=
  \norm{\mathrm{T}_{\tilde{\mu}}( (\mathrm{T}_{\mathcal{C}} f)^{\leq D} - \mathrm{T}_{\mathcal{C}}(f^{\leq D}))}\\
  &\leq
  \norm{(\mathrm{T}_{\mathcal{C}} f)^{\leq D} - \mathrm{T}_{\mathcal{C}}(f^{\leq D}))}\\
  &\leq O_{\alpha,D}(\zeta^{1/4})\norm{f}_2\\
  &\leq O_{\alpha,D}(\zeta^{1/4}),
  \end{align*}
  where we used Lemma~\ref{lem:coupling_preserve_deg}.
  Thus by the triangle inequality
  $\norm{g_1 - g_2}_{2} \leq O_{\alpha,D}((\zeta+\zeta')^{1/4})$.
  \end{proof}

  Combining all, we get that
  $\norm{\mathrm{T}_{\mathcal{C}}\mathrm{S}_{1-\beta} f - \mathrm{T}_{\tilde{\mu}} \mathrm{T}_{\mathcal{C}} f}_2\leq
  O_{\alpha,D}((\zeta+\zeta')^{1/4}) + 2(1+\Omega_{\beta,m}(1))^{-D}$. Choosing $D$ large enough
  so that $2(1+\Omega_{\beta,m}(1))^{-D}\leq \tau'/2$, and then $N$ large enough so that $\zeta'$ is small enough,
  and then $\zeta>0$ small enough, we get that
  $\norm{\mathrm{T}_{\mathcal{C}}\mathrm{S}_{1-\beta} f - \mathrm{T}_{\tilde{\mu}} \mathrm{T}_{\mathcal{C}} f}_2\leq \tau'$.
  Thus,
  \[
  \max_{i} I_i[\mathrm{T}_{\tilde{\mu}} \mathrm{T}_{\mathcal{C}} f]\leq
  2\max_{i} I_i[\mathrm{T}_{\mathcal{C}} \mathrm{S}_{1-\beta} f] +
  2\norm{\mathrm{T}_{\mathcal{C}}\mathrm{S}_{1-\beta} f - \mathrm{T}_{\tilde{\mu}} \mathrm{T}_{\mathcal{C}} f}_2
  = O(\tau').
  \]
  Note that by definition of
  $\tilde{\mu}$, on each coordinate $i$, the probability that ${\bf y}(1)_i = {\bf y}(2)_i$ is at least $1-\frac{\beta}{\alpha} \geq 1-\frac{1}{8d}$,
  so from Claim~\ref{claim:noisy_to_low_deg} it follows that
  $\max_{i} I_i^{\leq d}[\mathrm{T}_{\mathcal{C}} f]\leq O(\tau')$. Choosing $\tau' = c\tau$ for sufficiently small absolute $c>0$ finishes the proof.
\end{proof}

We are now ready to prove Lemma~\ref{lem:noisy_inf_coupling}.
\begin{proof}[Proof of Lemma~\ref{lem:noisy_inf_coupling}]
  Choose $\beta_0,\tau'',N'$ from Claim~\ref{claim:weird_noisy_to_noisy} for $\tau,\alpha,d$.
  For $0 < \beta\leq \beta_0$ we have by Claim~\ref{claim:sum_of_noisy_coupled_small},
  \[
  I[\mathrm{T}_{\mathcal{C}} \mathrm{S}_{1-\beta} f] = O_{m,\beta}(\log^2 n) + O_{m,\beta,\alpha}(\zeta n\log^2 n),
  \]
  so by Claim~\ref{claim:inf_move} we get that
  \begin{align*}
  \max_{i} I_i[\mathrm{T}_{\mathcal{C}} \mathrm{S}_{1-\beta} f]
  \leq
  \frac{2}{\alpha} I_i[f]
  +
  \frac{2}{\alpha^2 n} I[\mathrm{T}_{\mathcal{C}} f]
  &\leq
  \frac{2}{\alpha}\tau' +
  O_{m,\beta}\left(\frac{\log^2 n}{n}\right)
  +O_{m,\beta,\alpha}(\zeta\log^2 n)\\
  &\leq \frac{2}{\alpha}\tau' + O_{m,\beta,\alpha}\left(\frac{1}{\log n}\right),
  \end{align*}
  where we used the upper bound on $\zeta$.
  Choosing $\tau' = \tau''\alpha/4$ and $N'' = N''(\alpha,\beta,m,\tau,d)$ large enough, it follows that for $N = \max(N',N'')$,
  if $n\geq N$ then $\max_{i} I_i[\mathrm{T}_{\mathcal{C}} \mathrm{S}_{1-\beta} f]\leq \tau''$,
  and from Claim~\ref{claim:weird_noisy_to_noisy} that we conclude that
  $\max_i I_i^{\leq d}[\mathrm{T}_{\mathcal{C}} f]\leq \tau$
  (we also use the fact that $\zeta\leq 1/\log n\leq 1/\log N''$, so $\zeta$ is small enough
  to apply Claim~\ref{claim:weird_noisy_to_noisy} provided $N''$ is large enough).
\end{proof}

\subsubsection{Relation with projections}\label{sec:projections}
Suppose $n$ is even. We denote by $S_{n:n/2}$ the set of all $2$-to-$1$ maps
$\pi\colon [n]\to[n/2]$. For any $x\in[m]^{n/2}$ and $\pi\in S_{n:n/2}$, we define
the point $y = \pi^{-1}(x)\in [m]^n$ as $y_i = x_{\pi(i)}$ for all $i\in [n]$.
\begin{definition}\label{def:projections}
  Suppose $\mathcal{U}_{\vec{k}}$ is a multi-slice and $k_i$ is even for all $i$ (so in particular $n$ is even).
  Given $\pi\in S_{n:n/2}$, define the projection of
  $f\colon\mathcal{U}_{\vec{k}}\to\mathbb{R}$ along $\pi$
  as $f|_{\pi}\colon \mathcal{U}_{\vec{k}/2}\to[m]^{n/2}\to\mathbb{R}$, $f|_{\pi}(x) = f(\pi^{-1}(x))$.
\end{definition}

The main goal of this section is to prove the following lemma, asserting that if a function over the multi-slice
has a coordinate $i$ with large noisy influence, then for a random $2$-to-$1$ map $\bm{\pi}$, the coordinate
$\bm{\pi}(i)$ will have large noisy influence in $f|_{\bm{\pi}}$ with constant probability. More precisely:
\begin{lemma}\label{lem:projection_of_high_inf}
  For all $\tau',\alpha>0$, $m\in\mathbb{N}$
  there exist $\beta_1>0$, $\tau''>0$ and $N\in \mathbb{N}$ such that the following holds
  for $n\geq N$ and $0<\beta\leq \beta_1$. Suppose $\vec{k}$ is a vector of $m$ even integers summing up to $n$.
  If $f\colon\mathcal{U}_{\vec{k}}\to [-1,1]$ and $i\in[n]$ are such that $I_i^{(\beta)}[f]\geq \tau'$,
  then
  \[
  \Prob{\bm{\pi}\in S_{n:n/2}}{ I_{\bm{\pi}(i)}^{(\beta)}[f_{\bm{\pi}}]\geq \tau''}\geq \tau''.
  \]
\end{lemma}
The rest of this section is devoted to the proof of Lemma~\ref{lem:projection_of_high_inf}.

 It will be convenient for us to consider different notions of influence for functions on the multi-slice.
  For $i\in[n]$ and $g\colon\mathcal{U}_{\vec{k}}\to\mathbb{R}$, define
  $A_i^{g}(x) = \Expect{{\bf j}\in [n]}{g({}^{\pi_{i,{\bf j}}} {\bf x})}$ and then
  $M_i(g) = \Expect{\substack{{\bf x}\in\mathcal{U}_{\vec{k}}}}{
    \left(A_i^g({\bf x}) -g({\bf x})\right)^2}$.

 \begin{claim}\label{claim:lb_aux_inf_1}
 For all $g\colon\mathcal{U}_{\vec{k}}\to\mathbb{R}$ and $i\in[n]$ we have $M_i(g)\geq \frac{1}{6} I_i[g] - \frac{1}{2n} I[g]$.
 \end{claim}
 \begin{proof}
  Using $(a+b+c)^2\leq 3(a^2+b^2+c^2)$ we get
  \begin{align}
  I_i[g]
  =\Expect{{\bf x},{\bf s}}{(g(^{\pi_{i,{\bf s}}} {\bf x}) - g({\bf x}))^2}
  &\leq
  3\Expect{{\bf x},{\bf s}}{(A_i^g({\bf x}) - g({\bf x}))^2 + (A_i^g(^{\pi_{i,{\bf s}}}{\bf x}) - g(^{\pi_{i,{\bf s}}}{\bf x}))^2 +
  (A_i^g(^{\pi_{i,{\bf s}}}{\bf x})-A_i^g({\bf x}))^2} \notag\\\label{eq:15}
  &=6M_i(g) + 3\Expect{{\bf x},{\bf s}}{(A_i^g(^{\pi_{i,{\bf s}}}{\bf x})-A_i^g({\bf x}))^2},
  \end{align}
  and we upper bound the last expectation by $I[g]/n$ as follows.
  \[
  \Expect{{\bf x},{\bf s}}{(A_i^g(^{\pi_{i,{\bf s}}}{\bf x})-A_i^g({\bf x}))^2}
  =\Expect{{\bf x},{\bf s}}
  {
  \left(
  \Expect{{\bf j}}
  {g({}^{\pi_{i,{\bf j}}} {}^{\pi_{i,{\bf s}}}{\bf x})
  -g({}^{\pi_{i,{\bf j}}}{\bf x})}\right)^2}
  \leq
  \Expect{{\bf x},{\bf s},{\bf j}}
  {\left(g({}^{\pi_{i,{\bf j}}} {}^{\pi_{i,{\bf s}}}{\bf x})
  -g({}^{\pi_{i,{\bf j}}}{\bf x})\right)^2}.
  \]
  Make the change of variables ${\bf y} = ^{\pi_{i,{\bf j}}}{\bf x}$,
  and note that then we have that $^{\pi_{i,{\bf j}}} {}^{\pi_{i,{\bf s}}}{\bf x} = ^{\pi_{{\bf j},{\bf s}}} {\bf y}$.
  Thus
  \[
  \Expect{{\bf x},{\bf s}}{(A_i^g(^{\pi_{i,{\bf s}}}{\bf x})-A_i^g({\bf x}))^2}\leq
  \Expect{{\bf y},{\bf s},{\bf j}}{(g(^{\pi_{{\bf j},{\bf s}}} {\bf y}) - g({\bf y}))^2}
  =\frac{1}{n}I[g].
  \]
  Plugging this into~\eqref{eq:15} and rearranging finishes the proof.
\end{proof}

  We next define another notion of influence that will get us closer to the left hand of Lemma~\ref{lem:projection_of_high_inf}.
  Define
  \[
  B_i^g(x) = \cExpect{\substack{{\bf i'}\in [n]\setminus\set{i}\\ {\bf j}\in_R [n]\\ {\bf j'}\in [n]\setminus \set{i,{\bf i'},{\bf j}}}}
    {x_{\bf i'} = x_{\bf i}, x_{\bf j'} = x_{\bf j}}
    {
     g({}^{\pi_{{\bf i'},{\bf j'}}}{}^{\pi_{i,{\bf j}}}x)},
    \qquad
    Q_i(g) = \Expect{\substack{{\bf x}\in\mathcal{U}_{\vec{k}}}}{\left(B_i^g({\bf x}) - g({\bf x})\right)^2}.
    \]
  (we stress that the distribution over ${\bf j}$ is uniform, and then the distribution over
  ${\bf j'}$ is uniform among all $j'$ such that $x_{j'} = x_{{\bf j}}$ and different from $i, {\bf i'}, {\bf j}$)
  \begin{claim}\label{claim:lb_aux_inf_2}
  For all $g\colon\mathcal{U}_{\vec{k}}\to\mathbb{R}$ and $i\in[n]$ we have
    $Q_i(g)\geq \frac{1}{2}M_i(g) - O_{\alpha}\left(\frac{I[g]}{n}\right)$.
  \end{claim}
  \begin{proof}
     Using $(a+b)^2\geq \half a^2 - b^2$ we get $Q_i(g) \geq \half M_i(g) - \Expect{{\bf x}}{(A_i^g({\bf x}) - B_i^g({\bf x}))^2}$
  and we upper bound the last expectation by $O_{\alpha}\left(\frac{I[g]}{n}\right)$. Indeed, note that
  \begin{align*}
  \Expect{{\bf x}}{(A_i^{g}({\bf x}) - B_i^g({\bf x}))^2}
  &=\Expect{{\bf x}}
  {\left(\cExpect{\substack{{\bf i'}\in [n]\setminus\set{i}\\ {\bf j}\in_R\in [n]\\ {\bf j'}\in [n]\setminus \set{i,{\bf i'},{\bf j}}}}
    {{\bf x}_{\bf i'} = {\bf x}_{\bf i}, {\bf x}_{\bf j'} = {\bf x}_{\bf j}}
    {g({}^{\pi_{{\bf i'},{\bf j'}}}{}^{\pi_{i,{\bf j}}} {\bf x}) - g({}^{\pi_{i,{\bf j}}} {\bf x})}\right)^2}\\
  &\leq
  \Expect{{\bf x}}
  {\cExpect{\substack{{\bf i'}\in [n]\setminus\set{i}\\ {\bf j}\in_R\in [n]\\ {\bf j'}\in [n]\setminus \set{i,{\bf i'},{\bf j}}}}
    { {\bf x}_{\bf i'} = {\bf x}_{\bf i}, {\bf x}_{\bf j} = {\bf x}_{\bf j'}}
    {\left(g({}^{\pi_{{\bf i'},{\bf j'}}}{}^{\pi_{i,{\bf j}}} {\bf x}) - g({}^{\pi_{i,{\bf j}}} {\bf x})\right)^2}}\\
  &\leq
  \frac{n}{\alpha n - 3}\left(1+O\left(\frac{1}{n}\right)\right)\Expect{{\bf x}}
  {\Expect{{\bf i'}, {\bf j}, {\bf j'}}
    {\left(g({}^{\pi_{{\bf i'},{\bf j'}}}{}^{\pi_{i,{\bf j}}} {\bf x}) - g({}^{\pi_{i,{\bf j}}} {\bf x})\right)^2}}\\
    &=O\left(\frac{1}{\alpha}\right)
    \Expect{{\bf x}}
  {\Expect{{\bf i'}, {\bf j'}}
    {\left(g({}^{\pi_{{\bf i'},{\bf j'}}}{}{\bf x}) - g({\bf x})\right)^2}}\\
  &=O_{\alpha}\left(\frac{I[g]}{n}\right).\qedhere
  \end{align*}
  \end{proof}

  Finally, we relate the left hand side of Lemma~\ref{lem:projection_of_high_inf} directly to
  the $Q_i$ notion of influence.
  \begin{claim}\label{claim:expected_projected_noisy_inf}
  For all $\alpha>0$, there is $\beta_0>0$ such that if $0<\beta\leq \beta_1$, then
    $f\colon\mathcal{U}_{\vec{k}}\to [-1,1]$ and $i\in[n]$ we have
  \[
  \Expect{{\bm \pi}\in S_{n:n/2}}{I_{\bm{\pi}(i)}^{(\beta)}[f_{\bm{\pi}}]}
  \geq  \frac{1}{4}Q_i(\mathrm{S}_{1-\beta} f) - O\left(\frac{\beta}{\alpha}\right).
  \]
  \end{claim}
   \begin{proof}
  \begin{align*}
  \Expect{{\bm \pi}\in S_{n:n/2}}{I_{\bm{\pi}(i)}^{(\beta)}[f_{\bm{\pi}}]}
  =
  \Expect{{\bf x}\in \mathcal{U}_{\vec{k}/2}}
  {\Expect{{\bf j}\in [n/2]}{
  \Expect{{\bm \pi}\in S_{n:n/2}}{
  \left(
  (\mathrm{S}_{1-\beta} f|_{\bm{\pi}})(^{\pi_{i,{\bf j}}} {\bf x})
  -
  (\mathrm{S}_{1-\beta} f|_{\bm{\pi}})({\bf x})
  \right)^2
  }}}.
  \end{align*}
  Next, we introduce ${\bf z} = \bm{\pi}^{-1}({\bf x})$, condition on it and then use Cauchy-Schwarz to see that
  the above expression is at least
  \begin{equation}\label{eq:13}
  \Expect{{\bf z}\in \mathcal{U}_{\vec{k}}}
  {
  \left(
  \cExpect{\substack{{\bf x}\in \mathcal{U}_{\vec{k}/2}\\ \bm{\pi}\in S_{n:n/2}\\ {\bf j}\in [n/2]}}
  {\bm{\pi}^{-1}({\bf x}) = {\bf z}}
  {{(\mathrm{S}_{1-\beta} f|_{\bm{\pi}})(^{\pi_{\bm{\pi}(i),{\bf j}}} {\bf x})}}
  -
   \cExpect{\substack{{\bf x}\in \mathcal{U}_{\vec{k}/2}\\ \bm{\pi}\in S_{n:n/2}\\ {\bf j}\in [n/2]}}
  {\bm{\pi}^{-1}({\bf x}) = {\bf z}}
  {{(\mathrm{S}_{1-\beta} f|_{\bm{\pi}})({\bf x})}}
  \right)^2
  }.
  \end{equation}
  We say $\pi\in S_{n:n/2}$ is consistent with $z\in\mathcal{U}_{\vec{k}}$ if
  $z\in \pi^{-1}(\mathcal{U}_{\vec{k}/2})$.  Fix ${\bf z} = z$; in the second expectation, we first sample
  $\bm{\pi}$ consistent with $z$, then take ${\bf x} = \bm{\pi}(z)$, then take ${\bf x'}\sim \mathrm{S}_{1-\beta} {\bf x}$,
  and average $f_{\bm{\pi}}({\bf x'}) = f(\bm{\pi}^{-1}({\bf x'}))$. Note that the distribution of
  $\bm{\pi}^{-1}({\bf x'})$ is the same as $\mathrm{S}_{1-\beta} z$, so the second expectation is
  nothing but $\mathrm{S}_{1-\beta} f({\bf z})$.

  We move on to the first expectation. Fix ${\bf z} = z$; we first sample $\bm{\pi}$ consistent
  with $z$, then set ${\bf x} = \bm{\pi}(z)$, take ${\bf j}\in[n/2]$, look at the point ${\bf x'} = ^{\pi_{\bm{\pi}(i),{\bf j}}} {\bf x}$,
  then sample ${\bf x''}\sim \mathrm{S}_{1-\beta}{\bf x'}$ to yield ${\bf z'} = \bm{\pi}^{-1}({\bf x''})$. The first expectation
  is the average of $f({\bf z'})$ sampled this way, and we analyze the distribution of ${\bf z'}$.

  Note that the points ${\bf x}$, ${\bf x'}$ differ in either $0$ or $2$ coordinates. Thus, there is a natural coupling
  between $\mathrm{S}_{1-\beta}{\bf x}$ and $\mathrm{S}_{1-\beta}{\bf x'}$, call it $({\bf u},{\bf u'})$, in which
  ${\bf u},{\bf u'}$ always differ in either $0$ or $2$ coordinates. To see that, consider for each $a\in [m]$ the set
  of coordinate of ${\bf x}$ equal to $a$, and the set of coordinates ${\bf x}'$ equal to $a$. Then either:
  \begin{enumerate}
    \item these sets are the same for all $a$ -- in which case we can make the sampling procedure $\mathrm{S}_{1-\beta}$ to choose the
    same subsets of coordinates to change in each color class.
    \item Else, these sets are the same for all but $2$ of the $a$'s, say $a_1$ and $a_2$.
    This, the sets of coordinates corresponding to $a_1$ (as well as the sets of coordinates corresponding to $a_2$) have symmetric difference of size
    $1$. In this case, for each $a\neq a_1,a_2$
    we may make the sampling procedure $\mathrm{S}_{1-\beta}$ choose the subsets of coordinates,
    and for $a\in\set{a_1,a_2}$ we may choose subsets of coordinates that have symmetric difference
    at most $1$.
  \end{enumerate}

  Denote the set of coordinates in which ${\bf x'}$ and ${\bf x''}$ differ by ${\bf U}$. Thus, we get a coupling of $\mathrm{S}_{1-\beta} z$ and $\mathrm{S}_{1-\beta} {z'}$, call it
  ${\bf w}$, ${\bf w'}$, in which the points differ in either $0$ or $4$ coordinates; denote the set of these coordinates
  as ${\bf W} = {\pi}^{-1}({\bf U})$.
  Denote by $\pi_{{\bf W}}$ a composition of two transpositions on ${\bf W}$, so that
  ${\bf w}' = ^{\pi_{\bf W}}{\bf w}$ (or the identity if ${\bf W} = \emptyset$).
  Plugging this into~\eqref{eq:13}, we get that
  \[
  \eqref{eq:13}
  =
  \Expect{{\bf z}}
  {
  \left(
  \Expect{{\bf w}\sim \mathrm{S}_{1-\beta}{\bf z}}
  {
  \Expect{{\bf W}}
  {
  f({}^{\pi_{{\bf W}}}{\bf w}) -
  f({\bf w})}}
  \right)^2
  }
  =
  \Expect{{\bf z}}
  {
  \left(
  \Expect{{\bf W}}
  {
  \Expect{{\bf w}\sim \mathrm{S}_{1-\beta}{\bf z}}
  {
  f({}^{\pi_{{\bf W}}}{\bf w}) -
  f({\bf w})}}
  \right)^2
  },
  \]
  where in the last transition we used the fact that the distribution of ${\bf w}$ and
  ${\bf W}$ are independent. When ${\bf W}$ is the empty set we get $0$, so
  \[
  \eqref{eq:13}
  =
  \Expect{{\bf z}}
  {
  \left(
  \Expect{{\bf w}\sim \mathrm{S}_{1-\beta}{\bf z}}
  {
  \Expect{{\bf W}}
  {
  f({}^{\pi_{{\bf W}}}{\bf w}) -
  f({\bf w})}}
  \right)^2
  }
  =
  \Expect{{\bf z}}
  {
  \left(
  \Expect{{\bf W}}
  {
  1_{{\card{{\bf W}} = 4}}
  \Expect{{\bf w}\sim \mathrm{S}_{1-\beta}{\bf z}}
  {
  {}^{\pi_{{\bf W}}}f({\bf w}) -
  g({\bf w})}}
  \right)^2
  }.
  \]
  Next, we note that the probability that ${\bf W}$ is non-empty is
  $(1-O(\beta/\alpha))\geq \half$, and
  we need to inspect the distribution of ${\bf W}$ more closely when it is non-empty.
  In that case, ${\bf U}$ contains the coordinate $\bm{\pi}(i)$ along with a random coordinate
  ${\bf j}\in [n/2]$. Thus, the distribution of $\bm{\pi}^{-1}({\bf U})$ is
  $\set{i,{\bf i'}, {\bf j_1},{\bf j_2}}$, where:
  ${\bf i'}$ is uniform on $i'\neq i$ such that $z_{i'} = z_{i}$, ${\bf j_1}\in [n]$ is uniform, and conditioned
  on ${\bf j_1} = j_1$ the distribution of ${\bf j_2}$ is uniform among $j_2\neq j$ such that $z_{j_2} = z_{j_1}$.
  Thus, we get that
  \begin{align*}
  \eqref{eq:13}
  &\geq
  \frac{1}{4}
  \Expect{{\bf z}}
  {
  \left(
  \cExpect{{\bf i'}, {\bf j_1}\in [n],{\bf j_2}}
  {{\bf z}_{{\bf i'}} = {\bf z}_i, {\bf z}_{{\bf j_2}} = {\bf z}_{{\bf j_1}}}
  {
  \mathrm{S}_{1-\beta}({}^{\pi_{{\bf i'},{\bf j_2}}} {}^{\pi_{i,{\bf j_1}}} f)({\bf z}) -
  \mathrm{S}_{1-\beta}f({\bf z})}
  \right)^2
  }.
%  \\
%  &=
%  \frac{1}{4}
%  \Expect{{\bf z}}
%  {
%  \left(
%  \cExpect{{\bf i'}, {\bf j_1}\in [n],{\bf j_2}}
%  {{\bf z}_{{\bf i'}} = {\bf z}_i, {\bf z}_{{\bf j_2}} = {\bf z}_{{\bf j_1}}}
%  {
%  \mathrm{S}_{1-\beta}({}^{\pi_{{\bf i'},{\bf j_2}}} {}^{\pi_{i,{\bf j_1}}}g)({\bf z})
%  -
%  \mathrm{S}_{1-\beta} g({\bf z})}
%  \right)^2
%  }.
  \end{align*}
%  Using $(a+b)^2\geq \half a^2 - b^2$ we get that
%  \begin{align*}
%  \eqref{eq:13}
%  &\geq
%  \frac{1}{8}
%  \Expect{{\bf z}}
%  {
%  \left(
%  \Expect{{\bf j_1}\in [n]}
%  {
%  \mathrm{S}_{1-\beta}({}^{\pi_{i,{\bf j_1}}}g)({\bf z})
%  -
%  \mathrm{S}_{1-\beta} g({\bf z})}
%  \right)^2
%  }\\
%  &-
%  \frac{1}{4}
%  \Expect{{\bf z}}
%  {
%  \left(
%  \cExpect{{\bf i'}, {\bf j_1}\in [n],{\bf j_2}}
%  {{\bf z}_{{\bf i'}} = {\bf z}_i, {\bf z}_{{\bf j_2}} = {\bf z}_{{\bf j_1}}}
%  {
%  \mathrm{S}_{1-\beta}({}^{\pi_{{\bf i'},{\bf j_2}}} {}^{\pi_{i,{\bf j_1}}}g)({\bf z})
%  -
%  \mathrm{S}_{1-\beta}{}(^{\pi_{i,{\bf j_1}}}g)({\bf z})}
%  \right)^2
%  }.
%  \end{align*}
  Next, observe that for all $z$ and $\pi\in S_n$ which is a composition of two transpositions we have that
  $\card{\mathrm{S}_{1-\beta}{}({}^{\pi}f)(z) - \mathrm{S}_{1-\beta}{}f({}^{\pi}z)}\leq O\left(\frac{\beta}{\alpha}\right)$.
  This is seen by considering a natural coupling between $\mathrm{S}_{1-\beta}{}^{\pi}z$
  and ${}^\pi\mathrm{S}_{1-\beta} z$ similar to the coupling of ${\bf w},{\bf w'}$ above, and noting that
  the probability $\mathrm{S}_{1-\beta}$ touches a coordinate $\ell$ which is not a fixed point of $\pi$ is at
  most $O\left(\frac{\beta}{\alpha}\right)$, and if this doesn't happen then the sampled points are the same.
  Thus, the statistical distance between $\mathrm{S}_{1-\beta}{}^{\pi}z$ and ${}^\pi\mathrm{S}_{1-\beta} z$
  is $O\left(\frac{\beta}{\alpha}\right)$, and as $f$ is bounded our observation follows. Thus,
  \[
  \eqref{eq:13}\geq
  \frac{1}{4}
  \Expect{{\bf z}}
  {
  \left(
  \cExpect{{\bf i'}, {\bf j_1}\in [n],{\bf j_2}}
  {{\bf z}_{{\bf i'}} = {\bf z}_i, {\bf z}_{{\bf j_2}} = {\bf z}_{{\bf j_1}}}
  {
  \mathrm{S}_{1-\beta}f({}^{\pi_{{\bf i'},{\bf j_2}}} {}^{\pi_{i,{\bf j_1}}}{\bf z}) -
  \mathrm{S}_{1-\beta}f({\bf z})}
  \right)^2
  }
  -O\left(\frac{\beta}{\alpha}\right),
  \]
  which is equal to $\frac{1}{4}Q_i(\mathrm{S}_{1-\beta} g) - O\left(\frac{\beta}{\alpha}\right)$.
  \end{proof}

We are now ready to prove Lemma~\ref{lem:projection_of_high_inf}
\begin{proof}
Fix $\alpha,\tau'>0$ and $m\in\mathbb{N}$. We choose $\tau = \frac{\tau'}{1000}$, $\beta_1'$ from Claim~\ref{claim:expected_projected_noisy_inf},
then $\beta_1>0$ small enough, and finally $N = N(\alpha,\tau',m,\beta)$ large enough.

Assume in the setting of the lemma we have $I_i^{(\beta)}[f]\geq \tau'$. By Lemma~\ref{lem:sum_of_noisy_small}
we have $I^{(\beta)}[f] = O_{\beta}(1)$. Set $g = \mathrm{S}_{1-\beta} f$ and apply Claim~\ref{claim:lb_aux_inf_1}
to get that $M_i(g)\geq \frac{1}{6}\tau' - O_{\beta}\left(\frac{1}{N}\right)\geq \frac{1}{7}\tau'$ by the choice of $N$.
Thus, by Claim~\ref{claim:lb_aux_inf_2} we get $Q_i(g) \geq \frac{1}{14}\tau' - O_{\alpha}\left(\frac{1}{N}\right)\geq \frac{1}{20}\tau'$,
by choice of $N$. Thus, by Claim~\ref{claim:expected_projected_noisy_inf}
\[
\Expect{{\bm \pi}\in S_{n:n/2}}{I_{\bm{\pi}(i)}^{(\beta)}[f_{\bm{\pi}}]}
  \geq  \frac{1}{4}Q_i(g) - O\left(\frac{\beta}{\alpha}\right)
  \geq \frac{\tau'}{80} - O\left(\frac{\beta}{\alpha}\right)
  \geq \frac{\tau'}{100},
\]
provided $\beta_1$ is small enough. Now as $f$ is bounded, all of its projections are bounded and so
$I_{\pi(i)}^{(\beta)}[f_{\pi}]\leq 4$ for all $\pi$; thus
\[
\Prob{{\bm \pi}\in S_{n:n/2}}{I_{\bm{\pi}(i)}^{(\beta)}[f_{\bm{\pi}}]\geq \frac{\tau'}{1000}}
\geq \frac{\tau'}{1000}.\qedhere
\]
\end{proof}
\subsection{The reduction}
Let $\mathcal{P}$ be a weighted collection of $r$-ary CSPs over the alphabet $\Sigma = [m]$.
Given a $(1,s)$ dictatorship test for $\mathcal{P}$, denoted by $(w, \set{\mathcal{D}_{t}}_{\in\mathcal{T}}, p)$, in which
each $\mathcal{D}_t$ is connected, we describe a polynomial time reduction that takes as input an instance $\Phi$ of {\sf Gap-Rich-$2$-to-$1$}$_n[1,~\delta]$,
and outputs an instance $\Psi$ of CSP-$\mathcal{P}$ such that:
\begin{enumerate}
  \item Completeness: if ${\sf val}(\Phi)=1$, then ${\sf val}(\Psi) = 1$.
  \item Soundness: for all $\eps>0$, there is sufficiently small $\delta>0$ and sufficiently large even $n$, such that
  if ${\sf val}(\Phi)\leq \delta$, then ${\sf val}(\Psi)\leq s+\eps$.
\end{enumerate}
\subsubsection{Parameters}
Throughout the proof, we will have several parameters $\eps, r, m,\alpha,\beta_0,\beta_1,\beta,\tau,\tau',\tau'',\zeta,\delta$ and $n$, as
well as a finite set $\mathcal{T}'$ whose size we'll be interested in,
and we state the hierarchy between them. We denote by $0<a\ll b$ the relation that once we choose $a$, we take $b$ to be sufficiently large in comparison
to $a$. With this notation, our hierarchy is
\[
0< r,m,\eps^{-1},\alpha^{-1},\card{\mathcal{T}'}\ll\tau^{-1},\tau'^{-1},\tau''^{-1}\ll \beta_0^{-1}, \beta_1^{-1}\ll \beta^{-1} \ll \delta^{-1}\ll
N\ll n,
\]
and $\zeta = n^{-1/4}$. Also, we will assume that $\beta n/2$ is an integer
(otherwise we change $\beta$ slightly to arrange that, while keeping the hierarchy of the parameters).

\subsubsection{Setting up the multi-slices and the distributions $\mu_{t,P}$}
We may find a finite $\mathcal{T}'\subseteq\mathcal{T}$ such that $\sum\limits_{t\in \mathcal{T}\setminus\mathcal{T}'} p(t)\leq \eps/2$,
and we fix such $\mathcal{T}'$. First, as $n$ is an even integer, we may assume that for all $t\in\mathcal{T}'$, $P\in\mathcal{P}$ and
$\vec{a}\in {\sf supp}(\mathcal{D}(t,P))$,
we have that $n\cdot \mathcal{D}(t,P)(\vec{a})$ is an even integer. Otherwise, we may find distribution $(\mathcal{D}(t,P)')_{t\in\mathcal{T}'}$
such that $\mathcal{D}(t,P)',\mathcal{D}(t,P)$ have the same support and are $O_m(1/n^2)$-close in KL-divergence, and so
$\mathcal{D}(t,P)'^{\otimes n},\mathcal{D}(t,P)^{\otimes n}$ are $O_m(1/n)$-close in KL-divergence and by Pinsker's
inequality the statistical distance between them is $O_{m}(1/\sqrt{n}) = o(1)$.
In particular, this collection of distributions forms a $(1,s+o(1))$ dictatorship test, and we may proceed the argument with it.
Denote
\[
\alpha = \min\limits_{t\in\mathcal{T}', P\in\mathcal{P}}\min\limits_{\vec{a}\in {\sf supp}(\mathcal{D}(t,P))} (\mathcal{D}(t,P))(\vec{a})>0.
\]
For each $t\in\mathcal{T}'$, $P\in\mathcal{P}$ consider the distribution $\mathcal{D}(t,P)$ over $[m]^r$, and denote its marginal distributions
by $\mathcal{D}(t,P)_1,\ldots\mathcal{D}(t,P)_r$.
Denote $k(i,t,P)_a = n\cdot \mathcal{D}(t,P)_i(a)$, and note that $k(i,t,P)_a$ is an even integer.
We thus have that $\vec{k}(i,t,P) = (k(i,t,P)_1,\ldots,k(i,t,P)_m)$ is a sequence of even integers summing up to
$n$.

Thus, we may consider the $r\card{\mathcal{T}'}\card{\mathcal{P}}$-mutli slices $\left(\mathcal{U}_{\vec{k}(i,t,P)}\right)_{\substack{i\in [r]\\t\in\mathcal{T}',P\in\mathcal{P}}}\subseteq [m]^n$;
note that each one of them is $\alpha$-balanced. Next, we define for each $t\in \mathcal{T}'$, $P\in\mathcal{P}$ a distribution $\mu_{t,P}$ over
$\prod\limits_{i=1}^{r} \mathcal{U}_{\vec{k}(i,t,P)}$ that would ``simulate'' the dictatorship test $\mathcal{D}(t,P)^{\otimes n}$.
The distribution $\mu_{t,P}$ is uniform on $(x(1),\ldots,x(r))$ whose statistics match the expectation in $\mathcal{D}(t,P)^{\otimes n}$, i.e.~on
\[
\sett{(x(1),\ldots,x(r))\in\prod\limits_{i=1}^{r} \mathcal{U}_{\vec{k}(i,t,P)}}
{\forall \vec{a}\in[m]^r,~~\#\sett{i}{x(1)_i = a_1,\ldots,x(r)_i = a_r} = n\cdot(\mathcal{D}(t,P))(\vec{a})}.
\]
It is easily seen that $\mu_{t,P}$ is $\alpha$-admissible as per Definition~\ref{def:admissible},
and negatively correlated as per Definition~\ref{def:neg_correlated}. Also, we note that the product version of
$\mu_{t,P}$ as per Definition~\ref{def:product_ver}, i.e.~$\tilde{\mu_{t,P}}$, is nothing but $\mathcal{D}(t,P)^{\otimes n}$.

\begin{remark}\label{remark:far_dist_guess}
Below, we will assume that ${\mu_{t,P}}_i^{\otimes n}$ and ${\mu_{t,P}}_i^{\otimes n}$ are either identical, or are $1-o(1)$ far and
hence the distribution can be recovered by a single sample except with probability $o(1)$. This may be guaranteed provided that $n$ is
large enough, as follows.
We may find a constant $\beta>0$ such that for any $i,t,P, i',t',P'$, either $\mathcal{D}(t,P)_i$ and $\mathcal{D}(t,P)_{i'}$ are
identical, or else they are at least $\beta$ far in total variation distance. Thus, the distributions ${\mu_{t,P}}_i^{\otimes n}$ and
${\mu_{t,P}}_i^{\otimes n}$ are either identical, or are $1-o(1)$ far. This means that provided that $n$ is large enough, sampling ${\bf t}\sim\mathcal{T}$,
${\bf P}\sim w$ and $x\sim (\mu_{{\bf t},{\bf P}})_i^{\otimes n}$ for some $i$, and outputting $x$, one may recover $t', P'$ and $i'$ such that
${\mu_{{\bf t}, {\bf P}}}_i^{\otimes n} = {\mu_{t', P'}}_{i'}^{\otimes n}$ except with probability $o(1)$.
\end{remark}

\subsubsection{The PCP construction}
Write the Rich-$2$-to-$1$ Games instance $\Phi$ as $(L\cup R, E, [n], [n/2], \set{\phi_e}_{e\in E})$.
We replace each vertex $v\in R$ with copies of the multi-slices
$\left(\mathcal{U}_{\vec{k}(i,t,P)/2}\right)_{\substack{i\in [r]\\t\in\mathcal{T}'\\ P\in\mathcal{P}}}\subseteq [m]^{n/2}$;
we represent these as $(v,i,t,P,x)$ where $x\in \mathcal{U}_{\vec{k}(i,t,P)/2}$, and denote by $V$ the set of
all of these tuples. Our instance $\Psi$ of CSP-$\mathcal{P}$ will have a variable $z_{(v,i,t,x)}$ for each
$(v,i,t,P,x)\in V$. It will also  be useful for us to imagine that each $u\in L$ also has $r\card{\mathcal{T}'}\card{\mathcal{P}}$ multislices
$\left(\mathcal{U}_{\vec{k}(i,t,P)}\right)_{\substack{i\in [r]\\t\in\mathcal{T}'\\ P\in\mathcal{P}}}\subseteq [m]^{n}$
associated with it, but the reduction does not actually use these.

\paragraph{Folding.} We define the relation $\sim$ on $V$ by $(v(1),i(1),t(1),x(1)) \sim (v(2),i(2),t(2),x(2))$ if
there is a common neighbour $u\in L$ of $v(1), v(2)$, as well as $i\in[r]$, $t\in \mathcal{T}'$, $P\in\mathcal{P}$
and $x\in \mathcal{U}_{\vec{k}(i,t,P)}$, such that $x = \phi_{u,v(1)}^{-1}(x(1))$ and $x = \phi_{u,v(2)}^{-1}(x(2))$.
In words and slightly informally, we say two vertices are equivalent if they are the projection of the same point
$(u,i,t,P,x)$ in the imaginary copies of the multislices corresponding to $L$, along the corresponding constraint
$2$-to-$1$ maps $\phi_{u,v(1)}$ and $\phi_{u,v(2)}$. We extend $\sim$ to an equivalence relation on $V$, which,
by abusing notation, we also denote by $\sim$. Thus, we will look at the equivalence classes of $\sim$, and
denote the equivalence class of $(v,i,t,P,x)$ by $[(v,i,t,P,x)]$. For each equivalence class $C$ of $\sim$,
we choose a representative and identify all the variables corresponding to $(v,i,t,P,x)$ with the variable
of the representative. Formally, let $C$ be an equivalence class and suppose $(v,i,t,P,x)$ is its representative.
Then whenever we refer to a variable $z_{(v',i',t',P',x')}$ corresponding to $(v',i',t',P',x')$ from the
same equivalence class $C$, we replace its occurrence with $z_{(v,i,t,P,x)}$.

\paragraph{Constraints.}
The constraints of $\Psi$ are described by the following sampling procedure. Sample ${\bf t}\sim p$, and if
${\bf t}\not\in \mathcal{T}'$ put a trivial constraint (say, we add $r$ variables $\ell_1,\ldots,\ell_r$,
pick $P\in\mathcal{P}$, and put the constraint that $P(\ell_1,\ldots,\ell_r)$). Otherwise,
${\bf t}\in\mathcal{T}'$, and then we sample $u\in L$ with probability proportional to its degree, ${\bf P}\sim w$, and
$({\bf x}(1),\ldots,{\bf x}(r))\sim\mu_{{\bf t}, {\bf P}}$. Then for each $i=1,\ldots,r$, we sample ${\bf v}(i)$
a neighbour of $u$ conditioned on ${\bf x}(i)\in \phi_{u,{\bf v}(i)}^{-1}(\mathcal{U}_{\vec{k}(i,{\bf t}, {\bf P})/2})$.
Denote by ${\bf x}(i)'$ the unique point in $\mathcal{U}_{\vec{k}(i,{\bf t}, {\bf P})/2}$ such that
${\bf x}(i) = \phi_{u,{\bf v}(i)}^{-1}({\bf x}(i)')$, and add the constraint
\[
{\bf P}(z_{{\bf v}(1),1,{\bf t},{\bf x}(1)'},\ldots,z_{{\bf v}(r),r,{\bf t},{\bf x}(r)'}).
\]
\begin{remark}
  We note that in the sampling procedure above, one can take an arbitrary ${\bf v}(i)$ satisfying this
  property as opposed to a random one, without introducing any ambiguity. The reason is that any two
  possible such $v_i$'s would yield two tuples in $V$, between which the relation $\sim$ holds,
  and in that case by the folding they correspond to the same variable.
\end{remark}

The analysis of this reduction spans the following two subsections.
\subsection{Completeness of the reduction}
As is often the case, the completeness of the reduction is easy. Indeed, assume $\Phi$ is satisfiable, and let
$A_L\colon L\to[n]$, $A_R\colon R\to[n/2]$ be labelings that satisfy all of the constraints of $\Phi$. For each
$v$, we define the assignment $A'$ to the variables of $\Psi$ as
\[
A'(z_{v,i,t,x}) = x_{A_R(v)}.
\]
First, we note that the variables in each equivalence class of $\sim$ get the same value. For that, it suffices to observe
that for any $(v(1),i(1),t(1),x(1)) \sim (v(2),i(2),t(2),x(2))$ in the basic form of $\sim$, it holds that their variables
get the same value by $A'$. This is true, since if $(u,i,t,x)$ is the common neighbour demonstrating that the basic form
of $\sim$ holds, then as $x$ is the pullback of $x(1)$ along $\phi_{u,v(1)}$ we have that both coordinates in
$\phi_{u,v(1)}^{-1}(A_R(v(1)))$ have the same value in $x$, and since the constraint is satisfied we have
$A_L(u)\in \phi_{u,v(1)}^{-1}(A_V(v(1)))$. Thus, $x(1)_{A_R(v(1))} = x_{A_L(u)}$, and a similar argument shows that
$x(2)_{A_R(v(2))} = x_{A_L(u)}$. We thus get that
\[
A'(z_{v(1),i(1),t(1),x(1)})
=x(1)_{A_R(v)}
=x_{A_L(u)}
=x(2)_{A_R(v(2))}
=A'(z_{v(2),i(2),t(2),x(2)}).
\]
We may therefore think of the labeling $A'$ as also giving values to the imaginary multi-slices of vertices in $L$, where
$(u,i,t,x)$ gets the value $A'(z_{(v,i',t',x')})$ where $x = \phi_{u,v}^{-1}(x')$.
Next, we show that the constraints hold. Let $P(z_{v_1,1,t,x(1)},\ldots,z_{v_r,r,t,x(r)})$ be a constraint in the instance,
and let $u\in L$ be the vertex that samples $v_1,\ldots,v_r$. The constraint being satisfied amounts to
$P(x(1)_{A_R(v_1)},\ldots,x(r)_{A_R(v_r)}) = 1$, and as before we have that if $x(i)'$ is the pull-back of
$x(i)$ along $\phi_{u,v(i)}$, then $x(i)_{A_R(v_i)} = x(i)'_{A_L(u)}$, so we must show that
$P(x(1)'_{A_L(u)},\ldots,x(r)'_{A_L(u)})$. Considering the dictatorship function $f\colon [m]^n\to [m]$
defined as $f(y) = y_{A_L(u)}$, the constraint being satisfied is equivalent to
$P(f(x(1)'),\ldots,f(x(r)')) = 1$, and as $f$ is a dictatorship and $(x(1)',\ldots,x(r)')$ is in the support
$\mu_t$, and hence of $\mathcal{D}_t^{\otimes n}$, we have
that this is satisfied by the completeness property of the dictatorship test
(the first item of Definition~\ref{def:dictator_test}).

\subsection{Soundness analysis}
In this section we finish the analysis of the PCP construction by showing that if ${\sf val}(\Phi)\leq \delta$,
then ${\sf val}(\Psi)\leq s+\eps$. We prove that counter-positively.

Namely, suppose we have an assignment $A'$ to the variables satisfying at least $s+\eps$ fraction of the constraints; we show that
it is possible to find an assignment to $\Phi$ satisfying more than $\delta$ fraction of the constraints.

For each $v,i,t,P$ we will think of the assignment on the multi-slice of $v$ as
$f_{v,i,t,P}\colon\mathcal{U}_{\vec{k}(i,t,P)/2}\to [m]$. Let $u\in L$, since for all
$i,t,P,x$ such that $x\in\mathcal{U}_{\vec{k}(i,t,P)}$ we have that all projections
of $x$ along $\pi_{u,v}$ for neighbours $v$ of $u$ get the same value by $A'$, we may think
that this value is also given to $(u,i,t,P,x)$, and thus think of a function $f_{u,i,t,P}\colon\mathcal{U}_{\vec{k}(i,t,P)}\to [m]$
corresponding to these values. Thus, for all $u\in L$ and a neighbour $v$ of $u$, we have that $f_{v,i,t,P} = f_{u,i,t,P}|_{\phi_{u,v}}$.

Below, when we write ${\bf u}\in L$ we mean that we sample ${\bf u}$ with probability proportional to its degree.
\begin{claim}
  $\Expect{\substack{{\bf u}\in U \\ {\bf t}\sim p}}
  {
  \Expect{{\bf P}\sim w}{
  \Expect{({\bf x}(1),\ldots,{\bf x}(r))\sim\mu_{{\bf t},{\bf P}}}{
  {\bf P}(f_{{\bf u},1,{\bf t},{\bf P}}({\bf x}(1)),\ldots,f_{{\bf u},r,{\bf t},{\bf P}}({\bf x}(r)))}}}\geq s+\eps$.
\end{claim}
\begin{proof}
  By construction of the functions $f_{{\bf u}}$, this expression is equal to the fraction of constraints
  satisfied by $A'$.
\end{proof}
Note that ${\bf t}\not\in\mathcal{T}'$ with probability at most $\eps/2$, so we get that
\[
\Expect{\substack{{\bf u}\in U \\ {\bf t}\sim p}}
  {
  1_{{\bf t}\in\mathcal{T}'}
  \Expect{{\bf P}\sim w}{
  \Expect{({\bf x}(1),\ldots,{\bf x}(r))\sim\mu_{{\bf t}, {\bf P}}}{
  {\bf P}(f_{{\bf u},1,{\bf t},{\bf P}}({\bf x}(1)),\ldots,f_{{\bf u},r,{\bf t},{\bf P}}({\bf x}(r)))}}}\geq s+\half\eps.
\]
Denote
\[
{\sf Good}[L] =
\left\{
u\in L\middle|
  \Expect{\substack{{\bf t}\sim p}}
  {
  1_{{\bf t}\in\mathcal{T}'}
  \Expect{{\bf P}\sim w}{
  \Expect{\substack{({\bf x}(1),\ldots,{\bf x}(r))\\ \sim\mu_{{\bf t},{\bf P}}}}{
  {\bf P}(f_{u,1,{\bf t},{\bf P}}({\bf x}(1)),\ldots,f_{u,r,{\bf t},{\bf P}}({\bf x}(r)))}}}\geq s+\frac{1}{4}\eps\right\},
\]
then by an averaging argument $\Prob{{\bf u}\in L}{{\bf u}\in {\sf Good}[L]}\geq \frac{1}{4}\eps$. Next, we devise
a randomized labeling strategy for vertices in ${\sf Good}[L]$ and their neighbours.

Fix $u\in {\sf Good}[L]$, $t\in\mathcal{T}'$, $P\in\mathcal{P}$. Note that by Proposition~\ref{prop:coupling_construction} there
is a $(\Omega_{\alpha,m}(1),\zeta)$-coupling between $\mathcal{U}_{\vec{k}(i,t,P)}$ and $\nu_{\vec{k}(i,t,P)}$,
denote it by $\mathcal{C}_{i,t,P}$, and we also have a $(\Omega_{\alpha,r,m}(1),\zeta)$-coupling between
$\mu_{t,P}$ and $\tilde{\mu}_{t,P} = \mathcal{D}(t,P)^{\otimes n}$ (see Remark~\ref{remark:neg_assoc}).
We now think of the functions $f_{u,i,t}$ as receiving values in $\Delta_m$
(as in Section~\ref{sec:label_assignments}). Thus, we get from Theorem~\ref{thm:inv_label_assignments}
that
\[
\Expect{\substack{{\bf t}\sim p}}
  {
  1_{{\bf t}\in\mathcal{T}'}
  \Expect{{\bf P}\sim w}{
  \Expect{({\bf x}(1),\ldots,{\bf x}(r))\sim \mathcal{D}({\bf t}, {\bf P})^{\otimes n}}{
  \tilde{{\bf P}}(\mathrm{T}_{\mathcal{C}_{1,{\bf t},{\bf P}}}f_{u,1,{\bf t},{\bf P}}({\bf x}(1)),\ldots,\mathrm{T}_{\mathcal{C}_{1,{\bf t},{\bf P}}} f_{u,r,{\bf t},{\bf P}}({\bf x}(r)))}}}\geq s+\frac{1}{8}\eps.
\]

We now define a single function $g_u$ in order to appeal to the assumption regarding the dictatorship test. Upon receiving an input $x$, we guess $t',P'$ and $i$ such
that we believe $x$ was sampled from $\mathcal{D}({\bf t}, {\bf P})_i^{\otimes n}$ as in Remark~\ref{remark:far_dist_guess}. We then set
$g_u(x) = \mathrm{T}_{\mathcal{C}_{i,t',P'}}f_{u,i,t',P'}(x)$. Then the above inequality and the assertion of Remark~\ref{remark:far_dist_guess} implies that
\[
\Expect{\substack{{\bf t}\sim p}}
  {
  1_{{\bf t}\in\mathcal{T}'}
  \Expect{{\bf P}\sim w}{
  \Expect{({\bf x}(1),\ldots,{\bf x}(r))\sim \mathcal{D}({\bf t}, {\bf P})^{\otimes n}}{
  \tilde{{\bf P}}(g_{u}({\bf x}(1)),\ldots,g_{u}({\bf x}(r)))}}}\geq s+\frac{1}{8}\eps - o(1)\geq s+\frac{1}{16}\eps.
\]

Pick $d,\tau$ from the second item in Definition~\ref{def:dictator_test} for our dictatorship test for
$\gamma = \eps/16$ and $\eps/16$. Then from the soundness property of our dictatorship test,
we get that for each $u\in{\sf Good}[L]$, there are $t\in\mathcal{T}'$, $P\in\mathcal{P}$,
$i\in [r]$ and $j\in [n]$ such that $I_j^{\leq d}[g_u; \nu_{\vec{k}(i,t,P)}^{\otimes n}]\geq \tau$.
Thus, by Remark~\ref{remark:far_dist_guess} again we have that
\[
I_j^{\leq d}[\mathrm{T}_{\mathcal{C}_{i,t,P}} f_{u,i,t, P}; \nu_{\vec{k}(i,t,P)}^{\otimes n}]\geq \tau/2.
\]
Now recall that $\mathrm{T}_{\mathcal{C}_{i,t,P}} f_{u,i,t,P}$ is a function to $\Delta_m$, so it follows that one of its $m$ components has degree
$d$ influence at least $\frac{\tau}{2m}$; denote the $\ell$th components by
$(\mathrm{T}_{\mathcal{C}_{i,t,P}} f_{u,i,t,P})_{\ell}$, and note that by definition of the action of $\mathrm{T}_{\mathcal{C}_{i,t,P}}$ on vector
valued functions, this $\ell$th component is equal to $\mathrm{T}_{\mathcal{C}_{i,t,P}}((f_{u,i,t,P})_{\ell})$. Thus, taking $\beta_0,\tau'>0$ from Lemma~\ref{lem:noisy_inf_coupling} for $\frac{\tau}{2m}$,
we get that there is $j\in[n]$ such that  $I_j^{(\beta)}[(f_{u,i,t,P})_{\ell}]\geq \tau'$ (here we use the choice of $\zeta$).
Define
\[
{\sf List}[u] =
\left\{j\in [n]~|~\exists t\in\mathcal{T}', P\in\mathcal{P}, \ell\in [m], i\in [r]\text{ such that } I_j^{(\beta)}[(f_{u,i,t,P})_{\ell}]\geq \tau'\right\},
\]
then by the above reasoning ${\sf List}[u]\neq \emptyset$ for all $u\in{\sf Good}[L]$,
and by Lemma~\ref{lem:sum_of_noisy_small} and the union bound we have that
$\card{{\sf List}[u]}\leq O_{\beta,\card{\mathcal{T}'},m,r,\tau'}(1)$.

Take $\tau'',\beta_1>0$ from Lemma~\ref{lem:projection_of_high_inf}, and for each $v\in R$ denote
\[
{\sf List}[v] = \left\{j\in [n/2]~|~\exists t\in\mathcal{T}', P\in\mathcal{P}, \ell\in [m], i\in [r]\text{ such that } I_j^{(\beta)}[(f_{v,i,t,P})_{\ell}]\geq \tau''\right\}.
\]
As before, we have $\card{{\sf List}[v]}\leq O_{\beta,\card{\mathcal{T}'},m,r,\tau''}(1)$.
Fix $u\in {\sf Good}[L]$ and $j\in {\sf List}[u]$. Choosing a neighbour ${\bf v}$ of $u$ randomly, by the richness of the game,
the constraint $\phi_{u,{\bf v}}$ is distributed uniformly on $S_{n:n/2}$, and so by Lemma~\ref{lem:projection_of_high_inf}
we get that with probability at least $\tau''$, $I_{\phi_{u,{\bf v}}(j)}[(f_{u,i,t,P})_{\ell}|_{\phi_{u,{\bf v}}}] \geq \tau''$. But by construction
of the $f_{u,i,t,P}$ we have that $(f_{u,i,t,P})_{\ell}|_{\phi_{u,{\bf v}}} = (f_{{\bf v},i,t,P})_{\ell}$, so we get that
$\phi_{u,{\bf v}}(j)\in {\sf List}[{\bf v}]$ with probability at least $\tau''$. We have thus proved that for all $u\in {\sf Good}[L]$,
choosing a neighbour ${\bf v}$ of $u$ randomly, with probability at least $\tau''$
the lists ${\sf List}[u], {\sf List}[{\bf v}]$ contain a pair of labels that satisfy $\phi_{u,{\bf v}}$.

We thus use the following randomized strategy: define the assignment to $\Phi$ by labeling each $u\in L$ with an element $\sigma\in {\sf List}[u]$
chosen uniformly at random if this list is not empty (and otherwise an arbitrary $\sigma\in [n]$), and similarly pick the label of each $v\in R$ using
${\sf List}[v]$. It follows that the expected fraction of constraints satisfied by this strategy is at least
\[
\Prob{{\bf u}\in U}{{\bf u}\in {\sf Good}[L]}\cdot \Omega_{\beta,\card{\mathcal{T}'},m,r,\tau',\tau''}(1)
=\Omega_{\eps,\beta,\card{\mathcal{T}'},m,r,\tau',\tau''}(1) > \delta,
\]
so there is an assignment to $\Phi$ satisfying more than $\delta$ of the constraints, and we are done.

\begin{remark}\label{remark:extend}
  Using Theorem~\ref{thm:inv_annh} (or rather the analogue of Theorem~\ref{thm:inv_label_assignments} that follows from it)
  instead of Theorem~\ref{thm:inv_label_assignments}, one may relax the connectedness assumption in Theorem~\ref{thm:dictator_to_hard} as in Theorem~\ref{thm:thm_extend_hardness_convert} below.
  Roughly speaking, it is enough that for all $t\in\mathcal{T}, P\in\mathcal{P}$, the sequence of distributions $\set{\mathcal{D}(t,P)^{\otimes n}}_{n\in\mathbb{N}}$
  and its multi-slice analogue annihilate high degrees functions.
  \begin{thm}\label{thm:thm_extend_hardness_convert}
  Let $r\in\mathbb{N}$, $m\in\mathbb{N}$, and
  suppose $\mathcal{P}\subseteq \set{P\colon [m]^r\to\set{0,1}}$ is a collection of predicates.
  Suppose there is a $(1,s)$ dictatorship test, $(w,\set{\mathcal{D}(t,P)}_{t\in\mathcal{T}, P\in\mathcal{P}} ,p)$, for $\mathcal{P}$
  satisfying the following properties:
  \begin{enumerate}
    \item for all $t\in\mathcal{T}, P\in\mathcal{P}$, the collection of distributions $\set{\mathcal{D}(t,P)^{\otimes n}}_{n\in\mathbb{N}}$
    annihilates high degree functions as per Definition~\ref{def:ann};
    \item the multi-slice analogues of $\mathcal{D}(t,P)^{\otimes n}$ also annihilate high-degree functions. That is, for each $t\in\mathcal{T}$, $P\in\mathcal{P}$ and
    $n\in\mathbb{N}$, considering $\vec{k_n(t,P)} = (k_n(t,P)_{\vec{a}})_{\vec{a}\in[m]^r}$ a list of integers adding up to $n$ such that
    $\card{k_n(t,P)_{\vec{a}} - n\mathcal{D}(t,P)(\vec{a})} < 1$ for all $\vec{a}\in[m]^r$, the sequence of uniform distributions on the
    $\vec{k_n(t,P)}$ multi-slice of $([m]^r)^n$,
    that is $\left(\mathcal{U}_{\vec{k_n(t,P)}},{\sf Uniform}\right)_{n\in\mathbb{N}}$,
    annihilates high degree functions.
  \end{enumerate}
  Then, for all $\eps>0$, the problem {\sf Gap-$\mathcal{P}$}$[1,s+\eps]$ is NP-hard assuming Conjecture~\ref{conj:rich}.
  \end{thm}
  We omit the full proof, as it is a straightforward adaptation of the argument above.
\end{remark}

\subsection{Proof of Corollaries~\ref{cor:CSP},~\ref{cor:graph_col}}
In this section, we instantiate Theorem~\ref{thm:dictator_to_hard} to establish Corollaries~\ref{cor:CSP},~\ref{cor:graph_col}.
\begin{proof}[Proof of Corollary~\ref{cor:CSP}]
  Fix $\eps>0$. In~\cite[Theorem 1.1]{BKT}, the authors have constructed an $r$-ary predicate
  $P_r\colon\{-1,1\}^r\to\{0,1\}$ with $2r+1$ accepting assignments based on the Hadamard code,
  and constructed a discrete distribution $\mathcal{E}$ of $\delta$ and a distribution $\mathcal{D}_{r,\delta}$ for each $\delta$ in the support of $\mathcal{E}$, such that if
  \[
  \Expect{\delta\sim \mathcal{E}}{\Expect{(x(1),\ldots,x(r))\sim \mathcal{D}_{r,\delta}}{P(f(x(1)),\ldots, f(x(r)))}}\geq \frac{2r+1}{2^r} + \eps
  \]
  for $f\colon \{-1,1\}^n\to \{-1,1\}$ that is folded\footnote{I.e., $f(x) = -f(-x)$ for all $x$.},
  then $\max_{i\in [n]} I_i^{\leq d}[f]\geq \tau$ for $d\in\mathbb{N}$, $\tau>0$ depending only on $\eps$.
  Considering the multi-linear expansion of $P_r$ and thereby extending $P_r$ to $[-1,1]^r$, their proof actually shows the same conclusion for
  $f\colon \{0,1\}^n\to [-1,1]$ that is folded, and we will make use of this.
  We also record here that by~\cite[Observation 4.1]{BKT}, the marginal of $x(j)_i$ is uniform for all $i,j$.

  We now transform this into a dictatorship test as per Definition~\ref{def:dictator_test}.
  The measure $w$ is uniform over $\{-1,1\}^r$, and for each $a\in\{-1,1\}^r$ we consider the predicate $P_{r,a}(x) = P(a_1x_1,\ldots,a_rx_r)$.
  We take $\mathcal{T} = {\sf supp}(\mathcal{E})$, the measure $p$ is the same as the distribution $\mathcal{E}$, and
  the distribution $\mathcal{D}(\delta,P_{r,a})$ is defined as $a \mathcal{D}_{r,\delta}$, i.e.~$(a_1x(1),\ldots,a_rx(r))$ where
  $(x(1),\ldots,x(r))\sim \mathcal{D}_{r,\delta}$.

  As all dictators satisfy $P_r$ with probability $1$, it is easy to see that the completeness of the above test is $1$.

  As for the soundness, given a function $f\colon\{-1,1\}^n\to \{-1,1\}$ that passes the dictatorship test with probability at least $\frac{2r+1}{2^r} + \eps$, we have
  \begin{align*}
  \frac{2r+1}{2^r} + \eps
  &\leq
  \Expect{a, \delta}{\Expect{(x(1),\ldots,x(r))\sim a\mathcal{D}_{r,\delta}}{P_{a,r}(f(x(1)),\ldots,f(x(r)))}}\\
  &=
  \Expect{a, \delta}{\Expect{(x(1),\ldots,x(r))\sim \mathcal{D}_{r,\delta}}{P_r(a_1f(a_1x(1)),\ldots,a_rf(a_rx(r)))}}\\
  &=
  \Expect{\delta}{\Expect{(x(1),\ldots,x(r))\sim \mathcal{D}_{r,\delta}}{P_r(g(x(1)),\ldots,g(x(r)))}},
  \end{align*}
  where $g(x) = \frac{1}{2}(f(x) - f(-x))$. Note that $g\colon \{-1,1\}^n\to[-1,1]$ is folded. Thus, by the main result of~\cite{BKT} it follows
  that there is $i\in [n]$ such that $I_i^{\leq d}[g]\geq \tau$ for $d,\tau$ depending only on $\eps$. Looking at the Fourier expression for
  $I_i^{\leq d}[g]$, it follows that $I_i^{\leq d}[f]\geq \tau$, and thus we have shown that $(w,p,\mathcal{D}(\delta,P_{r,a}))$ forms a dictatorship
  test with completeness $1$ and soundness $\frac{2r+1}{2^r}$.

  The result now follows by Theorem~\ref{thm:dictator_to_hard}
\end{proof}

\begin{proof}[Proof sketch of Corollary~\ref{cor:graph_col}]
 Consider $\Sigma = \{0,1,2\}$ and $P\colon \Sigma^2\to\{0,1\}$ defined as $P(x,y) = 1_{x\neq y}$.
 The distributions $w, t$ are trivial, and the distribution $\mathcal{D}(t,P)$ is uniform over $(x,y)\in\{(x,y)\in \Sigma^2~|~x\neq y\}$.
 We execute the reduction in Theorem~\ref{thm:dictator_to_hard}
 to construct a graph $H$ whose vertices correspond to variables of the instance produced, and edges correspond to pairs of variables between which there
 is a constraint. The completeness is clear, as any dictatorships satisfies $P_a(f(x),f(y)) = 1$ if $(x,y)\in P_a^{-1}(1)^{\otimes n}$.

 As for the soundness, if the resulting graph has an independent set of size at least $\delta$ fraction of the vertices, then we may consider its indicator
 function $f$, and note that for every edge $(x,y)$ in $H$ we have that either $f(x)=0$ or $f(y)=1$. As in the analysis therein, we consider the set of vertices
 $\mathcal{U}$ of the Rich-$2$-to-$1$ Games instances for which the average of $f$ inside their cloud is at least $\delta/2$, so that the fractional size of
 $\mathcal{U}$ is at least $\delta/2$. In other words, letting $Q(x,y)=1_{x=y=1}$, we have that $\Expect{(x,y)}{Q(f_u(x),f_u(y))} = 0$
 for all $u\in \mathcal{U}$. Applying our invariance principle, it follows that there is a coupling $\mathcal{C}$ between the multi-slice of $u$ and $\Sigma^{n}$
 such that $\Expect{(x,y)\sim P^{-1}(1)^{\otimes n}}{Q(\mathrm{T}_{\mathcal{C}}f_u(x),\mathrm{T}_{\mathcal{C}}f_u(y))}=o(1)$. From this and~\cite[Theorem 3.1]{DMR}
 it follows that there is a coordinates $i$ such that $I_i^{\leq d}[\mathrm{T}_{\mathcal{C}}f_u]\geq \tau$ where $d\in\mathbb{N}$, $\tau>0$ depend only on $\delta>0$.
 From here the analysis in the proof of Theorem~\ref{thm:dictator_to_hard} continues as is.
\end{proof}

\section{Acknowledgements}
We thank an anonymous referee for several helpful comments, and in particular for bringing~\cite{negativeassoc,negativeassoc2} to our attention.
We thank another anonymous referee for detailed comments that led to improvements in the presentation, as well as for pointing out that the
operator $\mathcal{R}_{\vec{r}}$ is not necessarily self adjoint.
\bibliographystyle{plain}
\bibliography{ref}

\appendix
\section{Facts about multinomial coefficients}
Let $H(p_1,\ldots,p_m) = \sum\limits_{i} p_i\log(1/p_i)$ be the entropy function.
We need the following standard facts. The first estimates multinomial coefficients
with the entropy function and is easy to derive from Stirling approximation, and we omit the proof.
\begin{fact}\label{Fact:stirling_cor}
  For positive integers $v_1,\ldots,v_r$ that sum up to $n$ we have
  \[
  {n \choose v_1,\ldots,v_r} = \left(1/(2\pi)^{(r-1)/2}+o_n(1)\right)\sqrt{\frac{n}{v_1\cdots v_r}} ~2^{H(\frac{v_1}{n},\ldots,\frac{v_r}{n})\cdot n}
  \]
\end{fact}

Second, we need the following crude estimate on the difference between the entropy of two distributions that are close point-wise.
\begin{fact}\label{fact:entropy_close}
  Suppose $p_1,\ldots,p_m$ and $q_1,\ldots,q_m$ are distributions satisfying that
  $\card{p_i - q_i}\leq \eps$ for all $i$. Then
  \[
  \card{H(p_1,\ldots,p_m) - H(q_1,\ldots,q_m)}
  \leq 4m \eps \log(1/\eps).
  \]
\end{fact}
\begin{proof}
We prove that $H(p_1,\ldots,p_m) - H(q_1,\ldots,q_m)\leq 4\eps m\log(1/\eps)$, and the claim follows by symmetry between $p$ and $q$.
Let $X = \sett{i}{p_i\geq 2\eps}$. Then $\sum\limits_{i\not\in X}{p_i\log (1/p_i)}\leq 2m\eps\log(1/\eps)$, and for
any $i\in X$ we have that $q_i\geq \eps$. Thus,
\begin{align*}
H(p_1,\ldots,p_m) - H(q_1,\ldots,q_m)
&\leq
2m\eps\log\left(\frac{1}{\eps}\right)
+\sum\limits_{i\in X} p_i \log\left(\frac{1}{p_i}\right)-q_i \log\left(\frac{1}{q_i}\right)\\
&\leq 2m\eps\log\left(\frac{1}{\eps}\right)
+\sum\limits_{i\in X} p_i \log\left(\frac{1}{p_i}\right)-(p_i-\eps) \log\left(\frac{1}{q_i}\right)\\
&\leq 3m\eps\log\left(\frac{1}{\eps}\right)
+\sum\limits_{i\in X} p_i\log\left(\frac{q_i}{p_i}\right)\\
&\leq 3m\eps\log\left(\frac{1}{\eps}\right)
+\sum\limits_{i\in X} p_i\card{\frac{q_i}{p_i}-1}
\leq 4m\eps\log\left(\frac{1}{\eps}\right),
\end{align*}
where in the penultimate inequality we used $\log z\leq \card{z-1}$.
\end{proof}

\section{Facts about product spaces}\label{sec:prod_space_apx}
\subsection{Contraction of high degrees}
\begin{claim}\label{claim:cheeger}[Lemma 2.9 in~\cite{MosselGaussian}]
  Let $\alpha>0$. Suppose $\mu$ is a connected measure on $[m]\times [m]$ whose marginal distributions
  both are equal to $\nu$, and the probability of each atom is at least $\alpha$.
  Consider $\mathrm{T}_{\mu}: L^{2}([m],\nu)\to L^2([m],\nu)$ defined as
  \[
  \mathrm{T}_{\mu} f (y) = \cExpect{({\bf x},{\bf y})\sim \mu}{{\bf y} = y}{f({\bf x})}.
  \]
  Then $\lambda_2(\mathrm{T}_{\mu})\leq 1-\alpha^2/2$.
\end{claim}

\begin{claim}
  Let $\mu_1,\ldots,\mu_n$ be measures on $[m]\times [m]$,
  such that
  \begin{enumerate}
    \item For each $i$, the marginal distribution of $\mu_i$ on both coordinates is the same, denote it by $\nu_i$;
    \item $\lambda_2(\mathrm{T}_{\mu_i})\leq 1-\beta$.
  \end{enumerate}
  Let $n,d\in\mathbb{N}$, and consider  $\mathrm{T}: L^{2}([m]^n,\nu)\to L^2([m]^n,\nu)$
  defined as $\mathrm{T} = \mathrm{T}_{\mu_1}\otimes \ldots\otimes \mathrm{T}_{\mu_n}$, where $\nu = \nu_1\times \ldots\times \nu_n$.
  Then for all $f\in V_{>d}([m]^n,\nu^{\otimes n})$ it holds that
  \[
  \norm{\mathrm{T} f}_{2,\nu}\leq (1-\beta)^d\norm{f}_{2,\nu}.
  \]
\end{claim}
\begin{proof}
  Write $f$ by its Efron-Stein decomposition
  with respect to the measure $\nu$,
  $f = \sum\limits_{S\subseteq [n], \card{S}>d} f^{=S}$.
  Then we have that
  $\norm{\mathrm{T} f}_2^2 = \sum\limits_{S\subseteq [n], \card{S}>d}\norm{\mathrm{T} f^{=S}}_2^2$,
  so it is enough to show that $\norm{\mathrm{T} f^{=S}}_2^2\leq (1-\beta)^{2d}\norm{f^{=S}}_2^2$.

  Assume without loss of generality that $S = \set{1,\ldots,s}$. For each $j$, define
  $\mathrm{T}_{\leq j} = \bigotimes_{i\leq j}\mathrm{T}_{\mu_i} \otimes \mathrm{I}^{\otimes (n-j)}$,
  and $g_j = \mathrm{T}_{\leq j} f^{=S}$. We argue that $\norm{g_{j+1}}_{2,\nu}^2\leq (1-\beta)^2 \norm{g_j}_{2,\nu}^2$,
  which by induction finishes the proof. For each $j$ and $z\in [m]^{[n]\setminus \set{j+1}}$,
  let $g_j[z]\colon [m]\to\mathbb{R}$ be the restriction of the function $g_j$ wherein all coordinates,
  except $j+1$, have been fixed according to $z$.
  \[
  \norm{g_{j+1}}_{2,\nu}^2
  =\Expect{{\bf z}\sim\prod\limits_{i\neq j+1}\nu_i}{\norm{\mathrm{T}_{\mu_{j+1}}(g_j[{\bf z}])}_{2,\nu_{j+1}}^2}
  \leq
  \Expect{{\bf z}\sim\prod\limits_{i\neq j+1}\nu_i}{(1-\beta)^2\norm{g_j[{\bf z}]}_{2,\nu_{j+1}}^2}
  =(1-\beta)^2 \norm{g_j}_{2,\nu}^2.
  \]
  In the second transition, we used the fact that the average of $g_j[z]$ according to $\nu_{j+1}$ is $0$ for all $z$.
\end{proof}

Combining the two claims, we get the following lemma.
\begin{lemma}\label{lem:high_deg_dies_prod_connected}
  For all $\alpha>0$, there is $\eps>0$ such that the following holds.
  Let $\mu$ be a measure on $[m]\times [m]$, such that
  \begin{enumerate}
    \item The marginal distribution of $\mu$ on both coordinates is the same, denote it by $\nu$.
    \item The probability of each atom in $\mu$ is at least $\alpha$.
    \item $\mu$ is connected.
  \end{enumerate}
  Let $n,d\in\mathbb{N}$, and consider  $\mathrm{T}: L^{2}([m]^n,\nu^{\otimes n})\to L^2([m]^n,\nu^{\otimes n})$
  defined as $\mathrm{T} = \mathrm{T}_{\mu}^{\otimes n}$.
  Then for all $f\in V_{>d}([m]^n,\nu^{\otimes n})$ it holds that $\norm{\mathrm{T} f}_{2,\nu^n}\leq (1-\beta)^d\norm{f}_{2,\nu^n}$.
\end{lemma}

\subsection{Influences}\label{sec:influences_product}
In this section, we define influences in product spaces, and state several well-known statements.
Our exposition will be very brief and we defer the reader to~\cite[Chapter 8]{Od} for a more thorough presentation.
Let $([m]^n, \nu^{\otimes n})$ be a product space. For $f\colon [m]^n\to \mathbb{R}$, we define
the averaging operator $\mathrm{E}_i$ according to the $i$th coordinate as
$\mathrm{E}_i f(x) = \Expect{{\bf y}_i\sim\nu_{\vec{k}}}{f(x_{-i},{\bf y}_i)}$.
We then define the Laplacian operator of the $i$th coordinate as $\mathrm{L}_i f(x) = f(x) - \mathrm{E}_i f(x)$,
and the Laplacian $\mathrm{L}$ as $\mathrm{L} f(x) = \sum\limits_{i=1}^n \mathrm{L}_i f(x)$.

\begin{definition}
  The influence of variable $i$ on $f\colon ([m]^n, \nu^{\otimes n})\to\mathbb{R}$ is
  $I_i[f] = \norm{L_i f}_2^2$. The total influence of $f$ is $I[f] = \sum\limits_{i=1}^{n} I_i[f]$.
\end{definition}

\begin{fact}\label{fact:inf_laplacian}
  $I[f] = \inner{f}{\mathrm{L} f}$.
\end{fact}
%\begin{proof}
%First, we note that for all $i$ it holds that $\inner{\mathrm{E}_i f}{\mathrm{L}_i f} = 0$. Indeed, as $\mathrm{E}_i f(x)$ does not depend
%on $x_i$, we have that
%\begin{align*}
%\inner{\mathrm{E}_i f}{\mathrm{L}_i f}
%&=\Expect{{\bf x}\sim\nu_{\vec{k}}^{\otimes n}}
%{\mathrm{E}_i f({\bf x}) \mathrm{L}_i f({\bf x})}\\
%&=\Expect{{\bf x}_1,\ldots,{\bf x}_{i-1},{\bf x}_{i+1},\ldots,{\bf x}_n\sim\nu_{\vec{k}}}
%{\mathrm{E}_i f({\bf x}_1,\ldots,{\bf x}_{i-1},1,{\bf x}_{i+1},\ldots,{\bf x}_n)
%\Expect{{\bf x}_i\sim\nu_{\vec{k}}}{\mathrm{L}_i f({\bf x})}}=0.
%\end{align*}
%Thus,
%  \[
%  I[f]
%  = \sum\limits_{i=1}^{n} \norm{\mathrm{L}_i f}_2^2
%  = \sum\limits_{i=1}^{n} \inner{f}{\mathrm{L}_i f} - \inner{\mathrm{E}_i f}{\mathrm{L}_i f}
%  = \sum\limits_{i=1}^{n} \inner{f}{\mathrm{L}_i f}
%  = \inner{f}{\mathrm{L} f}.\qedhere
%  \]
%\end{proof}

We will also use the Efron-Stein decomposition of product spaces.
For $S$, we denote by
$V_{S}([m]^n,\nu^{\otimes n})$ the subspace of functions $f\colon [m]^n\to\mathbb{R}$ depending
only on the coordinates of $S$, and by $V_{=S}([m]^n,\nu^{\otimes n})\subseteq V_{S}([m]^n,\nu^{\otimes n})$
the subspace of functions that are also orthogonal to all $V_T([m]^n,\nu^{\otimes n})$ for $T\not\subseteq S$.
Then it is well known that any $f\colon [m]^n\to\mathbb{R}$ may be uniquely written as
$f = \sum\limits_{S\subseteq [n]} f^{=S}$, where $f^{=S} \in V_{=S}([m]^n,\nu^{\otimes n})$.
We also define the degree decomposition as $f^{=d} = \sum\limits_{\card{S} = d} f^{=S}$, and
$f^{\leq d} = f^{=0} + \ldots + f^{=d}$.
\begin{definition}
  The degree $d$ influence of variable $i$ on $f\colon ([m]^n, \nu^{\otimes n})\to\mathbb{R}$ is
  $I_i^{\leq d}[f] = \norm{(\mathrm{L}_i f)^{\leq d}}_2^2$.
\end{definition}

\begin{fact}\label{fact:influence_formula}
Let $f\colon [m]^n\to\mathbb{R}$, $i\in [n]$ and $d\in\mathbb{N}$. Then
$I_i[f] = \sum\limits_{S\ni i}{\norm{f^{=S}}_2^2}$, and
$I_i^{\leq d}[f] = \sum\limits_{\substack{S\ni i\\\card{S}\leq d}}{\norm{f^{=S}}_2^2}$.
\end{fact}

\begin{claim}\label{claim:noisy_to_low_deg}
  Suppose $d\in\mathbb{N}$, $\beta>0$ are such that $d\beta\leq \frac{1}{8}$, and let $\mu$ be a connected distribution
  on $[m]\times [m]$ whose marginals are both equal to $\nu(a)$, such that $\mu(a,a)\geq (1-\beta)\nu(a)$.
  Then for all $f\colon([m]^n,\nu^{\otimes n})\to\mathbb{R}$ and $i\in [n]$ it holds that
  $I_i^{\leq d}[f]\leq 2I_{i}[\mathrm{T}_{\mu} f]$.
\end{claim}
\begin{proof}
  Consider the spaces $V_{=S}([m]^n,\nu^{\otimes n})$ defining the Efron-Stein decomposition of $f$, and note that
  both $\mathrm{T}_{\mu}$ and $\mathrm{S}_{\mu} = \mathrm{T}_{\mu}^{*}\mathrm{T}_{\mu}$ preserves each one of them.
  We may therefore write $V_{=S}([m]^n,\nu^{\otimes n}) = \bigoplus V_{=S}^{\theta}([m]^n,\nu^{\otimes n})$, where
  $V_{=S}^{\theta}([m]^n,\nu^{\otimes n})$ is an eigenspace of $\mathrm{S}_{\mu}$ with eigenvalue $\theta$, and note that
  all eigenvalues $\theta$ are non-negative.
  Thus, we write $f = \sum\limits_{S\subseteq [n],\theta} f^{=S,\theta}$, where
  $f^{=S,\theta}\in V_{=S}^{\theta}([m]^n,\nu^{\otimes n})$ so by Fact~\ref{fact:influence_formula} and Parseval
  \begin{align}\label{eq:14}
  I_{i}[\mathrm{T}_{\mu} f]
  =
  \sum\limits_{S\ni i} \norm{\mathrm{T}_{\mu} f^{=S}}_2^2
  =
  \sum\limits_{S\ni i} \inner{f^{=S}}{\mathrm{S}_{\mu} f^{=S}}
  =
  \sum\limits_{S\ni i,\theta} \sum\limits_{\theta}\inner{f^{=S,\theta}}{\theta f^{=S,\theta}}
  &=
  \sum\limits_{S\ni i,\theta} \theta \norm{f^{=S,\theta}}_2^2\notag\\
  &
  \geq\sum\limits_{\substack{S\ni i,\theta\\ \card{S}\leq d}} \theta\norm{f^{=S,\theta}}_2^2,
  \end{align}
  and we show that $\theta\geq \half$ for all $\card{S}\leq d$ and $\theta$ such that $V_{=S}^{\theta}([m]^n,\nu^{\otimes n})\neq \set{0}$.
  Indeed, let $g\in V_{=S}^{\theta}([m]^n,\nu^{\otimes n})$ be non-zero, and let $x\in[m]^n$ be the point maximizing $\card{g(x)}$.
  Denote by $\mu'$ the distribution corresponding to $\mathrm{S}_{\mu}$ (i.e.~taking $({\bf a},{\bf b})\sim\mu$, then $({\bf a'},{\bf b'})\sim\mu$
  conditioned on ${\bf b'} = {\bf b}$ and outputting $({\bf a},{\bf a'})$).  Then as $g$ is a $S$-junta,
  we have that
  \begin{align*}
  &\theta \card{g(x)}
  =
  \card{\mathrm{S}_{\mu} g(x)}\\
  &\geq \card{ \cProb{({\bf x},{\bf y})\sim\mu'^{\otimes n}}{{\bf x} = x}{{\bf y}_S = x_S} g(x) +
  \cProb{({\bf x},{\bf y})\sim\mu'^{\otimes n}}{{\bf x} = x}{{\bf y}_S \neq x_S}\cExpect{({\bf x},{\bf y})\sim\mu'^{\otimes n}}{{\bf x} = x, {\bf y}_S\neq x_S}{g({\bf y})}}\\
  &\geq
  (1-2\cProb{({\bf x},{\bf y})\sim\mu'^{\otimes n}}{{\bf x} = x}{{\bf y}_S \neq x_S})\card{g(x)}.
  \end{align*}
  As by the union bound
  $\cProb{({\bf x},{\bf y})\sim\mu^{\otimes n}}{{\bf x} = x}{{\bf y}_S \neq x_S}\leq 2\beta\card{S}\leq 2\beta d\leq \frac{1}{4}$, it follows that
  $\theta\geq \half$. Plugging this into~\eqref{eq:14} and using
  \[
  \sum\limits_{\substack{S\ni i,\theta\\ \card{S}\leq d}} \norm{f^{=S,\theta}}_2^2
  = \sum\limits_{\substack{S\ni i\\ \card{S}\leq d}} \norm{f^{=S}}_2^2
  = I_i^{\leq d}[f],
  \]
  where the first transition is by Parseval and the second transition is by Fact~\ref{fact:influence_formula}, finishes the proof.
\end{proof}

\subsection{The noise operator}\label{sec:std_noise}
Let $L^{2}([m]^n,\nu_1\times\ldots\times\nu_n)$ be a product space.
\begin{definition}
  For $\rho>0$, the standard noise operator $\mathrm{T}_{\rho}$ is defined as follows.
  For each $x\in[m]^n$, we have a distribution $\mathrm{T}_{\rho} x$ defined as:
  to sample ${\bf y}\sim \mathrm{T}_{\rho} x$, for each $i\in[n]$ independently, take
  ${\bf y}_i = x_i$ with probability $\rho$, and otherwise sample ${\bf y}_i\sim\nu_i$.

  We then define the operator $\mathrm{T}_{\rho}\colon L^{2}([m]^n,\nu_1\times\ldots\times\nu_n)\to L^{2}([m]^n,\nu_1\times\ldots\times\nu_n)$
  as
  \[
  \mathrm{T}_{\rho} f(x) =
  \Expect{{\bf y}\sim \mathrm{T}_{\rho} x}{ f({\bf y})}.
  \]
\end{definition}

\begin{fact}\label{fact:noise_operator_ev}
  Let $f\in V_{=S}([m]^n,\nu_1\times\ldots\times\nu_n)$. Then
  $\mathrm{T}_{\rho} f = \rho^{\card{S}} f$.
\end{fact}

\section{Deferred proofs}\label{sec:deferred_pfs}
\subsection{Proof of Claim~\ref{claim:bd_inf_coup_low_deg}}
The main component in the proof of Claim~\ref{claim:bd_inf_coup_low_deg} is the following claim.
\begin{claim}\label{claim:upperbd_inf_junta_coup}
    Let $J\subseteq [n]$ be of size at most $d$, let $\mathcal{U}_{\vec{k}}\subseteq[m]^n$ be $\alpha$-balanced, and let $\mathcal{C}$ be a $(\alpha,\zeta)$-coupling between
    $\mathcal{U}_{\vec{k}}\subseteq [m]^n$ and $([m]^n, \nu_{\vec{k}}^{\otimes n})$. Then for all $g\in V_{=J}(\mathcal{U}_{\vec{k}})$,
    we have $I[\mathrm{T}_{\mathcal{C}} g]\leq \left(\frac{4d^2}{\alpha^2}\zeta n + d^2\right)\norm{g}_2^2$.
  \end{claim}
  \begin{proof}
  Fix $i$, and denote $h=\mathrm{T}_{\mathcal{C}} g$. For $i\in J$, we use the trivial bound
  $I_i[h]\leq \norm{h}_2^2\leq\norm{g}_2^2$, and the rest of the argument is devoted to bounding
  the influence of $i\not\in J$.

  By Cauchy-Schwarz we have $I_i[h]\leq \Expect{{\bf y},{\bf y'}\sim\nu_{\vec{k}}}{(h({\bf y}_{-i}, {\bf y'}_i) - h({\bf y}))^2}$.
  We sample ${\bf y}$, ${\bf y'}$, and then take ${\bf y''} = ({\bf y}_{-i}, {\bf y'}_i)$. We now sample
  ${\bf x}$ as $({\bf x},{\bf y}(0))\sim\mathcal{C}$ conditioned on ${\bf y}(0) = {\bf y}$
  and ${\bf x''}$ as $({\bf x''},{\bf y}(0))\sim\mathcal{C}$ conditioned on ${\bf y}(0) = {\bf y''}$. Then we get that
  \[
  I_i[h]\leq
  \Expect{{\bf y}, {\bf y'}}
  {\left(\cExpect{{\bf x}, {\bf x''}}{{\bf y}, {\bf y'}}{g({\bf x}) - g({\bf x''})}\right)^2}
  \leq
  \Expect{\substack{{\bf y}, {\bf y'}\\ {\bf x}, {\bf x''}}}
  {\left(g({\bf x}) - g({\bf x''})\right)^2}.
  \]
  Denote by ${\bf j}_1$, \ldots, ${\bf j}_{2\ell}$ the set of coordinates on which ${\bf x}$ and ${\bf x'}$ differ. We
  stress here that $\ell$ is also a random variable, and by the third property of couplings its expectation is at most
  $2n\zeta$.

  We may find a matching between them, say ${\bf j}_{2s+1}$ and ${\bf j}_{2s+2}$ for $s=0,\ldots,\ell-1$, such that
  performing the transpositions $\pi_{{\bf j}_{2s+1},{\bf j}_{2s+2}}$ in the order $s=0,\ldots,\ell-1$ on ${\bf x}$
  gives ${\bf x'}$. We then permute these pairs randomly.
  Denote ${\bf x}(0) = {\bf x}$, and inductively ${\bf x}(s+1) = {}^{\pi_{{\bf j}_{2s+1},{\bf j}_{2s+2}}}{\bf x}(s)$.
  We note that if ${\bf j}_{2s+1},{\bf j}_{2s+2}\not\in J$, then $g({\bf x}(s+1)) = g({\bf x}(s))$. We thus get
  \begin{align}
  \Expect{\substack{{\bf y}, {\bf y'}\\ {\bf x}, {\bf x''}}}
  {\left(g({\bf x}) - g({\bf x''})\right)^2}
  &=\Expect{\substack{{\bf y}, {\bf y'}\\ {\bf x}, {\bf x''}}}
  {
  \left(\sum\limits_{s=0}^{\ell-1}
  (g({\bf x}(s)) - g({\bf x}(s+1)))
  1_{J\cap\set{{\bf j}_{2s+1},{\bf j}_{2s+2}}\neq\emptyset}\right)^2}\notag\\
  &\leq
  \Expect{\substack{{\bf y}, {\bf y'}\\ {\bf x}, {\bf x''}}}
  {
  \left(\sum\limits_{s=0}^{\ell-1}(g({\bf x}(s)) - g({\bf x}(s+1)))^2\right)
  \left(\sum\limits_{s=0}^{\ell-1} 1_{J\cap\set{{\bf j}_{2s+1},{\bf j}_{2s+2}}\neq\emptyset}\right)}\notag\\
  &\leq
  d\Expect{\substack{{\bf y}, {\bf y'}\\ {\bf x}, {\bf x''}}}
  {\sum\limits_{s=0}^{\ell-1}(g({\bf x}(s)) - g({\bf x}(s+1)))^2}
  .\label{eq:19}
  \end{align}
  %Consider the first expectation, and condition on $\ell$; by symmetry
%  $\cExpect{}{\ell}{1_{J\cap\set{{\bf j}_{2s+1},{\bf j}_{2s+2}}}}\leq \frac{\card{J}}{n}$,
%  so the first expectation is at most $\frac{d}{n}\Expect{}{\ell} \leq 2d\zeta$.
  Conditioning, we get that
  \begin{align}
  ~\eqref{eq:19}
  \leq
  &d
  \Expect{\ell}
  {\sum\limits_{s=0}^{\ell-1}
  \cExpect{\substack{{\bf y}, {\bf y'}\\ {\bf x}, {\bf x''}}}
  {\ell, {\bf j}_{2s+1},{\bf j}_{2s+2}\neq i}
  {(g({\bf x}(s)) - g({\bf x}(s+1)))^2 }}\notag\\\label{eq:21}
  &+
  d\Expect{\ell}{
  \cExpect{\substack{{\bf y}, {\bf y'}\\ {\bf x}, {\bf x''}}}{\ell}{
  \sum\limits_{s=0}^{\ell-1}
  1_{i\in\set{{\bf j}_{2s+1},{\bf j}_{2s+2}}}(g({\bf x}(s)) - g({\bf x}(s+1)))^2}
  },
  \end{align}
  and we analyze each expectation separately. For the first expectation,
  conditioned on $\ell$, by symmetry, for each $s$, the marginal distribution of ${\bf x}(s)$ is uniform
  over $\mathcal{U}_{\vec{k}}$.
  Conditioned on ${\bf x}(s)$, ${\bf j}_{2s+1},{\bf j}_{2s+2}\neq i$, by symmetry, for all $j,j'$ we have that the probability that
  ${\bf j}_{2s+1}=j,{\bf j}_{2s+2}=j'$ is at most $\frac{1}{(\alpha n-1)(\alpha n-2)}$, thus the first conditional expectation
  in~\eqref{eq:21} is at most
  \[
  \frac{2}{\alpha^2}
  \Expect{\substack{{\bf j}\in [n],{\bf j'}\in[n] \\ {\bf x}}}{(g({}^{\pi_{{\bf j},{\bf j'}}} {\bf x}) - g({\bf x}))^2}
  =\frac{2}{n\alpha^2} I[g]
  \leq \frac{2d}{n\alpha^2}\norm{g}_2^2,
  \]
  where we used Claim~\ref{claim:trivial_inf_bd}. Thus, the contribution of the first term in~\eqref{eq:21}
  is at most $\frac{2d^2}{\alpha^2n}\norm{g}_2^2\Expect{}{\ell}\leq \frac{4d^2}{\alpha^2}\zeta\norm{g}_2^2$. As for the second term, for at most one $s$ we have that
  $1_{i\in\set{{\bf j}_{2s+1},{\bf j}_{2s+2}}} = 1$; denote this as a random variable ${\bf s}$, and note
  that by symmetry it is uniform over ${\ell}$, so we may as well assume that ${\bf s}=1$. Thus, this expectation is
  at most $I_i[g]$. We therefore get that
  $I_i[h]\leq ~\eqref{eq:21} \leq \frac{4d}{\alpha^2}\zeta\norm{g}_2^2 + dI_i[g]$, and summing over $i$ we get that
  \[
  I[h]\leq \frac{4d^2}{\alpha^2}\zeta n\norm{g}_2^2 + dI[g]\leq \left(\frac{4d^2}{\alpha^2}\zeta n + d^2\right)\norm{g}_2^2,
  \]
  where we used Claim~\ref{claim:trivial_inf_bd}.
  \end{proof}

  We are now ready to prove Claim~\ref{claim:bd_inf_coup_low_deg}.
  \begin{proof}[Proof of Claim~\ref{claim:bd_inf_coup_low_deg}]
  First, note that by Fact~\ref{fact:inf_laplacian}
    \[
    I[\mathrm{T}_{\mathcal{C}} g]
    = \inner{\mathrm{L}\mathrm{T}_{\mathcal{C}} g}{\mathrm{T}_{\mathcal{C}} g}
    =\inner{\mathrm{T}_{\mathcal{C}}^{*}\mathrm{L}\mathrm{T}_{\mathcal{C}} g}{g}.
    \]
  Denote $R = \mathrm{T}_{\mathcal{C}}^{*}\mathrm{L}\mathrm{T}_{\mathcal{C}}$. Note that $\mathcal{C}$ is symmetric,
  and so by Claim~\ref{claim:sym_implies_commute} commutes with $S_n$, and $\mathrm{L}$ also commutes with $S_n$, we
  get that $R\colon L^2(\mathcal{U}_{\vec{k}}) \to L^2(\mathcal{U}_{\vec{k}})$ commutes with the action of $S_n$, and
  therefore by Claim~\ref{claim:preserve_rt} preserves degrees. As $R$ is self-adjoint, we may then decompose
  $V_{d}(\mathcal{U}_{\vec{k}})$ as a sum of eigenspaces $\bigoplus_{\theta} V_{d}^{\theta}(\mathcal{U}_{\vec{k}})$.
  Thus, if $\theta^{\star}$ is the maximum eigenvalue
  of $R$ on $V_{d}(\mathcal{U}_{\vec{k}})$, then we get that $I[\mathrm{T}_{\mathcal{C}} g] \leq \theta^{\star}\norm{g}_2^2$.

  To bound $\theta^{\star}$, note that by Claim~\ref{claim:only_higher_degs} we get that $R$ preserves juntas and
  degrees, so $V_{=J}(\mathcal{U}_{\vec{k}})$ are invariant spaces under $R$. We may therefore
  write it as a sum of eigenspaces $\bigoplus_{\theta} V_{=J}^{\theta}(\mathcal{U}_{\vec{k}})$, and get as we have seen
  previously that the spaces $\bigoplus_{\card{J}\leq d} V_{=J}^{\theta}(\mathcal{U}_{\vec{k}})$ are the eigenspaces of $R$
  on $V_{d}(\mathcal{U}_{\vec{k}})$. Thus, there is $\card{J}\leq d$ such that $V_{=J}^{\theta^{\star}}(\mathcal{U}_{\vec{k}})\neq \set{0}$,
  and we take a non-zero $g^{\star}\in V_{=J}^{\theta^{\star}}(\mathcal{U}_{\vec{k}})$. We now get from Claim~\ref{claim:upperbd_inf_junta_coup} that
  $\theta^{\star}\leq\frac{I[\mathrm{T}_{\mathcal{C}} g^{\star}]}{\norm{g^{\star}}_2^2} \leq \frac{4d^2}{\alpha^2}\zeta n + d^2$, and we are done.
  \end{proof}
\end{document}